%% file: main.tex
\documentclass{lmcs} 


\input init

\input comm

\begin{document}


\title[Context-Bounded Model Checking for POWER]{Context-Bounded Model Checking for POWER}
\titlecomment{{\lsuper*} A preliminary version of this paper appeared as
at TACAS'17~\cite{abdullaABN17}.}

\author[P.A.~Abdulla]{Parosh Aziz Abdulla\rsuper a}	
\address{\lsuper{a,b}Uppsala University, Sweden}	
\email{\{parosh, mohamed\_faouzi.atig\}@it.uu.se}  
%
\author[M.F.~Atig]{Mohamed Faouzi Atig\rsuper b}	
\author[A.~Bouajjani]{Ahmed Bouajjani\rsuper c}	
\address{ \lsuper{c}IRIF Universit\'e Paris Diderot - Paris 7, France}	
\email{abou@liafa.univ-paris-diderot.fr}  

\author[T.P.~Ngo]{Tuan Phong Ngo\rsuper{d}}
\address{ \lsuper{d}Hanoi University of Science and Technology, Vietnam (Corresponding author)}	
\email{phong.ngotuan@hust.edu.vn}


\begin{abstract}
We propose an under-approximate reachability analysis algorithm for programs running under the POWER memory model, in the spirit of the work on context-bounded analysis initiated by Qadeer et al.\ in 2005 for detecting bugs in concurrent programs (supposed to be running under the classical SC model). To that end, we first introduce a new notion of context-bounding that is suitable for reasoning about computations under POWER, which generalizes the one defined by Atig et al.\ in 2011 for the TSO memory model. Then, we provide a polynomial size reduction of the context-bounded state reachability problem under POWER  to the same problem under SC: Given an input concurrent program $\mathcal{P}$, our method  produces a concurrent program $\mathcal{P}'$ such that, for a fixed number of context switches, running $\mathcal{P}'$ under SC yields the same set of reachable states as running $\mathcal{P}$ under POWER. The generated program $\mathcal{P}'$ contains the same number of processes as $\mathcal{P}$ plus two additional processes, and operates on the same data domain. By leveraging the standard model checker CBMC, we have implemented a prototype tool and applied it on a set of benchmarks, showing the feasibility of our approach.
\end{abstract}

\maketitle


\section{Introduction}
For performance reasons, modern multi-processors may reorder memory access operations. This is due to complex buffering and caching mechanisms that make the response memory queries (load operations) faster, and allow to speed up computations by parallelizing independent operations and computation flows. Therefore, operations may not be visible to all processors at the same time, and they are not necessarily seen in the same order by different processors (when they concern different variables). The only model where all operations are visible immediately to all processors is the Sequential Consistency (SC) model \cite{lamport-79} which corresponds to the standard interleaving semantics where the program order between operations of a same processor is preserved. Modern   architectures adopt  weaker models (in the sense that they allow more behaviours) due to the relaxation in various ways of the program order. Examples of such weak  models are TSO adopted in Intel x86 machines for instance, POWER adopted in PowerPC machines, or the model adopted in ARM machines.  

Apprehending the effects of all the relaxations allowed in such models is extremely hard. For instance, while TSO allows reordering stores past loads (of different variables) reflecting the use of store buffers, a model such as POWER allows reordering of all kinds of store and load operations under quite subtle conditions. A lot of work has been devoted to the definition of formal models that  accurately capture  the program semantics corresponding to models such as TSO~\cite{SSONM2010,DBLP:conf/tphol/OwensSS09} and POWER~\cite{DBLP:conf/pldi/SarkarSAMW11,DBLP:conf/pldi/SarkarMOBSMAW12,DBLP:journals/toplas/AlglaveMT14,DBLP:conf/cav/Mador-HaimMSMAOAMSW12}. Still, programming against weak memory models is a hard and error prone task. 
Therefore, developing formal verification approaches under weak memory models is of paramount importance. In particular, it is crucial in this context to have efficient algorithms for automatic bug detection. This paper addresses precisely this issue and presents an algorithmic approach for checking  state reachability in concurrent  programs running under the POWER semantics as defined in \cite{DBLP:conf/pldi/SarkarSAMW11,DBLP:conf/pldi/SarkarMOBSMAW12,DM14}. 

The verification of concurrent programs under weak memory models is known to be complex. Indeed, encoding the buffering and storage mechanisms used in these models leads in general to complex, infinite-state formal operational models involving unbounded data structures like FIFO queues (or more generally unbounded partial order constraints). For the case of TSO, 
efficient and precise encodings of the effects of its storage mechanism have been designed recently \cite{DBLP:conf/tacas/AbdullaACLR12,DBLP:conf/concur/AbdullaABN16,lmcs:4228}. However, it is not clear how to define such precise and practical encodings for POWER.  

In this paper, we consider an alternative approach. We investigate the issue of defining approximate analysis. Our approach consists in introducing a parametric under-approximation schema in the spirit of context-bounding \cite{DBLP:conf/tacas/QadeerR05,MQ07,DBLP:journals/fmsd/LalR09,DBLP:conf/cav/TorreMP09,ABP2011}. Context-bounding has been proposed in \cite{DBLP:conf/tacas/QadeerR05} as a suitable approach for efficient bug detection in multithreaded programs. Indeed, for concurrent programs, a bounding concept that provides both good coverage and scalability must be based on aspects related to the interactions between concurrent components. It has been shown experimentally that concurrency bugs usually show up  after a small number of context switches \cite{MQ07}.

In the context of weak memory models, context-bounded analysis has been extended in \cite{ABP2011} to the case of programs running under TSO. The work we present here aims at extending this approach to the case of POWER. This extension is actually very challenging due to the complexity of POWER and requires developing new techniques that are  different from, and much more involved than, the ones used for the case of TSO. First, we introduce a new concept of bounding that is suitable for POWER. Intuitively, the architecture of POWER is  similar to a distributed system with a replicated memory, where each processor has its own replica, and where operations are propagated between replicas according to some specific protocols. Our bounding concept is based on this architecture. We consider that a computation is divided in a sequence of ``contexts'', where a context is a computation segment for which there is precisely one {\em active} processor. All actions within a context are either operations issued by the active processor, or propagation actions performed by its storage subsystem.  
Then, in our analysis,  we consider  only computations that have a number of contexts that is less or equal than some given bound. Notice that while we bound the number of contexts in a computation, we do not put any bound on the lengths of the contexts, nor on the size of the storage system.

 We prove that for every bound  $\contextnum$, and for every concurrent program $\prog$, it is possible to construct, using code-to-code translation, another concurrent
program $\sprog$ such  that for every $\contextnum$-bounded computation $\run$ in $\prog$ running
under the POWER semantics there is a corresponding $\contextnum$-bounded computation
$\srun$ of $\sprog$ running under the SC semantics that reaches the same set of states and vice-versa.
Thus,  the context-bounded state reachability problem for $\prog$ can be reduced to the context-bounded state reachability problem for $\sprog$ under SC.
 We show that the program $\sprog$ has the same number of processes as $\prog$ plus two additional processes, and only  $O(\sizeof\procset\cdot\sizeof\varset\cdot\contextnum+\sizeof\regset )$ additional  shared variables and local registers compared to  $\prog$, where $\sizeof\procset$ is the number of processes, $\sizeof\varset$ is the number of shared variables, and $\sizeof\regset$ is the number of local registers in $\prog$. Furthermore, the obtained program has the same type of data structures and variables  as the original one. As a
consequence, we obtain for instance that for finite-data programs, the context-bounded analysis of  programs running under the POWER semantics is decidable.
 Moreover, our code-to-code translation   allows to leverage existing verification tools for concurrent programs to carry out verification of safety properties under POWER.

To show the applicability of our approach, we have implemented our reduction in a prototyping tool, namely {\sf Power2SC}. We have used \textsf{CBMC} version 5.1  \cite{DBLP:conf/tacas/ClarkeKL04} as the backend tool for solving SC reachability queries. We have carried out several experiments showing the efficiency of our approach. Our experimental results confirm the assumption that concurrency bugs manifest themselves within small bounds of context switches. They also confirm that our approach based on context-bounding is more efficient and scalable than approaches based on bounding sizes of computations and of storage systems.

\paragraph{\bf Related work.}
There has been a lot of work on automatic  verification  of programs running under weak memory models, based on precise, under-approximate, and abstract analyses, e.g.,~\cite{DBLP:conf/pldi/LiuNPVY12,KVY2010,KVY2011,ABP2011,eps402285,fmcad16,DBLP:conf/sas/DanMVY13,DBLP:conf/tacas/AbdullaACLR12,DBLP:conf/esop/AbdullaAP15,BM2008,BSS2011,DBLP:conf/esop/BouajjaniDM13,BAM07,yang-gopalakrishnan-PDPS04,tacas15:tso,Zhang:pldi15,DBLP:conf/oopsla/DemskyL15,AlglaveKT13,DBLP:conf/ictac/TravkinW16,DBLP:conf/fm/LahavV16,Dan201762,AbdullaAJN18,Kokologiannakis18,NorrisD16,Huang016,LeonFHM17,LeonFHM18,GavrilenkoLFHM19,DBLP:journals/pacmpl/AbdullaAJN18,DBLP:conf/pldi/AbdullaAAK19,DBLP:journals/pacmpl/RaadDRLV19,DBLP:conf/pldi/Kokologiannakis19,DBLP:conf/pldi/LahavM19,DBLP:conf/ppopp/OuD17,DBLP:journals/toplas/NorrisD16}. While most of these  works concern TSO, only  a few  of them address the safety verification problem under POWER (e.g., \cite{DBLP:conf/cav/AbdullaAJL16,AlglaveKT13,eps402285,DBLP:conf/esop/AlglaveKNT13,DBLP:journals/toplas/AlglaveMT14,LeonFHM17,LeonFHM18,GavrilenkoLFHM19}). The paper~\cite{DM14}   addresses the different issue of checking robustness against POWER, i.e., whether a program has the same (trace) semantics for both POWER and SC.  

The \textsf{Goto-Instrument}~\cite{DBLP:conf/esop/AlglaveKNT13,gotoinstrument} extends the \textsf{CBMC} framework by taking into account  weak memory models including TSO and POWER. While this approach uses reductions to SC analysis, it is conceptually and technically  different from ours. 
\textsf{Goto-Instrument} uses 
axiomatic model for POWER~\cite{DBLP:journals/toplas/AlglaveMT14}
while 
we use the operational one.
Using the axiomatic model, 
\textsf{Goto-Instrument}
builds all abstract event structures that  contain potential cycles breaking the memory model.  
The potential cycles are then instrumented and validated under SC.
Instead of detecting the appearance of these cycles,
our approach checks state reachability problem (c.f.\ Section~\ref{bounded:reachability:section}).
The work in~\cite{AlglaveKT13} develops a verification technique combining partial orders with bounded model checking, that is applicable to various weak memory models including  TSO and POWER. However, these techniques are not anymore supported by the latest version of \textsf{CBMC}. The work in~\cite{DBLP:conf/cav/AbdullaAJL16} develops stateless model checking techniques under POWER. 
In Section~\ref{experiment:section}, we compare the performances of our approach with those of~\cite{DBLP:conf/esop/AlglaveKNT13} and \cite{DBLP:conf/cav/AbdullaAJL16}. The  tool \textsf{PPCMEM}~\cite{DBLP:conf/pldi/SarkarSAMW11} operates on small litmus tests under the POWER semantics. Our tool can handle in an efficient and precise way such litmus tests. 

The \textsf{Cseq} tool~\cite{TomascoI0TP15a, InversoT0TP14, fmcad16,TomascoN0TP17,Nguyen0TP16,eps402285} presents   a new verification approach, based on  code-to-code translations, for programs running under  SC, TSO, and PSO.   Our approach  and the ones proposed in \cite{fmcad16,eps402285,TomascoI0TP15a} are orthogonal since we are using different bounding parameters. To be more precise, we bound the number of contexts that follows the spirit of Qadeer et al.~\cite{Qadeer08}
while   Tomasco et al.~\cite{fmcad16,eps402285,TomascoI0TP15a} bound the number of write operations. 
Although they  discuss the extension of their approach to programs running under  POWER~\cite{eps402285}, the detailed formalization and the implementation of their extension   are kept for future work.

Recently, \textsf{DARTAGNAN}~\cite{GavrilenkoLFHM19} and \textsf{PORTHOS}~\cite{LeonFHM17} implement  new approaches for efficiently verifying programs running under weak memory models using SMT encoding.
These tools can handle different memory models such as TSO, POWER,  and ARM.
Similar to our approach, \textsf{DARTAGNAN}~\cite{GavrilenkoLFHM19} checks state reachability problem. Meanwhile, given two memory models, \textsf{PORTHOS}~\cite{LeonFHM17} tries to find a state that can be reachable in one model but unreachable in the other.
Unfortunately, we were not able  to compare our tool with    \textsf{DARTAGNAN} and \textsf{PORTHOS}.The reason is that  our tool accepts C/Pthreads input programs
while  \textsf{DARTAGNAN} and \textsf{PORTHOS} do not.


\section{Concurrent Programs and Semantics}
\label{syntax:semantic:section}
In this section, we  first introduce some notations and definitions that we will use throughout this paper. Then,  
we present   the syntax  we use for {\it concurrent programs} and the POWER operational semantics including the transition system it induces as    in~\cite{DM14,DBLP:conf/pldi/SarkarSAMW11}.
Finally, we give
our definition
of context-bounding  and
an example of a context-bounded computation under the POWER semantics.

\subsection{Preliminaries}
Consider sets $\aset$ and $\bset$.
We use $\mapingsover\aset\bset$ to denote the set
of (partial) functions from $\aset$ to $\bset$, and write
$f:\aset\rightarrow\bset$ to indicate that $\fun\in\mapingsover\aset\bset$.
We write $\fun(\aelem)=\myundef$ to denote that
$\fun$ is undefined for $\aelem$.
We use 
$\fun[\aelem\assigned\belem]$ to denote the function
$\gfun$ such that $\gfun(\aelem)=\belem$ and
$\gfun(\xx)=\fun(\xx)$ if $\xx\neq\aelem$.
We will use a function $\gen$ which, for a given set $\aset$,
returns an arbitrary element $\genof\aset\in\aset$.
For integers $\ii,\jj$, we use $[\ii..\jj]$ 
to denote the set $\set{\ii,\ii+1,\ldots,\jj}$.
We use $\wordsover\aset$ to denote the set
of finite words over $\aset$.
For words $\word_1,\word_2\in\wordsover\aset$, we use $\word_1\app\word_2$
to denote the concatenation of $\word_1$ and $\word_2$.

\subsection{Syntax}
Fig.~\ref{program_syntax} gives the grammar for a small but 
general language that we use for defining
concurrent programs.
A similar grammar has been widely used 
in several related work (e.g., \cite{InversoT0TP14,TomascoI0TP15a,ABP2011}).

A program $\prog$ first declares  a set
$\varset$ of (shared) variables followed by the code of a set 
$\procset$ of processes.
Each process $\proc$ has a finite set $\regsetof\proc$ of (local) {\it registers}.
We assume w.l.o.g. that the sets of  registers
of the different processes are disjoint,
and define $\regset:=\cup_\proc\regsetof\proc$.
The code of each process $\proc\in\procset$ 
starts by  declaring a set of  
registers followed by  a sequence of instructions.
For the sake of simplicity, we assume that
the data domain of both the shared variables and  registers 
is a single set $\dataset$.
We assume a special element $\zero\in\dataset$
which is the initial value of each shared variable or register.

\begin{figure}
 \center
 \begin{tikzpicture}[codeblock/.style={line width=0.5pt, inner xsep=0pt, inner ysep=0pt}]
\node[codeblock, font=\normalsize] (init) at (current bounding box.north west) {
\small{
$
\begin{array}{rcl}
\prog &::= & \keyword{vars:} \; \xvar^*\\
		&& \keyword{procs:}\; \proc^*\\
\proc &::= &\keyword{regs:} \; \reg^*\\
		&& \keyword{instrs:}\;\instr^* \\
\instr &::= & \lbl : \stmt;\\
\stmt  &::=& \xvar\!\assigned\!\expr \\
		&& |~ \reg\!\assigned\!\xvar \\
		&& |~ \reg\!\assigned\!\expr \\
		 && |~ \keyword{if} \; \expr ~\keyword{then} \; \instr^* \; \keyword{else} \; \instr^*\\
		&& |~\keyword{while} \; \expr ~\keyword{do} \; \instr^*\\
		&& |~\keyword{assume}~\expr\\
		&& |~\keyword{assert}~\expr\\    
		&& |~\keyword\terminated
\end{array}
$
}
};
\end{tikzpicture}
\caption{{Syntax of concurrent programs.}}
\label{program_syntax}
\end{figure}

Each  instruction $\instr$ is of the form $\lbl\!:\!\stmt$ 
 where  $\lbl$ is a unique
label (across all processes)  and $\stmt$ is a statement.
We define $\labelingof\instr:=\lbl$ and
$\stmtof\instr:=\stmt$.
We define $\instrsetof\proc$ to be the set of instructions
occurring in $\proc$, and
define $\instrset:=\cup_{\proc\in\procset}\instrsetof\proc$.
We assume that $\instrsetof\proc$ contains a designated
{\it initial} instruction $\initinstrof\proc$ from which $\proc$
starts its execution.

There are several types of instructions.
A {\it write} instruction has a statement of the form
$\xvar\assigned\expr$ where $\xvar\in\varset$ is a variable and
$\expr$ is an {\it expression}.
A {\it read} instruction in a process $\proc\in\procset$
has a statement of the form $\reg\assigned\xvar$, where $\reg$
is a register in $\proc$ and $\xvar\in\varset$ is a variable.
An {\it assign} instruction in a process $\proc\in\procset$
has a statement of the form $\reg\assigned\expr$, where $\reg$
is a register in $\proc$ and $\expr$ is an expression.
We will assume a set of expressions
containing a set of operators applied to  constants and registers,
but not referring  to the content of memory (i.e., the set of  variables). 
Conditional, iterative, assume, and assert instructions
(collectively called {\it aci} instructions) can be explained in
a similar manner.
The statement $\keyword\terminated$ will cause the process
to terminate its execution.
We assume that $\keyword\terminated$ occurs only once in the
code of a process $\proc$ and that it has the label
$\termlblof\proc$.

We give a number of definitions that we will use in the definition of the POWER operational semantics.
\begin{enumerate}
\item
For a write instruction $\instr$
where $\stmtof{\instr}$ is of the form
 $\xvar\assigned\expr$
or a read instruction $\instr$ 
where $\stmtof{\instr}$ is of the form
$\reg\assigned\xvar$,
we define $\varof\instr:=\xvar$.
For an instruction $\instr$ that is neither  read nor  write,
 we define $\varof\instr:=\myundef$.
In other words,
the variable function $\varof\instr$
returns the variable in  $\instr$.
\item
For a write instruction $\instr$
where $\stmtof{\instr}$ is of the form
$\xvar\assigned\expr$,
an assign instruction 
$\instr$
where $\stmtof{\instr}$ is of the form
$\reg\assigned\expr$,
or an aci instruction $\instr$
where $\stmtof{\instr}$ is of the form
$\keyword{if} \; \expr ~\keyword{then} \; \instr^* \; \keyword{else} \; \instr^*$,
$\keyword{while} \; \expr ~\keyword{do} \; \instr^*$,
 $\keyword{assume}~\expr$, 
or $\keyword{assert}~\expr$,
we define $\exprof\instr:=\expr$.
For an instruction that is not write, assign, or  aci,
we define $\exprof\instr:=\myundef$.
In other words,
the expression function $\exprof\instr$
returns the expression in $\instr$.
\item
For an expression $\expr$, we use $\regsetof\expr$ to denote the
set of registers that occur in $\expr$.
Then, for an instruction $\instr$,
we define $\regsetof\instr:=\regsetof{\exprof\instr}$.
Note that $\regsetof\instr=\emptyset$ if $\exprof\instr=\myundef$.
\end{enumerate}

For an instruction
$\instr\in\instrsetof\proc$,
we define $\nextof\instr$ to be the set of
instructions that may follow $\instr$ in a run of a process.
Notice that this set contains two elements if
$\instr$ is an aci instruction 
(in the case of an assume or assert instruction, we assume that 
if the condition evaluates to $\false$, then the process
moves to $\keyword\terminated$),
no element if $\instr$ is a  terminating instruction, and 
a single element otherwise.
We define $\tnextof\instr$ (resp.\ $\fnextof\instr$) to be the (unique)
instruction to which the process execution moves 
in case the condition in the statement
of $\instr$ evaluates to $\true$ (resp.\ $\false$).

In Section~\ref{full_syntax:section}, 
we will describe how to deal with  address operators
in read and write instructions and  the synchronization primitives.
We note that by following~\cite{DM14,DBLP:conf/pldi/SarkarSAMW11},
we 
do not include the load-reserve/store-conditional primitives~\cite{DBLP:conf/pldi/SarkarMOBSMAW12} (also known as load-linked/store-conditional or load-exclusive/store-exclusive) for POWER in our grammar.
These instructions are normally used to implement the compare-and-swap (CAS) instructions.
We leave load-reserve/store-conditional instructions to future work.

\subsection{Configurations}
\label{semantics:section}
We will assume an infinite set $\eventset$ of {\it events},
and will use an event to represent a single execution of an instruction in
a process.

A given instruction may be executed several times during a run
of the program (for instance, when it is in the body of a loop).
In such a case, the different executions are represented by different events.
An event $\event$ is executed in several steps.
In general, any event must be {\it fetched} and {\it committed}.
Between fetching  and committing steps,
a read, write, or assign event must  be 
 {\it initialized}. 
Furthermore, a write event may be {\it propagated} to the other processes.

A {\it configuration} $\conf$ is a tuple
$\tuple{\events,\eorder,\ilabeling,\status,\rfrom,\propagated,\corder}$,
defined as follows.

\subsubsection{Events}
We use 
$\events\subseteq\eventset$ to denote a finite set of {\it events}, namely
the events that have been created up to the current point in the execution
of the program.
We use also  $\ilabeling:\events\mapsto\instrset$ to denote a function that maps an
event $\event$ to the instruction $\instrof\event$
that $\event$ is executing.

We partition the set $\events$ into disjoint sets
$\eventsof\proc$, for $\proc\in\procset$, where 
$\eventsof\proc:=
\setcomp{\event\in\events}{\instrof\event\in\instrsetof\proc}$, i.e.,
for a process $\proc\in\procset$, the set $\eventsof\proc$ 
contains the events whose instructions belong to $\proc$.
For an event $\event\in\eventsof\proc$, we define
$\procof\event:=\proc$.

We say that $\event$ is a {\it write} event
if $\instrof\event$ is a write instruction.
We use $\wevents$ to denote the set of write events.
Similarly, we define the set $\revents$ of {\it read}
events, 
the set $\aevents$ of {\it assign}
events, 
and the set
$\acievents$ of
{\it aci} events
whose instructions are either
assume, assert, conditional, or iterative.
We define $\weventsof\proc$,  $\reventsof\proc$, $\aeventsof\proc$, and 
$\acieventsof\proc$,  to be the restrictions of 
the above sets to $\eventsof\proc$.
For each variable $\xvar\in\varset$, we assume 
a special write event $\initeventof\xvar$, called the \textit{initializer} event for $\xvar$.
This event is not performed by any of the processes in $\procset$,
and writes the value $0$ to $\xvar$.
Finally, we define $\initeventset:=\setcomp{\initeventof\xvar}{\xvar\in\varset}$
to be a set disjoint from the set of events $\eventset$ that contains all the {initializer} events.

\subsubsection{Program Order}
The {\it program-order} relation $\eorder\subseteq\events\times\events$ is an 
irreflexive  partial order that describes, for a process $\proc\in\procset$,
the order in which events are fetched from the code
of $\proc$.

We require that
\begin{enumerate}
\item
$\event_1\not\eorder\event_2$
if $\procof{\event_1}\neq\procof{\event_2}$, i.e., 
$\eorder$ only relates events belonging to the same process, and 
\item 
$\eorder$ is a total order on $\eventsof\proc$.
\end{enumerate}

\subsubsection{Status}
The function
$\status:\events\mapsto\set{\fetch,\init,\commit}$ defines, for
an event $\event$, the current {\it status} of $\event$, i.e.,
whether it has been fetched, initialized, or committed.

\subsubsection{Propagation}
The function
$\propagated:\procset\times\varset\mapsto\wevents\cup\initeventset$ defines,
for a process $\proc\in\procset$ and variable $\xvar\in\varset$,
the latest write event on $\xvar$ that has been propagated to $\proc$.

\subsubsection{Read-From}
The function $\rfrom:\revents\mapsto\wevents\cup\initeventset$ defines, 
for a read event $\event\in\revents$, 
 the write event $\rfromof\event$ from which $\event$ gets its value.

\subsubsection{Coherence Order}
All  processes share a global view about the order
in which write events are propagated.
This is described by the {\it coherence order} relation $\corder$ that is
a partial order
 on $\wevents$  such that
$\event_1\corder\event_2$ only if
$\varof{\event_1}=\varof{\event_2}$, i.e., 
it relates only events that write on identical
variables.
If a write event $\event_1$ is propagated to a process
before another write event $\event_2$ and both
events write on the same variable, then $\event_1\corder\event_2$ 
holds.
Furthermore, 
the events cannot be propagated to any other process in the reverse order.
As a consequence,
a write event is never propagated to a given process
if the process has already seen a coherence successor of this event.

\subsubsection{Dependencies}
We introduce a number of dependency orders on events that we
will use in the definition of the POWER semantics.

\begin{enumerate}
\item
We define the {\it per-location program-order}
$\plorder\subseteq\events\times\events$ such that
$\event_1\plorder\event_2$ if $\event_1\eorder\event_2$
and $\varof{\event_1}=\varof{\event_2}\in\varset$, i.e., it is the restriction
of the program order relation $\eorder$ to events with identical variables.
\item
We define the {\it data dependency} order $\dataorder$ 
such that $\event_1\dataorder\event_2$ if
\begin{enumerate}[(i)]
\item
$\event_1\in\revents \cup \aevents$, i.e., $\event_1$ is a read or assign event;
\item
$\event_2\in\wevents\cup \aevents\cup\acievents$,
i.e., $\event_2$ is  a write, assign, or  aci event; 
\item
$\event_1\eorder\event_2$;
\item
$\stmtof{\instrof{\event_1}}$ is of the form
$\reg\assigned\xvar$ or $\reg\assigned\expr$; 
\item
$\reg\in\regsetof{\instrof{\event_2}}$;
and 
\item
 there is no event
$\event_3\in\revents\cup\aevents$
such that 
$\event_1\eorder\event_3\eorder\event_2$
and
$\stmtof{\instrof{\event_3}}$ is of the form
$\reg\assigned\yvar$ or $\reg\assigned\expr'$.
Intuitively, the value loaded to register $\reg$ by $\event_1$ 
is used to compute the value of the expression   
$\exprof{\instrof{\event_2}}$.
\end{enumerate}
\item
We define the {\it control dependency} order $\ctrlorder$
such that $\event_1\ctrlorder\event_2$  if
$\event_1\in\acievents$ and
$\event_1\eorder\event_2$.
\end{enumerate}

\subsubsection{Committed and Initial Configurations}
We say that
$\conf$ is {\it committed}
if 
$\statusof\event=\commit$ for all events $\event$ in the event set  of $\conf$. 
The {\it initial configuration}
$\initconf$ is defined by
$$\tuple{\emptyset,\emptyset,\lambda \event.\bot,\lambda \event.\bot,\lambda \event.\bot,\lambda\proc.\lambda\xvar. \initeventof\xvar,\emptyset}$$
We use $\confset$ to denote the set of all configurations.

\subsubsection{Evaluation Functions}
Given a configuration $\conf$,  an  event $\event$, and an  expression $\expr$,  we first  define a function
$\valof{\conf}{\event,\expr}$
that returns the value of the 
expression $\expr$
when evaluated 
at the event $\event$
in the configuration $\conf$.
We define 
$\valof\conf\event:=\valof{\conf}{\event,\exprof{\instrof\event}}$.
Note that $\valof\conf\event=\myundef$ if $\exprof{\instrof\event}=\myundef$.

Let  $\conf=\tuple{\events,\eorder,\ilabeling,\status,\rfrom,\propagated,\corder}$ be a configuration.
Formally,
we define $\valof{\conf}{\event,\expr}$ recursively,
depending on the type of the  expression $\expr$:
\begin{enumerate}
\item
If $\expr$ 
is a constant $c$, then  $\valof{\conf}{\event,\expr}:=c$.
\item
 If $\expr$ is $f(\expr_1,\cdots,\expr_n)$
for some
 function
$f$ and expressions $\expr_1,\cdots,\expr_n$,
 then 
$\valof{\conf}{\event,\expr} := f(\valof{\conf}{\event,\expr_1},\cdots,\valof{\conf}{\event,\expr_n})$.
Note that
if $\valof{\conf}{\event,\expr_i}=\myundef$ for some $i: 1\leq i \leq n$,
then $ f(\valof{\conf}{\event,\expr_1},\cdots,\valof{\conf}{\event,\expr_n}):=\myundef$.
\item
If $\expr$ is
$\reg$
for some  register $\reg\in\regset$,
then let $\event'\in\events$ be  the closest
read or assign  event that precedes  $\event$ in the program order $\eorder$
and 
loads
a value
to the register $\reg$.
\begin{enumerate}[(i)]
\item 
If there is no such event $\event'$,
then 
$\valof\conf{\event,\expr} := 0$.
\item
If there is  such event $\event'$, $\event'\in\revents$,
and there is  a write event
$\event''\in\events\cup\initeventset$ 
such that 
$\rfromof{\event'}=\event''$.
\begin{itemize}
\item
If $\event''\in\initeventset$, then $\valof\conf{\event,\expr} := 0$.
\item
If $\event''\notin\initeventset$, then
let $\expr''=\exprof{\instrof{\event''}}$.
We define
$$\valof\conf{\event,\expr} :=\valof\conf{\event'',\expr''}$$
\end{itemize}
\item
If there is  such an event $\event'$, $\event'\in\revents$,
and
there is no such write event
$\event''\in\events\cup\initeventset$  
such that 
$\rfromof{\event'}=\event''$, i.e.\ $\rfromof{\event'}=\myundef$,
then
$\valof\conf{\event,\expr} := \myundef$.
\item
If there is  such an event $\event'$ and $\event'\in\aevents$,
then
let $\expr'=\exprof{\instrof{\event'}}$.
We define
$$\valof\conf{\event,\expr} :=\valof\conf{\event',\expr'}$$
\end{enumerate}
\end{enumerate}

\bgroup
\begin{table}[tb]
\caption{Definitions of predicates.}
\label{predicate:table}
\centering
\footnotesize
\begin{tabular}{ | C{2.1cm} | C{8.1cm} | L{4cm} | }
\hline
 {\bf Predicate} & {\bf Definition} & \multicolumn{1}{|c|}{{\bf Meaning}} \\ 
 \hline\hline
 	$\event\in\revents:\readcndof\conf\event$ 
 	&  
 		$\begin{aligned}
		\forall\event'&\in\revents:  \\
		&
			\Big(\big(\event'\plorder\event\big)   \implies 
			\big(\rfromof{\event'}\cordereq\rfromof{\event}\big)\Big)
		\end{aligned}$
	&  	For all read event $\event'$
		preceding the read event $\event$ in  $\plorder$, the write event
		from which $\event$ reads its value is not a coherence  predecessor
		of
		the write event for $\event'$. \\ 
\hline
	$\event\in\events:\commitcndof\conf\event$
 	&  $\begin{aligned}
		& \\
		&\forall\event'\in\events: \\
		&\left(\begin{gathered}
			\Big(\big(\event'\dataorder\event\big)\;  \vee\;
			\big(\event'\ctrlorder\event\big)\;   \vee\;
			\big(\event'\plorder\event\big) \Big)  \\
			 \implies \\
			\big(\statusof{\event'}=\commit\big) 
		  \end{gathered} \right) \\
	     	&
		\end{aligned}$
	&  	All events  preceding the event $\event$
		in $\dataorder$, $\ctrlorder$, or
		$\plorder$
		 have already been committed. \\ 
\hline
	$\event\in\wevents\cup\aevents:\writeinitcndof\conf\event$
 	&  $\begin{aligned}
		& \\
 		&\forall\event'\in\revents\cup\aevents: \\
		&\left(\begin{gathered}
		\big(\event'\dataorder\event\big) \\
		\implies\\
		\Big(\big(\statusof{\event'}=\init\big) \;\vee \; \big(\statusof{\event'}=\commit\big)\Big)
		 \end{gathered} \right) \\
	     	&
	  \end{aligned}$
	&  	All events  preceding the write or assign $\event$ in  $\dataorder$  have already been initialized.\\ 
\hline
	$\event\in\acievents:\validcndof\conf\event$
 	&  $\begin{aligned}
 		& \forall \event'\in\events: \\
		&\left(\!
		\begin{gathered}
		 \Big(\big(\event\eorder\event'\big)
		\;\wedge\;
		\big(\nexists\event''\in\events:\; \event\eorder\event''\eorder\event'\big)\Big) \\
		\implies \\
		\left(\!\begin{gathered}
		\Big(\big(\valof\conf\event\!=\!\true\big)\!\wedge\!
		\big(\instrof{\event'}\!=\!\tnextof{\instrof\event}\big)\Big) \\
		\vee \\
		\Big(\big(\valof\conf\event\!=\!\false \big)\!\wedge\!
		\big( \instrof{\event'}\!=\!\fnextof{\instrof\event}\big) \Big)
		\end{gathered}\!
		\right) 
		\end{gathered} \!\right)\\
	     	&
	  \end{aligned}$
	&  	If there exists an event $\event'$ that was fetched immediately
		after the aci event $\event$,
		$\event'$ is consistent with
		the value $\valof\conf\event$.\\ 
\hline
 
\end{tabular}
\end{table}
\egroup

\subsection{Transition Relation}
\label{transition:relation:section}
We define the transition relation as a relation
$\movesto{}\subseteq\confset\times\procset\times\confset$.
For configurations $\conf_1,\conf_2\in\confset$ and
a process $\proc\in\procset$, we write
$\conf_1\movesto\proc\conf_2$ to denote that
$\tuple{\conf_1,\proc,\conf_2}\in\movesto{}\!$.
Intuitively, this means that $\proc$  moves from  the 
current configuration  $\conf_1$ to $\conf_2$.
The relation $\movesto{}$ is defined through the set of inference rules
shown in Figure~\ref{rules:fig}.
Below we will explain these inference rules.
Table~\ref{predicate:table} gives some predicates used in the transition system.

\begin{figure}[tb]
\center
\begin{tikzpicture}
\node(dummy){};

\node(n1)[font=\footnotesize] at (dummy) 
{
$\event\not\in\events$,\;\;\;
$\eorder'=\eorder\cup\setcomp{\tuple{\event',\event}}{\event'\in\eventsof\proc}$,\;\;\;
$\instr\in{\maxinstrof\conf\proc}$};
\node(n2)[font=\footnotesize,anchor=north] at (n1.south) 
{
$\conf\movesto\proc\tuple{\events\cup\set{\event},\eorder',\ilabeling[\event\assigned\instr],\status[\event\assigned\fetch],\rfrom,\propagated,\corder}$
};
\draw (n2.north west) -- (n2.north east);
\node(l)[font=\footnotesize,anchor=west] at  ($(n2.north east)+(5pt,0pt)$)
{
{\tt Fetch}
};
\end{tikzpicture}

\begin{tikzpicture}
\node(dummy){};

\node(n1)[font=\footnotesize] at (dummy) 
{
$\event\in\reventsof\proc$,\;\;\;
$\statusof{\event}=\fetch$,\;\;\;
$\closestwriteof\conf\event=\event'$,\;\;\;
$\statusof{\event'}=\init$
};
\node(n2)[font=\footnotesize,anchor=north] at (n1.south) 
{
$\conf\movesto\proc\tuple{\events,\eorder,\ilabeling,\status[\event\assigned\init],\rfrom[\event\assigned\event'],\propagated,\corder}$
};
\draw (n1.south west) -- (n1.south east);
\node(l)[font=\footnotesize,anchor=west] at  ($(n1.south east)+(5pt,0pt)$)
{
{\tt InitReadFromLocal}
};
\end{tikzpicture}

\begin{tikzpicture}
\node(dummy){};

\node(n1)[font=\footnotesize] at (dummy) 
{
$\event\in\reventsof\proc$,\;\;\;
$\statusof{\event}\!=\!\fetch$,\;\;\;
$(\closestwriteof\conf\event=\myundef) \vee (\closestwriteof\conf\event\!=\!\event' \wedge \statusof{\event'}\!=\!\commit)$
};

\node(n2)[font=\footnotesize,anchor=north] at (n1.south) 
{
$\conf\movesto\proc\tuple{\events,\eorder,\ilabeling,\status[\event\assigned\init],\rfrom[\event\assigned\propagatedof\proc{\varof\event}],\propagated,\corder}$
};
\draw ($(n1.south west)+(0pt,0pt)$) -- ($(n1.south east)+(0pt,0pt)$);
\node(l)[font=\footnotesize,anchor=west] at  ($(n1.south east)+(5pt,0pt)$)
{
{\tt InitReadFromProp}
};
\end{tikzpicture}

\begin{tikzpicture}
\node(dummy){};

\node(n1)[font=\footnotesize] at (dummy) 
{
$\event\in\reventsof\proc$,\;\;\;
$\statusof{\event}=\init$,\;\;\;
$\commitcndof\conf\event$,\;\;\;
$\readcndof\conf\event$
};
\node(n2)[font=\footnotesize,anchor=north] at (n1.south) 
{
$\conf\movesto\proc\tuple{\events,\eorder,\ilabeling,\status[\event\assigned\commit],\rfrom,\propagated,\corder}$
};
\draw (n1.south west) -- (n1.south east);
\node(l)[font=\footnotesize,anchor=west] at  ($(n1.south east)+(5pt,0pt)$)
{
{\tt ComRead}
};
\end{tikzpicture}

\begin{tikzpicture}
\node(dummy){};

\node(n1)[font=\footnotesize] at (dummy) 
{
$\event\in\weventsof\proc$,\;\;\;
$\statusof{\event}=\fetch$,\;\;\;
$\writeinitcndof\conf\event$
};
\node(n2)[font=\footnotesize,anchor=north] at (n1.south) 
{
$\conf\movesto\proc\tuple{\events,\eorder,\ilabeling,\status[\event\assigned\init],\rfrom,\propagated,\corder}$
};
\draw (n2.north west) -- (n2.north east);
\node(l)[font=\footnotesize,anchor=west] at  ($(n2.north east)+(5pt,0pt)$)
{
{\tt InitWrite}
};
\end{tikzpicture}

\begin{tikzpicture}
\node(dummy){};

\node(n11)[font=\footnotesize] at (dummy) 
{
$\event\in\weventsof\proc$,\;\;\;
$\statusof{\event}=\init$,\;\;\;
$\commitcndof\conf\event$,
};
\node(n12)[font=\footnotesize] at ($(n11.south)+(0pt,-3pt)$) 
{
$\corder'=\corder\cup
\setcomp{\tuple{\event',\event}}
{\event'\cordereq\propagatedof{\proc}{\varof\event}}$
};
\node(n2)[font=\footnotesize,anchor=north] at (n12.south) 
{
$\conf\movesto\proc\tuple{\events,\eorder,\ilabeling,\status[\event\assigned\commit],\rfrom,\propagated[\tuple{\proc,\varof\event}\assigned\event],\corder'}$
};
\draw (n2.north west) -- (n2.north east);
\node(l)[font=\footnotesize,anchor=west] at  ($(n2.north east)+(5pt,0pt)$)
{
{\tt ComWrite}
};
\end{tikzpicture}

\begin{tikzpicture}
\node(dummy){};

\node(n11)[font=\footnotesize] at (dummy) 
{
$\qproc\in\procset$,\;\;\;
$\event\in\weventsof\proc$,\;\;\;
$\statusof{\event}=\commit$,\;\;\;
$\propagatedof{\qproc}{\varof\event}\corder\event$,
};
\node(n12)[font=\footnotesize] at ($(n11.south)+(0pt,-3pt)$) 
{
$\corder'=\corder\cup
\setcomp{\tuple{\event',\event}}
{\event'\cordereq\propagatedof{\qproc}{\varof\event}}$
};
\node(n2)[font=\footnotesize,anchor=north] at (n12.south) 
{
$\conf\movesto\proc\tuple{\events,\eorder,\ilabeling,\status,\rfrom,\propagated[\tuple{\qproc,\varof\event}\assigned\event],\corder'}$
};
\draw ($(n11.south west)+(0pt,-12pt)$) -- ($(n11.south east)+(0pt,-12pt)$);
\node(l)[font=\footnotesize,anchor=west] at  ($(n11.south east)+(5pt,-12pt)$)
{
{\tt PropWrite}
};
\end{tikzpicture}

\begin{tikzpicture}
\node(dummy){};

\node(n1)[font=\footnotesize] at (dummy) 
{
$\event\in\aeventsof\proc$,\;\;\;
$\statusof{\event}=\fetch$,\;\;\;
$\writeinitcndof\conf\event$
};

\node(n2)[font=\footnotesize,anchor=north] at (n1.south) 
{
$\conf\movesto\proc\tuple{\events,\eorder,\ilabeling,\status[\event\assigned\init],\rfrom,\propagated,\corder}$
};
\draw (n1.south west) -- (n1.south east);
\node(l)[font=\footnotesize,anchor=west] at  ($(n1.south east)+(5pt,0pt)$)
{
{\tt InitAssign}
};
\end{tikzpicture}

\begin{tikzpicture}
\node(dummy){};

\node(n1)[font=\footnotesize] at (dummy) 
{
$\event\in\aeventsof\proc$,\;\;\;
$\statusof{\event}=\init$,\;\;\;
$\commitcndof\conf\event$
};
\node(n2)[font=\footnotesize,anchor=north] at (n1.south) 
{
$\conf\movesto\proc\tuple{\events,\eorder,\ilabeling,\status[\event\assigned\commit],\rfrom,\propagated,\corder}$
};
\draw (n1.south west) -- (n1.south east);
\node(l)[font=\footnotesize,anchor=west] at  ($(n1.south east)+(5pt,0pt)$)
{
{\tt ComAssign}
};
\end{tikzpicture}

\begin{tikzpicture}
\node(dummy){};

\node(n1)[font=\footnotesize] at (dummy) 
{
$\event\in\acieventsof\proc$,\;\;\;
$\statusof{\event}=\fetch$,\;\;\;
$\commitcndof\conf\event$,\;\;\;
$\validcndof\conf\event$
};
\node(n2)[font=\footnotesize,anchor=north] at (n1.south) 
{
$\conf\movesto\proc\tuple{\events,\eorder,\ilabeling,\status[\event\assigned\commit],\rfrom,\propagated,\corder}$
};
\draw (n1.south west) -- (n1.south east);
\node(l)[font=\footnotesize,anchor=west] at  ($(n1.south east)+(5pt,0pt)$)
{
{\tt ComACI}
};
\end{tikzpicture}
\caption{Inference rules defining the relation $\movesto\proc{}$
where $\proc\in\procset$. We assume that $\conf$ is of the form $\tuple{\events,\eorder,\ilabeling,\status,\rfrom,\propagated,\corder}$.}
\label{rules:fig}

\end{figure}

The rule ${\tt Fetch}$ chooses the next instruction to be executed
in the code of a process $\proc\in\procset$.
This instruction should be a possible successor of the instruction
that was last executed by $\proc$.
To satisfy this condition, we define $\maxinstrof\conf\proc$ to be a
set of instructions  as follows:
\begin{enumerate}
\item
If $\eventsof\proc=\emptyset$ then define
$\maxinstrof\conf\proc:=\set{\initinstrof\proc}$,
i.e., the first instruction fetched by $\proc$ is $\initinstrof\proc$.
%
\item 
If $\eventsof\proc\neq\emptyset$, 
let $\event'\in\events_\proc$ be the
maximal event of $\proc$ (w.r.t.\ $\eorder$)
in the configuration $\conf$
and then
define  $\maxinstrof\conf\proc:=\nextof{\instrof{\event'}}$.
\end{enumerate}
In other words, we consider 
the instruction $\instr'=\instrof{\event'}\in\instrsetof\proc$,
and take its possible successors.
The possibility of choosing any of the 
(syntactically) possible successors corresponds to 
{\it speculatively} fetching statements.
As seen below, whenever we commit an 
aci
event, we check whether the made speculations are correct or not.
We create a new event $\event$, label it by $\instr\in\maxinstrof\conf\proc$,
and make it larger than all the other events of $\proc$
w.r.t.\ $\eorder$.
In such a way, we maintain the property that the order on
the events of $\proc$ reflects the order in which they are fetched
in the current run of the program.

There are two ways in which read events get their values, namely either
from {\it local} write events  that are performed by the process itself, 
or from write events that are {\it propagated} to the process.

The first case is covered by the rule ${\tt InitReadFromLocal}$
in which the process $\proc$ initializes a read
event $\event\in\revents_\proc$ on a variable
(say $\xvar$), where $\event$ has already been fetched.
Here, the event $\event$ is made to read its value from a
local write event $\event'\in\weventsof\proc$ on $\xvar$ 
such that 
\begin{enumerate}
\item
 $\event'$ 
has been initialized but not yet committed, and such that
\item
$\event'$ is the closest write event that precedes $\event$ in the order
$\plorder$.
\end{enumerate}
Notice that, by condition (2) $\event'$ is unique if it exists.
To formalize this, we define the {\it Closest Write} function
$\closestwriteof\conf\event:=\event'$ 
where $\event'$ is the unique event such that
\begin{enumerate}
\item
$\event'\in\wevents_\proc$, 
\item
$\event'\plorder\event$, 
and
\item
there is no event $\event''$ such that
$\event''\in\wevents_\proc$ and $\event'\plorder\event''\plorder\event$.
\end{enumerate}
Notice that such an event $\event'$ may not exist, i.e., it may be the case that
$\closestwriteof\conf\event=\myundef$.
If $\event'$ exists and it has been initialized but not committed, we initialize $\event$ 
and update the read-from relation appropriately.

The second case is  the case where such an event $\event'$ in the  rule ${\tt InitReadFromLocal}$
does not exist, i.e.,
if there is no write event on $\xvar$ before $\event$ by $\proc$, or if
the closest write event on $\xvar$ before $\event$ by $\proc$
has already been committed. 
We use the rule ${\tt InitReadFromProp}$ to let
$\event$ fetch its value from the latest write event on $\xvar$
that has been propagated to $\proc$.
Notice this event is the value of
$\propagatedof\proc\xvar$. 

To commit an initialized read event $\event\in\reventsof\proc$, we use the rule 
${\tt ComRead}$.
The rule can be performed if $\event$ satisfies two predicates
in $\conf$, namely $\readcndof\conf\event$ and $\commitcndof\conf\event$.

To initialize a fetched write event $\event\in\weventsof\proc$, we use the rule 
${\tt InitWrite}$ that requires all events that precede $\event$ in 
$\dataorder$ should have been initialized.
This condition is formulated by
the predicate
$\writeinitcndof\conf\event$.
When a write event in a process $\proc\in\procset$ is committed, it is also 
immediately propagated to $\proc$ itself.
To maintain the coherence order, the semantics keeps
the invariant that the latest write event on a variable
$\xvar\in\varset$ that has been propagated
to a process $\proc\in\procset$ is the largest
in coherence order among all write events on $\xvar$ that have been
propagated to $\proc$ up to now  in the run.
This invariant is maintained in ${\tt ComWrite}$ by requiring that the event
$\event$ (that is being committed) is strictly larger in coherence order
than the latest write event on the same variable as $\event$
that has been propagated to 
$\proc$.

Write events are propagated to other processes by the rule ${\tt PropWrite}$.
A write event $\event$ on a variable $\xvar$ is propagated to
a process $\qproc$ only if it has a coherence order that is strictly
larger than the coherence of any event that has been to propagated to $\qproc$ up
to now.
Notice that this is given by coherence order of 
$\propagatedof\qproc\xvar$ which is the latest write
event on $\xvar$ that has been propagated to $\qproc$.

To initialize and commit a fetched assign event $\event\in\aeventsof\proc$,
we use the rules 
${\tt InitAssign}$
and 
${\tt ComAssign}$
respectively.
When the assign event is initialized,
we require that all events that precede $\event$ in 
$\dataorder$ should have been initialized
by the predicate
$\writeinitcndof\conf\event$.
Similarly, when the assign event is committed,
we must satisfy the predicate 
$\commitcndof\conf\event$.

When committing an aci event by the rule ${\tt ComACI}$, we  require
that we verify any potential speculation that have been
made when fetching the subsequent events.
We formulate this requirement by the 
predicate $\validcndof\conf\event$.

\subsection{Bounded State Reachability}
\label{bounded:reachability:section}
We give our definitions of a run, a context, and a $\contextnum$-bounded run.
Then, we define the state reachability problem and $\contextnum$-bounded state reachability problem.

\begin{defi}[Run]
A {\it run} $\run$ is a sequence of transitions
$\conf_0\movesto{\proc_1}\conf_1\movesto{\proc_2}\conf_2\cdots\conf_{\nn-1}\movesto{\proc_\nn}\conf_\nn$.
\end{defi}
Given a run $\run ) \conf_0\movesto{\proc_1}\conf_1\movesto{\proc_2}\conf_2\cdots\conf_{\nn-1}\movesto{\proc_\nn}\conf_\nn$, we write $\conf_0\movesto{\run}\conf_\nn$.
Moreover, we define $\lastof\run:=\conf_\nn$.
We also define $\eprocof\run:=\proc_1\proc_2\cdots\proc_\nn$,
i.e., it is the sequence of processes performing the transitions
in $\run$.

\begin{defi}[Complete configuration]
A configuration $\conf$ is {\it complete} if 
\begin{enumerate}
\item
 $\conf$ is committed, and
\item
there is no configuration $\conf'$
such that
$\conf\movesto{\proc}\conf'$ for all $\proc\in\procset$
by allowing $\proc$ to execute any initializing, committing, or propagating inference rule.
\end{enumerate}
\end{defi}
It should be the case 
that all fetched instructions are
committed,
and all fetched write instructions
have been propagated or cannot be propagated to a process in the system.

\begin{defi}[Complete run]
A run $\run$ is {\it complete}  if 
$\lastof\run$ is complete.
\end{defi}

\begin{defi}[Context]
A sequence $\procseq=\proc_1\proc_2\cdots\proc_\nn\in\wordsover\procset$
is a {\it context} if there is a process $\proc\in\procset$ such that
$\proc_\ii=\proc$ for all $\ii:1\leq\ii\leq\nn$. %
\end{defi}

\begin{defi}[$\contextnum$-bounded run]
For a given natural number $\contextnum$,
a run $\run$ is  {\it $\kk$-bounded}
if 
$\run\uparrow=\procseq_1\app\procseq_2\app\cdots\procseq_\kk$
where $\procseq_\ii$ is a context for all $\ii:1\leq\ii\leq\kk$.
\end{defi}

For  $\conf\in\confset$ and  $\proc\in\procset$,
we define the set of {\it reachable labels} of the configuration $\conf$ as follows.
Let $\event_\proc\in\events_\proc$ be the maximal event of $\proc$ (w.r.t.\ $\eorder$)
in $\conf$.
We define
$\lblof{\conf,\proc}:={{\tt lbl}}(\instrof{\event_\proc})$,
i.e.\ 
process $\proc$ reaches the label of the maximal event $\event$ of $\proc$ (w.r.t.\ $\eorder$) in $\conf$.
Observe that in the case such an event $\event_\proc$
does not exist,
we define $\lblof{\conf,\proc}=\undefined$.
We define
$\lblof{\conf}:=
\{\lblof{\conf,\proc})\;|\;\proc\in\procset\}$.

\begin{defi}[State reachability problem]\label{reachability:problem:defi}
In the {\it state reachability problem}, we are given a
label $\lbl$ 
and asked whether there is a complete
run $\run$ and a configuration $\conf$ such that
$\initconf\movesto\run\conf$ where $\lbl\in\lblof\conf$.
\end{defi}

\begin{defi}[$\contextnum$-bounded state reachability problem]
For a given natural number $\contextnum$, 
the {\it $\contextnum$-bounded state reachability problem} is defined 
by requiring  that the run $\run$
in Definition~\ref{reachability:problem:defi} is $\contextnum$-bounded.
\end{defi}

\begin{figure}[t]
 \center
\begin{tikzpicture}
[codeblock/.style={line width=0.5pt, inner xsep=0pt, inner ysep=0pt}]

\node[codeblock,font=\normalsize] (process1) at (current bounding box.north west) 
{
\begingroup
\small
{
$
\begin{array}{rcl}
&&\keyword{vars:}\; \xvar, \yvar\\
&&\keyword{procs:}\; \proc_1, \proc_1 
\\
\\
\\
&& 
\proc_1 \\
&&\;\;0:\; \hspace{2mm}  \xvar\assigned1;\\
&&\;\;1:\; \hspace{2mm}  \yvar\assigned1;\\
&&\;\; 2:\; \hspace{2mm} \keyword\terminated;\\
\end{array}
$
}
\endgroup
};

\node[codeblock,font=\normalsize] (process2) at 
($(process1.east)+(18mm,0mm)$)
{
\begingroup
\small
{
$
\begin{array}{rcl}
&& 
\proc_2\\\
&&\hspace{1mm}\; \keyword{regs:}\; \reg_1,\reg_2\\
&&\;\; 3: \hspace{2mm}\reg_1\!\assigned\!\yvar;\\
&&\;\; 4:\; \hspace{2mm} \reg_2\!\assigned\!\xvar;\\
&&\;\; 5:\; \hspace{2mm} \keyword{assume}\;\; \reg_1=1;\\
&&\;\; 6:\; \hspace{2mm} \keyword{assume}\;\; \reg_2=0;\\
&&\;\; 7:\; \hspace{2mm} \text{/* empty line */}\\
&&\;\; 8:\; \hspace{2mm} \keyword\terminated;\\
\end{array}
$
}
\endgroup
};
\end{tikzpicture}
\caption{{
A  variant of the MP (Message Passing)  program~\cite{DBLP:journals/toplas/AlglaveMT14}.
}
}
\label{message-passing}
\end{figure}

\begin{figure}[tb]
 \centering
\hspace{-5mm}
\begin{minipage}[]{0.45\linewidth}
\centering
  \subfloat[]
    {
\begin{tabular}{| c |  c |}
\hline
 {{\bf Event}} & {\bf Instruction} \\ \hline\hline
 $\event_1$ & $0:\xvar\!\assigned\!1$ \\ 
$\event_2$ & $1:\yvar\!\assigned\!1$ \\ 
$\event_3$ & $3:\reg_1\!\assigned\!\yvar$ \\ 
$\event_4$ & $4:\reg_2\!\assigned\!\xvar$ \\ 
\hline
\end{tabular}
\label{Power_run_a}
    } 
\end{minipage}

 \begin{minipage}[]{0.6\linewidth}
   \centering
    \subfloat[]
    {
     \label{Power_run_b}
	\includegraphics{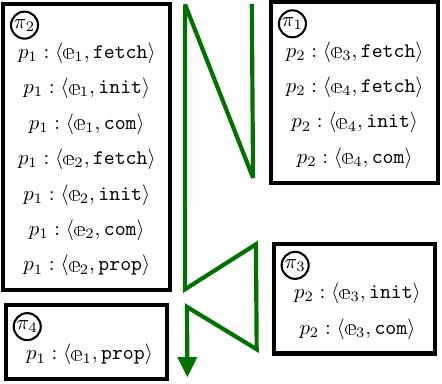}
       }
  \end{minipage}

\caption{
{
A complete run satisfies the state reachability problem of the program in Figure~\ref{message-passing}:
(A) read and write events and
(B) four contexts of the run containing  read and write events. The notion $\tuple{\event,\statusof{\event}}$ gives the recent status of an event $\event$.
}
}
\label{Power_run:fig}
\end{figure}

\subsection{Example}
We will give an example of a small concurrent program that has different behaviours under the SC and POWER semantics.
We  first intuitively explain  the program 
 and its behaviours under the SC semantics.
Then, we give a specific state reachability problem which the program cannot satisfy under SC.
 Then, we explore a context-bounded run of the  program under POWER that gives a positive answer for the state reachability problem.

 Figure~\ref{message-passing} illustrates 
a  program that is written following the syntax in Figure~\ref{program_syntax}.
The program
has two processes $\procset\!=\!\set{\proc_1, \proc_2}$ communicating through two shared   variables $\varset\!=\!\set{\xvar, \yvar}$.
Moreover, process $\proc_2$ has two  registers $\regset\!=\!\set{\reg_1,\reg_2}$.
At the beginning, all the variables and registers are initialized to $0$.
Process $\proc_1$ has two write instructions that set $\xvar$ and $\yvar$ to $1$. 
Process $\proc_2$ loads the values of  $\yvar$ and $\xvar$ into $\reg_1$ and $\reg_2$ respectively.
Then $\proc_2$
checks whether
the value of  $\reg_1$ is $1$ (line 5)
and the value of $\reg_2$ is $0$ (line 6).

The state reachability problem  under SC or POWER checks 
whether
  $\proc_2$ reaches  the label of line 7. 
  Note that $\proc_2$ can only reach  line 7 if it has executed the instructions 
  in lines 5 and  6 and it has evaluated these instructions to {\it true}.
Therefore, to satisfy this state reachability problem,
$\proc_2$ must read 1 from $\yvar$, 
and
while it is reading $\yvar$
it should not see that $\xvar$ has been set to $1$. 
Since at the beginning, 
all variables are 0, 
the value $1$ for $\yvar$
observed by $\proc_2$
must be written by
process $\proc_1$.

The state reachability problem
has a negative answer
 under SC semantics.
 The reason is that 
 the program order between
 two write instructions to $\xvar$ and $\yvar$  
requires process $\proc_1$
to set $\xvar$ and $\yvar$ to $1$ in order.
As a consequence, 
when
$\proc_2$ reads 1 from $\yvar$,
 it must see that $\xvar$ has been set to $1$. 

However, the complete run $\run$ given in Figure~\ref{Power_run:fig} shows that  the state reachability problem is satisfiable under POWER.
For the sake of simplicity, we only show the part of the run relating to the read and write events.
The run $\run$ can be decomposed into 4 contexts: $\pi_1$, $\pi_2$, $\pi_3$, and $\pi_4$.
In the first context $\pi_1$, 
$\proc_2$ fetches the two read instructions from $\yvar$ and $\xvar$, described by $\event_3$ and $\event_4$ respectively. 
After that, it initializes the fetched event $\event_4$ and loads $0$ from $\xvar$ into register $\reg_1$,
and then commits $\event_4$.
In the second context $\pi_2$,
$\proc_1$ fetches the write instruction on $\xvar$,  described by $\event_1$, in order to initialize and commit,
but delay propagating $\event_1$ to $\proc_2$.
Then, $\proc_1$ fetches the write instruction on $\yvar$, described by $\event_2$.
At this time, it initializes, commits, and propagates  $\event_2$ to  $\proc_2$.
In the third context $\pi_3$, $\proc_2$ resumes its execution by  initializing the fetched event $\event_3$  to load $1$ from $\yvar$ that is the value  just propagated from $\proc_1$,
and then committing  $\event_3$. 
Then, $\proc_2$ fetches two assume events $\event_5$ and $\event_6$ (that are not shown in Figure~\ref{Power_run:fig})
 corresponding to the instructions  
``$5:\keyword{assume}\;\; \reg_1=1$'' and ``$6:\keyword{assume}\;\; \reg_2=0$''   respectively
in order to commit them and terminates.
Finally, in the fourth context $\pi_4$, $\proc_1$ terminates by propagating $\event_1$  to $\proc_2$.
The run $\run$ is complete and $4$-bounded, and it satisfies the  state reachability problem.  


\section{Translation}
\label{translation:section}

In this section, we give an algorithm
that reduces, for a given number $\contextnum$, the
$\contextnum$-bounded  state reachability problem under POWER to the corresponding 
problem under SC.
Given an input concurrent program
$\prog$, the algorithm constructs an output concurrent
program $\sprog$ whose size is polynomial
 in $\prog$ and $\contextnum$,
such that for each $\contextnum$-bounded run $\run$ in $\prog$ 
under  POWER  there is a corresponding $\contextnum$-bounded run
$\srun$ of $\sprog$ under
 SC  that reaches the same set of process labels.
Below, we first present a scheme for the translation of $\prog$, and 
mention some of the challenges that arise due to the POWER
semantics.
Then, we give a detailed description of the data structures we use
in $\sprog$.
Finally, we describe the codes of the processes
in $\sprog$.

\begin{figure}
 \centering

  \begin{tikzpicture}[codeblock/.style={line width=0.5pt, inner xsep=0pt, inner ysep=0pt}]
\node[codeblock, font=\normalsize] (init) at (current bounding box.north west) {
\small{
$
\begin{array}{rcl}
\translateof{\prog}{\contextnum} &\myeq & \keyword{vars:}\; \hcancel[red]{\xvar^*}\;  \tuple{{\tt addvars}}_\contextnum\\
						&& \keyword{procs:}\; (\translateof{\proc}\contextnum)^*\; \tuple{{\tt initProc}}_\contextnum\; \tuple{{\tt verProc}}_\contextnum \\
\translateof{\proc}\contextnum
						&\myeq& \keyword{regs:}\;\reg^*\\
						&& \keyword{instrs:}\; (\translateof{\instr}\contextnum^\proc)^*\\
\translateof{\instr}\contextnum^\proc 	&\myeq& \lbl : \tuple{{\tt activeCnt}}_\contextnum^\proc\;\; \translateof{\stmt}\contextnum^\proc\;\; \tuple{{\tt closeCnt}}_\contextnum^\proc\\

\translateof{\xvar\assigned\expr}\contextnum^\proc											&\myeq& \translateof{\xvar\assigned\expr}\contextnum^{\proc, {\tt Write}}\\
\translateof{\reg\assigned\xvar}\contextnum^\proc											&\myeq& \translateof{\reg\assigned\xvar}\contextnum^{\proc,{\tt Read}}\\
\translateof{\reg\assigned\expr}\contextnum^\proc											&\myeq& \translateof{\reg\assigned\expr}\contextnum^{\proc,{\tt Assign}}\\
\llbracket\keyword{if} \; \expr\; \keyword{then} \; \instr^* &\myeq&\keyword{if} \; \expr\; \keyword{then} \; (\translateof{\instr}\contextnum^\proc)^*\\
\; \keyword{else} \; \instr^* \;\;\;\;\;\;\;   \rrbracket_\contextnum^\proc  && \keyword{else} \; (\translateof{\instr}\contextnum^\proc)^*;\;  \tuple{{\tt control}}_\contextnum^\proc\\																			
\translateof{\keyword{while} \; \expr\;\keyword{do} \; \instr^*}\contextnum^\proc						&\!\myeq\!&\keyword{while} \; \expr\; \keyword{do} \; (\translateof{\instr}\contextnum^\proc)^*;\;  \tuple{{\tt control}}_\contextnum^\proc\\
\translateof{\keyword{assume}~\expr}\contextnum^\proc										&\myeq&\keyword{assume}~\expr; \; \tuple{{\tt control}}_\contextnum^\proc\\
\translateof{\keyword{assert}~\expr}\contextnum^\proc										&\myeq&\keyword{assert}~\expr; \; \tuple{{\tt control}}_\contextnum^\proc\\
\translateof{\keyword\terminated}\contextnum^\proc											&\myeq&\keyword\terminated\\	
\tuple{{\tt addvars}}_\contextnum&\myeq & \cvalof{\sizeof\procset}{\sizeof\varset}\contextnum\; \initcvalof{\sizeof\procset}{\sizeof\varset}\contextnum\\
						&& \utstof{\sizeof\procset}{\sizeof\varset}\contextnum\; \gtstof{\sizeof\procset}{\sizeof\varset}\contextnum\\
						&&\lvalof{\sizeof\procset}{\sizeof\varset}\\
						&& \ireadof{\sizeof\procset}{\sizeof\varset}\; \creadof{\sizeof\procset}{\sizeof\varset}\\
						&& \iwriteof{\sizeof\procset}{\sizeof\varset}\; \cwriteof{\sizeof\procset}{\sizeof\varset}\\
						&& \iregof{\sizeof\regset}\; \cregof{\sizeof\regset}\\
						&& \ctrlof{\sizeof\procset} \\
						&& \activeof\contextnum\\
						&& \ccontext\\
\tuple{{\tt activeCnt}}_\contextnum^\proc 		&\myeq&\assume{\activeof\ccontext=\proc}\\
\tuple{{\tt closeCnt}}_\contextnum^\proc		&\myeq&  \ccontext\assigned\ccontext+\genof{[0..\contextnum-1]};\\
					&&\keyword{assume}(\ccontext \leq \contextnum)\\
\tuple{{\tt control}}_\contextnum^\proc													&\myeq& \ctrlof\proc\!\assigned\!\ctrlof\proc + \genof{[0..\contextnum\!-\!1]}\!;\\
					&&\keyword{assume}(\ctrlof\proc \leq \contextnum)\\
\end{array}
$
}
};
\end{tikzpicture}
\caption{{Translation map $\translateof{.}\contextnum$. We omit  the label of an intermediary  instruction when it is irrelevant.}
}
\label{translation-map}
\end{figure}

\subsection{Scheme}
Our construction is based on a code-to-code translation scheme that transforms the program $\prog$ into the program $\sprog$ following the map function $\translateof{.}\contextnum$ given in Figure~\ref{translation-map}. Let $\procset$ and $\varset$ be the sets of processes
and (shared) variables in $\prog$.
The map  $\translateof{.}\contextnum$ {\it replaces} the variables of $\prog$ by $O(\sizeof\procset\cdot \contextnum)$ copies
of the set $\varset$, in addition to
a finite set of  {\it finite-data} structures (which will be formally defined in Section~\ref{data:structures:section}).

The map function  $\translateof{.}\contextnum$ declares  two additional processes ${\tt initProc}$ and ${\tt verProc}$ that will be used to initialize the data structures and to check the state reachability problem at the end of the run of   $\sprog$. The formal definitions of ${\tt initProc}$ and ${\tt verProc}$  will  be given in Section~\ref{initializing:process:section} and Section~\ref{verifying:process:section}.

Furthermore, 
the map function  $\translateof{.}\contextnum$  transforms the code of each process $\proc \in \procset$ to a corresponding process $\sproc$  
that will simulate the moves of $\proc$. The processes $\proc$ and $\sproc$ will have the same set of registers. For each instruction   $\instr$ appearing in the code of the process $\proc$, the map $\translateof{\instr}\contextnum^\proc$  transforms it to a sequence of instructions as follows: First, it  adds the code   defined by  ${{\tt activeCnt}}$ to check if the process $\proc$ is active during the current context, then  it transforms the statement $\stmt$ of the instruction $\instr$ into a sequence of instructions following the map $\translateof{\stmt}\contextnum^\proc$, and finally it adds  the sequence of instructions defined by   ${{\tt closeCnt}}$ to  guess the occurrence of a context switch. 
The translations of write, read, and assign statements will be described in Section~\ref{write:instructions:section}, Section~\ref{read:instructions:section},
and Section~\ref{assign:instructions:section}
respectively.
The translation of an aci
statement keeps the same statement and 
adds ${\tt control}$ to guess the contexts when the corresponding event will be committed. 
The terminating 
statement remains the same by the map function  $\translateof{\keyword\terminated}\contextnum^\proc$.

\subsection{Challenges}
There are two {\it aspects} of the POWER semantics
(cf.\ Section~\ref{semantics:section})
that make it difficult to simulate  the run $\run$
under the SC semantics, namely {\it non-atomicity} and
{\it asynchrony}.

First, events are  executed {\it non-atomically}.
In fact, an event
is first fetched and initialized before it is committed.
In particular, an event may be fetched in one context
and be initialized and committed only in later contexts.
Since there is no bound on the number of events that may be
fetched in a given context, our simulation should be able
to handle unbounded numbers of pending events.

Second, write events of one process
are propagated in an {\it asynchronous} manner to the other processes.
This implies that we may have unbounded numbers of 
``traveling'' events that are committed in one context and propagated
to other processes only in subsequent contexts.
This creates two {\it challenges} in the simulation.
On the one hand, we need to keep track of the coherence
order among the different write events.
On the other hand,
since write events are not distributed to  different processes at the same
time, the processes may have different views of the values of a given
variable at a given point of time.

Since it is not feasible to record the initializing, committing, and
propagating contexts of an unbounded number of events in the SC runs
of a finite-state program,
our algorithm will instead predict 
the {\it summary} of effects of arbitrarily long sequences
of events that may occur in a given context.
This is implemented using  a  scheme that
first {\it guesses} and then {\it checks} these summaries.
Concretely, each event $\event$ in the run $\run$ is simulated by
a sequence of instructions in $\srun$.
This sequence of instructions will be executed
atomically (without interruption from
other processes and events).
More precisely, if $\event$ is fetched in a context
$\kk:1\leq\kk\leq\contextnum$, then the corresponding sequence of instructions
will be executed in  the same context $\kk$ in $\srun$.
Furthermore, we let $\srun$ {\it guess} 
\begin{enumerate}
\item
the contexts where
$\event$ will be initialized, committed, and propagated to the 
other processes, and
\item
the values of variables that are seen by read operations.
\end{enumerate}
Then, we {\it check} whether the guesses made by $\srun$
are valid  according to the POWER semantics.
As we will see below, these checks are done both on-the-fly
during $\srun$, as well as at the end of $\srun$.

To implement the guess-and-check scheme, we use a number
of data structures, described below.

\subsection{Data Structures}
\label{data:structures:section}
We will introduce the data structures  used in our simulation
in order to deal with the above asynchrony and non-atomicity challenging aspects.

\subsubsection{Asynchrony}
In order to keep track of the coherence order, we  associate a
{\it timestamp} with each write event.

A timestamp $\tst$ is a mapping  
$\procset\mapsto\contextnum^{\one\two}$ where $\contextnum^{\one\two}:=\contextnum^{\one}\cup\contextnum^{\two}$,
$\contextnum^{\one}:=\set{1}\times[1..\contextnum]$ and $\contextnum^{\two}:=\set{2}\times[1..\contextnum]$.
For
a process $\proc\in\procset$,  if the value of $\tstof\proc$  is of the form
$\tuple{1,\kk}$ where $\kk\in[1..\contextnum]$, i.e.\ $\tstof\proc\in\contextnum^{\one}$,
then $\tstof\proc$
represents
that 
the associated event is propagated to $\proc$
in the context $\kk$.
If the value of $\tstof\proc$ is of the form $\tuple{2,\kk}$ where $\kk\in[1..\contextnum]$, i.e.\
$\tstof\proc\in\contextnum^{\two}$, then 
$\tstof\proc$ represents that
\begin{enumerate}
\item
the associated event will not be
propagated to $\proc$, and 
\item
the maximal context of all coherence predecessors of the event is $\kk$.
\end{enumerate}

For a timestamp $\tst$ in  the form $\tuple{1,\kk}$ or $\tuple{2,\kk}$,
we define $\projectof{\tstof\proc}:=\kk$.
We use $\tstset$ to denote the set of timestamps.
We define an order $\tstorder$ on $\tstset$  such that
$\tst_1\tstorder\tst_2$ if
$\projectof{\tst_1(\proc)} \leq \projectof{\tst_2(\proc)}$
for all processes
$\proc\in\procset$.
If $\tst_1\tstorder\tst_2$ and 
there is a process $\proc\in\procset$ such that
$\projectof{\tst_1(\proc)}<\projectof{\tst_2(\proc)}$, then
we write $\tst_1\ststorder\tst_2$.
Note that 
if 
${\tst_1(\proc)}\tstorder{\tst_2(\proc)}$
and $\tst_1\not\ststorder\tst_2$ then both
$\tst_1\tstorder\tst_2$  and $\tst_2\tstorder\tst_1$.

The coherence order $\corder$ on write events will be reflected by
the order $\tstorder$ on their timestamps.
In particular, for two events $\event_1$ and
$\event_2$ with timestamps
$\tst_1$ and $\tst_2$ respectively, 
if  $\tst_1\ststorder\tst_2$ then $\event_1$ precedes $\event_2$
in coherence order (following the definition of $\ststorder$).
Moreover, if both $\tst_1\tstorder\tst_2$  and $\tst_2\tstorder\tst_1$
then the two associated events
are from the same process,
and the coherence order between them can be reflected by 
the program order. 

Given two timestamps $\tst_1$ and $\tst_2$, we define
the {\it summary} of $\tst_1$ and $\tst_2$, denoted by
$\tst_1\summary\tst_2$, to be the timestamp
$\tst$ as follows. 
\begin{enumerate}
\item
If $\projectof{\tst_1(\proc)} > \projectof{\tst_2(\proc)}$
then $\tst(\proc):={\tst_1(\proc)}$.
 \item
 If $\projectof{\tst_2(\proc)} > \projectof{\tst_1(\proc)}$
then $\tst(\proc):={\tst_2(\proc)}$.
\item
If $\projectof{\tst_1(\proc)} = \projectof{\tst_2(\proc)} = \kk$
and ($\tst_1(\proc)\in\contextnum^{\two} \vee \tst_2(\proc)\in\contextnum^{\two}$)
then $\tst(\proc):=\tuple{2,\kk}$.
\item
If $\projectof{\tst_1(\proc)} = \projectof{\tst_2(\proc)} = \kk$
and ($\tst_1(\proc)\in\contextnum^{\one} \wedge \tst_2(\proc)\in\contextnum^{\one}$)
then $\tst(\proc):=\tuple{1,\kk}$.
\end{enumerate}

Our simulation observes the sequence of write events received 
by a process in each context. 
In fact, the simulation 
will initially {\it guess} and later {\it verify}
the summaries of the timestamps of such a sequence.
This is done using  the data structures $\gtst$ and $\utst$.

The mapping
$\gtst:\procset\times\varset\times[1..\contextnum]\mapsto\mapingsover\procset\ncontextnum$ stores, 
for a process $\proc\in\procset$, a variable $\xvar\in\varset$,
and a context $\kk:1\leq\kk\leq\contextnum$, 
an {\it initial guess} $\gtstof\proc\xvar\kk$ of the summary  
of the timestamps of the sequence of write events
on  $\xvar$ propagated to $\proc$ up to the 
{\it start} of the context $\kk$.

Starting from a given initial guess for a given context $\kk$, the time
stamp is updated successively using the sequence of write events on $\xvar$ propagated to $\proc$
in $\kk$. 
The result is stored using the mapping
 $\utst:\procset\times\varset\times[1..\contextnum]\mapsto\mapingsover\procset\ncontextnum$.
More precisely, we initially set the value of
$\utst$  to $\gtst$.
Each time a new write event $\event$ on $\xvar$
is is executed by $\proc$ 
in a context $\kk$, 
we guess the timestamp $\ntst$ of $\event$, and then
update $\utstof\proc\xvar\kk$ by computing its summary
with $\ntst$.
Thus, given a point in a context $\kk$, $\utstof\proc\xvar\kk$ contains the summary
of the timestamps of the whole sequence
of write events on $\xvar$  that
have been propagated to $\proc$ 
up to that point.
At the end of the simulation, we {\it verify}, for each
context $\kk:1\leq\kk<\contextnum$, that 
the value of $\utst$ at the end of the context $\kk$ is equal to the value
of $\gtst$ for the next context $\kk+1$.

Furthermore, we use three data structures for storing the values of variables.
The mapping
$\initcval:\procset\times\varset\times[1..\contextnum]\mapsto\dataset$ stores, 
for a process $\proc\in\procset$, a variable $\xvar\in\varset$,
and a context $\kk:1\leq\kk\leq\contextnum$, 
an {\it initial guess} $\initcvalof\proc\xvar\kk$ of the value of the latest
write event
on  $\xvar$ propagated to $\proc$ up to the 
{\it start} of the context $\kk$.
The mapping $\cval:\procset\times\varset\times[1..\contextnum]\mapsto\dataset$ 
stores, 
for a process $\proc\in\procset$, a variable
$\xvar\in\varset$, 
and a point in a context $\kk:1\leq\kk\leq\contextnum$,
the value $\cvalof\proc\xvar\kk$ of the latest write
event on $\xvar$ that has been propagated to $\proc$ up to that point.
Moreover,
the mapping $\lval:\procset\times\varset\mapsto\dataset$ stores,
for a process $\proc\in\procset$ and a variable
$\xvar\in\varset$, 
the latest value $\lvalof\proc\xvar$ that has been written on 
$\xvar$ by 
$\proc$.

\subsubsection{Non-atomicity}
In order to satisfy  dependencies between events,
we need to keep track of the contexts where they are 
initialized and committed.
One aspect of our translation is to only keep
track of the {\it context} where the {\it latest}
read or write event on a given variable in a given process
is initialized or committed.

The mapping $\iwrite:\procset\times\varset\mapsto[1..\contextnum]$ defines,
for  $\proc\in\procset$ and $\xvar\in\varset$,
the context $\iwriteof\proc\xvar$ 
where the latest write event on $\xvar$ by $\proc$
is initialized.
The mapping $\cwrite:\procset\times\varset\mapsto[1..\contextnum]$ is defined 
in a similar manner for committing (rather than initializing) write events.
Furthermore, we define similar mappings $\iread$ and $\cread$ for read  events.

The mapping $\ireg:\regset\mapsto[1..\contextnum]$
gives, for a register
$\reg\in\allregs$, the initializing context $\iregof\reg$
of the latest read or assign event loading a value  to $\reg$.
For an expression $\expr$, we define
$\iregof\expr:={\tt max}\setcomp{\iregof\reg}{\reg\in\regsetof\expr}$.
The mapping  $\creg:\regset\mapsto[1..\contextnum]$
gives the context for committing (rather than initializing) of the
read and assign events.
We extend $\creg$ from registers to expressions in a similar manner
to $\ireg$.

Finally, the mapping $\ctrl:\procset\mapsto[1..\contextnum]$
gives, for a process $\proc\in\procset$, the committing context $\ctrlof\proc$
of the latest aci event in $\proc$.
Variables $\myactive$ and $\ccontext$ will be described in 
Section~\ref{initializing:process:section}.

\begin{figure}[tb]
\begin{minipage}{0.55\textwidth}
\small{
\begin{algorithm}[H]
\For{\mbox{$\proc\in\procset \wedge \xvar\in\varset$}}{\label{diverseinit}
$\ireadof\proc\xvar\assigned1$;
$\creadof\proc\xvar\assigned1$;
$\iwriteof\proc\xvar\assigned1$\;
$\cwriteof\proc\xvar\assigned1$;
$\lvalof\proc\xvar\assigned\zero$;
$\cvalof\proc\xvar1\assigned\zero$\;
  \lFor {\mbox{$\qproc\in\procset$}}{
     $\utstof\proc\xvar1(\qproc)\assigned\tuple{2,1}$
  }
}
\For {\mbox{$\proc\in\procset$}}{
$\ctrlof\proc\assigned1$;
}

\For {\mbox{$\reg\in\regset$}}{
$\iregof\reg\assigned1$;
$\cregof\reg\assigned1$;
}

\For {\mbox{$\proc\in\procset  \wedge \xvar\in\varset \wedge \kk\in[2..\contextnum]$}}{\label{gtstinit}
    \For {\mbox{$\qproc\in\procset$}}{
      $\gtstof\proc\xvar\kk(\qproc)\assigned\genof{\contextnum^{\one\two}}$;
    }
$\utstof\proc\xvar\kk\assigned\gtstof\proc\xvar\kk$\;
$\initcvalof\proc\xvar\kk\assigned\genof{\dataset}$\;
$\cvalof\proc\xvar\kk\assigned\initcvalof\proc\xvar\kk$\;
}

\For {$\kk\in[1..\contextnum]$}{\label{activeproc}
$\activeof\kk\assigned\genof\procset$;
}
$\ccontext\assigned 1$\;
\caption{{$\tuple{{\tt initProc}}_\contextnum$.}}
\label{init:fig}
\end{algorithm}
}
\end{minipage}
\end{figure}

\subsection{Initializing Process}
\label{initializing:process:section}
Algorithm~\ref{init:fig} shows the initializing process.
The for-loop of lines~\ref{diverseinit}, 5, and 7 define the values of the initializing and committing
data structures for the variables and  registers together with 
$\lvalof\proc\xvar$, 
$\cvalof\proc\xvar1$, $\utstof\proc\xvar1$, and $\ctrlof\proc$ for all $\proc\in\procset$ and $\xvar\in\varset$.
The for-loop of line~9 
defines the initial values of $\utst$ and $\cval$ at the start of each context $k\geq2$
(as described above).
The for-loop of line 15
chooses an {\it active} process to
execute in each context.
This information is stored in 
variables
$\activeof{\kk}$ for all $\kk\in\contextnum$.
The {\it current context} variable $\ccontext$ is initialized to $1$.

\begin{figure}[tb]
\begin{minipage}{.75\linewidth}
\small{
\begin{algorithm}[H]
\tcp{Guess}
$\iwriteof\proc\xvar\assigned\genof{[1..\contextnum]}$\;\label{geniw}
$\oldcwrite\assigned\cwriteof\proc\xvar$\;
$\cwriteof\proc\xvar\assigned\genof{[1..\contextnum]}$\;\label{gencw}
\For {$\qproc\in\procset$}{\label{genwtst1}
  $\ntstof\qproc\assigned\genof{\ncontextnum}$\;\label{genwtst2}
} 

\tcp{Check}
$\assume{\iwriteof\proc\xvar\geq \ccontext}$\;\label{iwccontext1}
$\assume{\activeof{\iwriteof\proc\xvar}=\proc}$\;\label{iwccontext2}
$\assume{\iwriteof\proc\xvar\geq\iregof\expr}$\;\label{iwdwrinitcnd}
$\assume{\cwriteof\proc\xvar\geq\iwriteof\proc\xvar}$\;\label{cwiw1}
${\tt assume}(\cwriteof\proc\xvar\geq \text{\hspace{0.5em}}{\tt max}\{\cregof\expr,\ctrlof\proc,\creadof\proc\xvar,\oldcwrite\})$\;
\For {$\qproc\in\procset$}{\label{genwcheck}
  \If{$\qproc=\proc$}{\label{ntstcw}
    $\assume{\ntstof\qproc\in\contextnum^{\one} \wedge \projectof{\ntstof\qproc}=\cwriteof\proc\xvar}$\;
    }
  \If{$\qproc\neq\proc$}{\label{ntstcwother}
    ${\tt assume}(\ntstof\qproc\!\in\contextnum^{\one}\!\implies\!\projectof{\ntstof\qproc}\geq \cwriteof\proc\xvar)$\;
    }
  \If{$\ntstof\qproc\in\contextnum^{\one}$}{\label{coherencecw}
    $\assume{\utstof\qproc\xvar{\projectof{\ntstof\qproc}}\tstorder\ntst}$\;
    ${\tt assume}(\activeof{\projectof{\ntstof\qproc}}=\proc)$\;
    } 
    \lElse {\label{skipcoherencecw}
    $\assume{\exists k: 1\!\leq\!k\!\leq\!\contextnum: \ntst\!\tstorder\!\utstof\qproc\xvar{k}}$
    }
}

\tcp{Update}
\For {$\qproc\in\procset$}{\label{genwupdate}
  \If{$\ntstof\qproc\in\contextnum^{\one}$}{
    $\utstof\qproc\xvar{\projectof{\ntstof\qproc}}\assigned\utstof\qproc\xvar{\projectof{\ntstof\qproc}}\summary\ntst
    $\;
$\cvalof\qproc\xvar{\projectof{\ntstof\qproc}}\assigned\expr$\;\label{valupdate}
 }}
 $\lvalof\proc\xvar\assigned\expr$\;\label{lvalupdate}
 
\caption{{$\translateof{\xvar\assigned\expr}\contextnum^{\proc,{\tt Write}}$.}}
\label{write:fig}
\end{algorithm}
}
\end{minipage}
\end{figure}

\subsection{Write Instructions}
\label{write:instructions:section}
Consider a write instruction $\instr$ of
a process $\proc\in\procset$ whose $\stmtof\instr$ is of the form
$\xvar\assigned\expr$.
The translation of this instruction is shown in Algorithm~\ref{write:fig}.
The code simulates an event $\event$ executing $\instr$, by
encoding the effects of the inference rules
${\tt InitWrite}$,  ${\tt ComWrite}$, and ${\tt PropWrite}$
that initialize, commit, and propagate a write event respectively.
The translation consists of three parts, namely
{\it guessing}, {\it checking}, and {\it update}.

\subsubsection{Guessing}
We guess the initializing and committing contexts 
for the event $\event$, together with its timestamp.
In line~\ref{geniw}, we guess the context where the event
$\event$ will be initialized, and
store the guess in $\iwriteof\proc\xvar$.
Similarly,
in line~\ref{gencw}, we guess the context where the event $\event$ will be committed,
and store the guess in $\cwriteof\proc\xvar$
(having stored its old value in the previous line).
In the for-loop of line~\ref{genwtst1}, 
we guess a timestamp for $\event$ and store it in $\ntst$.
This means that, for each process
$\qproc\in\procset$, we guess the context where the event $\event$ will
be propagated to $\qproc$ and 
we store this guess in $\ntstof\qproc$.

\subsubsection{Checking}
We perform sanity checks on the guessed values
in order to verify that they are consistent with
the POWER semantics.

Lines 6 -- 8 perform the sanity checks
for $\iwriteof\proc\xvar$.
In lines~6,  we verify that
the initializing context of the event $\event$
is not smaller than the current context.
This captures the fact that initialization happens after fetching
of $\event$.
Line 7 verifies that initialization happens in
a context where $\proc$ is active.
In line~8, we check whether $\writeinitcnd$ 
in the rule ${\tt InitWrite}$ is satisfied.
To do that, we verify that the data dependency order $\dataorder$ holds.
More precisely, we find, for each register $\reg$ that occurs
in $\expr$, the initializing context of the latest 
read or assign event loading to $\reg$.
We make sure that the initializing context of $\event$ is later than
the initializing contexts of all these read and assign events.
By definition, the largest of 
all these contexts is 
stored in $\iregof\expr$.

Lines 9 -- 10
perform the sanity checks for $\cwriteof\proc\xvar$.
In  
line 9,
we check the committing context of the event
$\event$ is at least as large as its initializing context.
%
In 
line 10,
we check that $\commitcnd$ 
in the rule ${\tt ComWrite}$ is satisfied.
To do that, we check that the committing context is larger
than 
\begin{enumerate}
\item
the committing context of all the read and assign events from
which the registers in the expression $\expr$ fetch their values
(to satisfy the data dependency order $\dataorder$, in
a similar manner to that described for initialization above),
\item
the committing contexts of the latest
read and write events on $\xvar$ in $\proc$, i.e., $\creadof\proc\xvar$ and
$\cwriteof\proc\xvar$
(to satisfy the per-location program order $\plorder$), and
\item
the committing context of the latest
aci event in $\proc$, i.e., $\ctrlof\proc$
(to satisfy the control order $\ctrlorder$).
\end{enumerate}

The for-loop of 
line 11
performs three sanity checks on 
$\ntst$.
In 
line 12, 
we verify that 
the event $\event$ is propagated to $\proc$ in the same context as
the one where it is committed.
This is consistent with the rule  ${\tt ComWrite}$
which requires that when a write event is committed
then it is immediately propagated to the committing process.
In 
line 14, 
we verify that if the event $\event$ is propagated
to a process $\qproc$ (different from $\proc$), then the propagation
takes place in a context later than or equal to the one where $\event$ is committed.
This is to be consistent with the fact that a write event is 
propagated to other processes only after it has been committed.
In 
line 17, 
we check that guessed timestamp
of the event $\event$ does not cause a violation of the coherence order $\corder$.
To do that, we consider each process $\qproc\in\procset$
to which $\event$ will be propagated (i.e., $\ntstof\qproc\in\contextnum^{\one}$).
The timestamp of $\event$ should be larger than the timestamp
of any other write event $\event'$ on $\xvar$ 
that has been propagated to $\qproc$ up to the current point
(since $\event$ should be larger in coherence order than $\event'$).
Notice that by construction the timestamp of the largest such event
$\event'$ is currently
stored in $\utstof\qproc\xvar{\ntstof\qproc}$.
Moreover, in line 18, we check that the event is propagated to $\qproc$ in the context where $\proc$ is active. 
Line 19 checks that for the case the event is never   propagated to $\qproc$ 
(i.e. $\ntstof\qproc\in\contextnum^{\two}$), $\qproc$ will receive
a coherence successor of this event in some context.

\subsubsection{Updating}
The for-loop of line 20 
uses the values guessed
above for updating the global data structure $\utst$.
More precisely, if the event $\event$ is propagated to a process $\qproc$,
i.e., $\ntstof\qproc\in\contextnum^{\one}$, then  we add
$\ntst$ to the
summary of the timestamps of the sequence of write operations 
on $\xvar$ propagated to $\qproc$ 
up to the current point in the context $\ntstof\qproc$.
Lines 23 -- 24
 assign the value $\expr$ to
$\cvalof\proc\xvar{\ntstof\qproc}$ and
$\lvalof\proc\xvar$ respectively.
Recall that the former stores the value defined by the latest
write event on $\xvar$ propagated to $\qproc$ 
up to the current point in the context $\ntstof\qproc$,
and the latter stores the value defined by the latest write on
$\xvar$ by
 $\proc$.

\begin{figure}[tb]
\begin{minipage}{.8\linewidth}
\small{
\begin{algorithm}[H]
\tcp{Guess}
$\oldiread\assigned\ireadof\proc\xvar$\;\label{storeir}
$\ireadof\proc\xvar\assigned\genof{[1..\contextnum]}$;
$\iregof\reg\assigned\ireadof\proc\xvar$\;\label{genir}
$\oldcread\assigned\creadof\proc\xvar$\;\label{storecr}
$\creadof\proc\xvar\assigned\genof{[1..\contextnum]}$;
$\cregof\reg\assigned\creadof\proc\xvar$\;\label{gencr}

\tcp{Check}
$\assume{\ireadof\proc\xvar\geq\ccontext}$\;\label{irccontext1}
$\assume{\activeof{\ireadof\proc\xvar}=\proc}$\;\label{irccontext2}
$\assume{\ireadof\proc\xvar\geq\iwriteof\proc\xvar}$\;\label{iriwcnd}
${\tt assume}(\ireadof\proc\xvar\geq\cwriteof\proc\xvar \implies  \utstof\proc\xvar\oldiread
\tstorder
\utstof\proc\xvar{\ireadof\proc\xvar})$\;\label{rdcwrcnd}
$\assume{\creadof\proc\xvar\geq\ireadof\proc\xvar}$\;\label{crir1}
$\assume{\activeof{\creadof\proc\xvar}=\proc}$\;\label{crir2}
${\tt assume}(\creadof\proc\xvar\geq 
{\tt max}\set{\ctrlof\proc,\oldcread,\cwriteof\proc\xvar})$\; \label{crcmcnd}
\tcp{Update}
 \lIf{$\ireadof\proc\xvar\!<\!\cwriteof\proc\xvar$}{\label{localread}
    $\reg\assigned\lvalof\proc\xvar$
    } \lElse {
    $\reg\assigned\cvalof\proc\xvar{\ireadof\proc\xvar}$
    }

\caption{{$\translateof{\reg\assigned\xvar}\contextnum^{\proc,{\tt Read}}$.}}
\label{read:fig}
\end{algorithm}
}
\end{minipage}
\end{figure}

\subsection{Read Instructions}
\label{read:instructions:section}
Consider a read instruction $\instr$ in
a process $\proc\in\procset$ whose  $\stmtof\instr$ is of the form
$\reg\assigned\xvar$.
The translation of this instruction is shown in Algorithm~\ref{read:fig}.
The code simulates an event $\event$ executing $\instr$
by encoding the three inference
rules ${\tt InitReadFromLocal}$, ${\tt InitReadFromProp}$, and ${\tt ComRead}$.
In a similar manner to a write instruction,
the translation scheme for a read instruction consists of 
 guessing,  checking, and  update parts.
Notice however that the initialization of the read event is carried out through
two different inference rules.

\subsubsection{Guessing}
In line~1, we store the old value of $\ireadof\proc\xvar$.
In line~2, we guess the context where 
the event $\event$ will be initialized, and
store the guessed context both in $\ireadof\proc\xvar$
and  $\iregof\reg$. 
Recall that the latter records the
initializing context of the latest read or assign event loading a value  to $\reg$.
In lines~3 -- 4, we execute similar instructions
for committing (rather than initializing).

\subsubsection{Checking}
Lines~5 -- 9 perform the sanity checks
for $\ireadof\proc\xvar$.
Lines~5 -- 6 check that 
the initializing context for the event $\event$
is not smaller than the current context
and that the initialization happens in a context where $\proc$ is active.
Line~7 ensures that at least
one of the two inference rules ${\tt InitReadFromLocal}$
and ${\tt InitReadFromProp}$ is satisfied, by
checking that the closest write event $\closestwriteof\conf\event$
(if it exists)
has been initialized or committed.
In line~8,
we satisfy $\readcnd$ 
in the rule ${\tt ComRead}$.
Lines 9 -- 11
perform the sanity checks for $\creadof\proc\xvar$
in a similar manner to the corresponding
instructions for write events (see above).

\subsubsection{Updating}
The purpose of the update part
(the if-statement of 
line 12)
is to ensure that the correct
read-from relation is defined as described by the 
inference rules ${\tt InitReadFromLocal}$ and ${\tt InitReadFromProp}$.
If $\ireadof\proc\xvar<\cwriteof\proc\xvar$,  
then this means that the latest write event $\event'$ on
$\xvar$ by
 $\proc$ 
is not committed and hence, according to ${\tt InitReadFromLocal}$,
 the event $\event$ reads its value from  that event.
Recall that this value is stored in $\lvalof\proc\xvar$.
On the other hand, if $\ireadof\proc\xvar\geq\cwriteof\proc\xvar$
then  the event $\event'$ has been committed and hence,
according to ${\tt InitReadFromProp}$, the event $\event$ reads its value from  
the latest write event on $\xvar$ propagated to $\proc$
in the context where $\event$ is initialized.
We notice that this value is stored in 
$\cvalof\proc\xvar{\ireadof\proc\xvar}$.

\begin{figure}[tb]
\begin{minipage}{.58\linewidth}
\small{
\begin{algorithm}[H]
\tcp{Guess}
$\iregof\reg\assigned\genof{[1..\contextnum]}$\;\label{genireg}
$\cregof\reg\assigned\genof{[1..\contextnum]}$\;\label{gencreg}

\tcp{Check}
$\assume{\iregof\reg\geq\ccontext}$\;\label{iregccontext1}
$\assume{\activeof{\iregof\reg}=\proc}$\;\label{iregccontext2}
$\assume{\iregof\reg\geq\iregof\expr}$\;\label{iregwrinitcnd}
%
%
$\assume{\cregof\reg\geq\iregof\reg}$\;\label{cregireg1}
$\assume{\activeof{\cregof\reg}=\proc}$\;\label{cregireg2}
${\tt assume}(\creadof\proc\xvar\geq  {\tt max}\set{\cregof\expr, \ctrlof\proc})$\; \label{cregcmcnd}
\tcp{Update}
$\reg \assigned \expr$\;\label{lvalupdate} 
\caption{{$\translateof{\reg\assigned\expr}\contextnum^{\proc,{\tt Assign}}$.}}
\label{assign:fig}
\end{algorithm}
}
\end{minipage}
\end{figure}

\subsection{Assign Instructions}
\label{assign:instructions:section}
Consider an assign instruction $\instr$ in
a process $\proc\in\procset$ whose  $\stmtof\instr$ is of the form
$\reg\assigned\expr$.
The translation of this instruction is shown in Algorithm~\ref{assign:fig}.
The code simulates an event $\event$ executing $\instr$
by encoding the two inference
rules ${\tt InitAssign}$ and ${\tt ComAssign}$.
In a similar manner to a write or read instruction,
the translation scheme for an assign instruction consists of 
 guessing,  checking, and  update parts.

\subsubsection{Guessing}
In line~1, we guess the context where 
the event $\event$ will be initialized, and
store the guessed context  in $\iregof\reg$. 
%
%
In line~2, we execute a similar instruction
for committing.

\subsubsection{Checking}
Lines~3 -- 5 perform the sanity checks
for $\iregof\reg$.
Lines~3 -- 4 check that 
the initializing context for the event $\event$
is not smaller than the current context
and that the initialization happens in a context where $\proc$ is active.
In line~5, we check whether $\writeinitcnd$ 
in the rule ${\tt InitAssign}$ is satisfied
in a similar 
manner to the corresponding instructions for write events (see Section~\ref{write:instructions:section}).
Lines 6 -- 8
perform  a  sanity checks for $\cregof\reg$.

\subsubsection{Updating}
Line 9 simply loads the value of $\expr$ 
to the register $\reg$.

\begin{figure}
\removelatexerror
\begin{minipage}{0.55\textwidth}
\small
\begin{algorithm}[H]
\For {\mbox{$\proc\in\procset \wedge \xvar\in\varset \wedge \kk\in[1..\contextnum-1]$}}{\label{gtstverfiy}
$\assume{\utstof\proc\xvar\kk=\gtstof\proc\xvar{\kk+1}}$\;
$\assume{\cvalof\proc\xvar\kk=\initcvalof\proc\xvar{\kk+1}}$\;
}
\lIf {$\lbl$ is reachable} {
 {\it  error}
 }
\caption{{$\tuple{{\tt verProc}}_\contextnum$.}}
\label{verify:fig}
\end{algorithm}
\end{minipage}
\end{figure}

\subsection{Verifying Process}
\label{verifying:process:section}
The verifying process makes sure that the updated value $\utst$ of the
timestamp 
at the end of a given context $k: 1\leq k \leq \contextnum-1$ is equal
to the corresponding guessed value $\gtst$ at the start of the next context.
It also performs the corresponding test for the values
written to  variables (by comparing $\cval$ and $\initcval$).
Finally, it checks whether we reach an error label $\lbl$ (given in the state reachability problem) or not.


\section{Extending the Semantics: Address Operators and Synchronisation Instructions}
\label{full:seciton}
In this section, 
we give the syntax of concurrent programs and the POWER operational semantics 
while taking into account address operators  and synchronization instructions 
as formalized  
in~\cite{DM14,DBLP:conf/pldi/SarkarSAMW11}.
We also give an example of a small program that illustrates
how synchronization instructions work under the POWER semantics.

\begin{figure}
 \center
 \begin{tikzpicture}[codeblock/.style={line width=0.5pt, inner xsep=0pt, inner ysep=0pt}]
\node[codeblock, font=\normalsize] (init) at (current bounding box.north west) {
\small{
$
\begin{array}{rcl}
\prog &::= & \keyword{vars:} \; \xvar^*\\
		&& \keyword{procs:}\; \proc^*\\
\proc &::= &\keyword{regs:} \; \reg^*\\
		&& \keyword{instrs:}\;\instr^* \\
\instr &::= & \lbl : \stmt;\\
\stmt  &::=& \xvar\!\assigned\!\expr  \\
	&& |~  \color{blue}{[\expr']\!\assigned\!\expr}\\
 	&&|~ \reg\!\assigned\!\xvar \\
	&& |~ \color{blue}{\reg\!\assigned\![\expr]}\\
	&&|~ \reg\!\assigned\!\expr \\
	&& |~ \keyword{if} \; \expr ~\keyword{then} \; \instr^* \; \keyword{else} \; \instr^*\\
	&& |~\keyword{while} \; \expr ~\keyword{do} \; \instr^*\\
  	&&|~\keyword{assume}~\expr\\  
	&&|~\keyword{assert}~\expr\\
	&&|~ {\color{blue}{\syncinstr}} \\
	&&|~  {\color{blue}{\lwsyncinstr}} \\
	&&|~ {\color{blue}{\isyncinstr}}\\
	&& |~\keyword\terminated
\end{array}
$
}
};
\end{tikzpicture}
\caption{{Syntax of concurrent programs includes the address operators and sychronization instructions. The additional statements are written in blue.}}
\label{full_program_syntax}
\end{figure}

\subsection{Syntax}
\label{full_syntax:section}
Figure~\ref{full_program_syntax}
gives the grammar containing 
address operators 
and synchronisation instructions. The additional statements are written   in blue. 

The address operators  are used in read and write instructions.
We assume that all shared variables  have unique addresses.
Memory accessing instructions use the notation 
$[\expr]$ to denote
the memory location where the address is given by 
the value of the expression $\expr$.
A read statement of the form $\reg\!\assigned\![\expr]$
loads the value stored in the memory location given by the value
of the expression $\expr$ to the register $\reg$.
A write statement of the form $[\expr']\assigned\expr$
stores the value of the expression $\expr$
to the memory location given by the value
of the expression $\expr'$.

There are three kinds 
of synchronisation (or fence or memory barrier) statements, namely {\it sync},  {\it lwsync}, and {\it isync}.
Intuitively, the synchronization instructions are used to enforce 
the committing order between  read and/or write instructions
or 
the propagation ordering
between write instructions. 
We will  explain in detail the semantics of 
the synchronisation instructions in Section~\ref{configurations:section},  Section~\ref{transition:relation:full:section}, and Section~\ref{synchonization:example:section}.

We recall and extend several definitions
that we will use in  the extended POWER operational semantics.  

We keep the definitions 
of the instruction set $\instrset$,
$\labelingof\instr$,   
$\stmtof\instr$, 
$\regsetof\instr$, 
$\nextof\instr$,
$\tnextof\instr$, and
 $\fnextof\instr$
  as in Section~\ref{syntax:semantic:section}.

We extend the definitions of  the functions $\varof\instr$ 
and $\exprof\instr$ 
to cover the address operators. First, we define $\varof\instr$ as follows.
\begin{enumerate}
\item
For a write instruction $\instr$
where $\stmtof\instr$ is of the form
$\xvar\assigned\expr$ or
a read instruction $\instr$
where $\stmtof\instr$ is of the form
$\reg\assigned\xvar$,
we define $\varof\instr:=\xvar$.
\item
For a write instruction $\instr$
where $\stmtof\instr$ is of the form
$[\expr']\assigned\expr$
or
a read instruction $\instr$
where $\stmtof\instr$ is of the form
 $\reg\assigned[\expr]$,
we define $\varof\instr:=\undetermined$.
Intuitively, this means that the variable in $\stmtof\instr$ is {\it undetermined}. 
\item
For  an instruction $\instr$ that is neither write nor read,
we define  $\varof\instr:=\myundef$.
\end{enumerate}
Next,  we define $\exprof\instr$.
\begin{enumerate}
\item
For a write instruction $\instr$
where $\stmtof\instr$ is of the form
$\xvar\assigned\expr$ 
or 
$[\expr']\assigned\expr$,
an assign instruction $\instr$
where $\stmtof\instr$ is of the form
$\reg\assigned\expr$, 
or an aci instruction $\instr$
where $\stmtof\instr$ is of the form
$\keyword{assume}~\expr$,
$\keyword{assert}~\expr$,
$\keyword{if} \; \expr ~\keyword{then} \; \instr^* \; \keyword{else} \; \instr^*$,
or
$\keyword{while} \; \expr ~\keyword{do} \; \instr^*$,
we define $\exprof\instr:=\expr$.
\item
For an instruction $\instr$ that is neither write, assign, nor aci,
we define $\exprof\instr:=\myundef$.
\end{enumerate}

Given an instruction $\instr$, we define   $\addrof\instr$ to be the address function in the instruction as follows. 
\begin{enumerate}
\item
For a write instruction $\instr$
where $\stmtof\instr$ is of the form
$[\expr']\!\assigned\!\expr$
we define  $\addrof\instr:=\expr'$.
\item
 For a read instruction $\instr$ 
where $\stmtof\instr$ is of the form
$\reg\!\assigned\![\expr]$,
we define  $\addrof\instr:=\expr$.
\item
For a write instruction $\instr$
where $\stmtof\instr$ is of the form
 $\xvar\!\assigned\!\expr$
 or a read instruction $\instr$ 
where $\stmtof\instr$ is of the form
$\reg\!\assigned\!\xvar$,
we define $\addrof\instr$ to be a constant that is the address of the  variable $\xvar$.
\item
For an instruction $\instr$
that is neither  write nor  read,
we define $\addrof{\instr}:=\myundef$.
\end{enumerate}

\subsection{Configurations}
\label{configurations:section}
We assume that
the set $\eventset$  contains synchonization events.
Similar to the semantics in Section~\ref{syntax:semantic:section}, 
we present the execution of an instruction
by an event through several steps,
namely
{\it fetching}, {\it initializing}, {\it committing}, and {\it propagating}.
For the special case of a synchronization instruction, it is first
 fetched and then committed without being initialized.
Furthermore, after a sync or lwsync instruction is committed, 
it will be propagated to the other processes.

A {\it configuration} $\conf$ is a tuple
$$
\langle\events,\eorder,\ilabeling,\status,\rfrom,\propagated,\synpropagated,\astoregroup,\asyngroup,\corder\rangle
$$ defined as follows.

\subsubsection{Events}
We keep the definitions of 
$\events$,
$\eventsof\proc$, 
$\ilabeling(\event)$, 
$\procof\event$, 
 $\wevents$,
 $\revents$, 
  $\aevents$,  
$\acievents$, 
$\weventsof\proc$, $\reventsof\proc$,
$\aeventsof\proc$,
 and  
$\acieventsof\proc$ 
as 
in Section~\ref{syntax:semantic:section}.
Moreover, we use 
$\syncevents$, $\lwsyncevents$, and $\isyncevents$
to denote 
the set  of {\it sync} events,
{\it lwsync} events,
and 
 {\it isync} events respectively.
We define $\synceventsof\proc$, $\lwsynceventsof\proc$,
and $\isynceventsof\proc$ to be the restrictions of 
the above sets to $\eventsof\proc$.

\subsubsection{Program Order, Status, Propagation, Read-From, Coherence Order.}
We keep the definitions of $\eorder$,
$\status$,
$\propagated$,
 $\rfrom$, and
 $\corder$
  as in 
  Section~\ref{syntax:semantic:section}.

\subsubsection{Synchronisation Propagation}
The function $\synpropagated:\procset\mapsto 2^{\syncevents\cup\lwsyncevents}$ defines,
for a process $\proc\in\procset$,
the {\it set} of sync  and lwsync events
propagated to $\proc$.
In contrast to a write event, 
there is no global view 
about the order in which 
 sync  and lwsync events are propagated
(that is presented by the coherence order for write events).
Moreover, a  sync or lwsync event will be 
propagated to all processes in the system.

\subsubsection{Seen Writes}
The function $\astoregroup:(\syncevents\cup\lwsyncevents)\times\varset\mapsto {\wevents}$ defines,
for a  sync  or lwsync  event and a variable $\xvar\in\varset$,
the last write event on $\xvar$
that has been propagated to the process committing the synchonization event
\footnote{In~\cite{DBLP:conf/pldi/SarkarSAMW11},
for a sync or lwsync event $\event$,
the set $\set{\astoregroup(\event,\xvar)|\xvar\in\varset}$
is called  
the Group A writes of the event $\event$.}. 

\subsubsection{Seen Synchronisations}
The function $\asyngroup:\wevents\mapsto 2^{\syncevents\cup\lwsyncevents}$ defines,
for a write event,
the set of   sync  and lwsync events that have been propagated to the
process committing the write event.

\subsubsection{Dependencies}
To formalize the POWER operational semantics, we need to define the dependency 
orders on the set of events.
We keep the definition of 
$\ctrlorder$
as in Section~\ref{syntax:semantic:section}.
Below we extend the orders
$\plorder$
and 
$\dataorder$.
We also introduce the
address dependency order $\addressorder$.

\begin{enumerate}
\item
We define the {\it per-location program-order}
$\plorder\subseteq\events\times\events$ such that
$\event_1\plorder\event_2$ if $\event_1\eorder\event_2$,
and 
$\varof{\instrof{\event_1}}=\varof{\instrof{\event_2}}\in\varset$ or $\varof{\instrof{\event_1}}=\undetermined$ or $\varof{\instrof{\event_2}}=\undetermined$, 
i.e. it is the restriction
of the program order relation $\eorder$ to  events with identical or undetermined variables.%
\item
We define the {\it data dependency} order $\dataorder$ 
such that $\event_1\dataorder\event_2$ if
\begin{enumerate}[(i)]
\item
$\event_1\in\revents\cup\aevents$, i.e., $\event_1$ is a read or assign event;
\item
$\event_2\in\wevents\cup\aevents\cup\acievents$,
i.e., $\event_2$ is a write, assign,  or aci event; 
\item
$\event_1\eorder\event_2$;
\item
$\stmtof{\instrof{\event_1}}$ is of the form
$\reg\assigned\xvar$,
$\reg\assigned[\expr]$,
or $\reg\assigned\expr$; 
\item
$\stmtof{\instrof{\event_2}}$ is of the form
$\xvar\assigned\expr'$,
$[\expr'']\assigned\expr'$,
$\keyword{if}\; \expr'  \;\keyword{then}\; {\instr}^* \;\keyword{else}\; {\instr}^*$,
or $\keyword{while}\; \expr' \;\keyword{do}\;  {\instr}^*$
and $\reg\in\regsetof{\expr'}$;
and 
\item
there is no
$\event_3\in\revents\cup\aevents$
such that 
$\event_1\eorder\event_3\eorder\event_2$
and
$\stmtof{\instrof{\event_3}}$ is of the form
$\reg\assigned\yvar$, $\reg\assigned[\expr'']$, or $\reg\assigned\expr''$.
\end{enumerate}
\item
We define the {\it address dependency} order $\addressorder$ 
such that $\event_1\addressorder\event_2$ if
\begin{enumerate}[(i)]
\item
$\event_1\in\revents\cup\aevents$, i.e., $\event_1$ is a read or assign event;
\item
$\event_2\in\revents\cup\wevents$,
i.e., $\event_2$ is either a read or write event; 
\item
$\event_1\eorder\event_2$;
\item
$\stmtof{\instrof{\event_1}}$ is of the form
$\reg\assigned\xvar$, $\reg\assigned[\expr]$, or $\reg\assigned\expr$; 
\item
$\stmtof{\instrof{\event_2}}$ 
is of the form $\reg'\assigned[\expr']$ such that $\reg\in\regsetof{\expr'}$
or 
 of the form
$[\expr'']\assigned\expr'$
such that $\reg\in\regsetof{\expr''}$;
and 
\item
there is no
$\event_3\in\revents\cup\aevents$
such that 
$\event_1\eorder\event_3\eorder\event_2$
and
$\stmtof{\instrof{\event_3}}$ is of the form
$\reg\assigned\yvar$, $\reg\assigned[\expr''']$, or $\reg\assigned\expr'''$.
\end{enumerate}
Intuitively, the loaded value by $\event_1$ 
is used to compute the address  
$\addrof{\instrof{\event_2}}$.
\end{enumerate}

\begin{table}
\caption{Definitions of predicates. We omit the predicates that are identical to Section~\ref{syntax:semantic:section}.}
\label{predicate:table:full}
\centering
\footnotesize
\begin{tabular}{ | C{2.7cm} | C{7.7cm} | L{4.5cm} | }
\hline
 {\bf Predicate} & {\bf Definition} & \multicolumn{1}{|c|}{{\bf Meaning}} \\ 
 \hline\hline
 	$\event\in\revents:\readinitcndof\conf\event$ 
 	&  
		$\forall\event'\in\revents\cup\aevents:  
		\big(\event'\addressorder\event\big)
		\implies
		\big(\statusof{\event'}=\init\big)$
	&  	All read and assign events preceding  $\event$ in  $\addressorder$ 
		have already been initialized. \\ 
\hline
 	$\event\in\wevents\cup\aevents:\writeinitcndof\conf\event$ 
 	&  
 		$\begin{aligned}
		\forall&\event'\in\revents\cup\aevents:  \\
		&	\Big(\!\big(\event'\dataorder\event\big) \! \vee \!
			\big(\event'\addressorder\event\big)
			\implies
			\big(\statusof{\event'}\!=\!\init\big)\!\Big)
		\end{aligned}$
	&  	All read and assign events preceding  $\event$ in  $\dataorder$ or $\addressorder$
		have already been initialized. \\ 
\hline
	$\event\in\revents:\readcndof\conf\event$ 
 	&  
 		$\begin{aligned}
		\forall\event'&\in\revents:  \\
		&\left(\begin{gathered}
			\big(\event'\plorder\event\big) \; \wedge \; 
				\big(\newvarof\conf{\event'}=\newvarof\conf\event\big) \\
				 \implies \\ 
			\big(\rfromof{\event'}\cordereq\rfromof{\event}\big)
			\end{gathered} \right)
		\end{aligned}$
	&  	For all read event $\event'$
		 preceding the read $\event$ in  $\plorder$
		(with the same defined variable), the write event
		from which $\event$ reads its value is not a coherence  predecessor
		the write event for $\event'$. \\ 
\hline
	$\event\in\events:\commitcndof\conf\event$
 	&  $\begin{aligned}
		&\forall\event'\in\events: \\
		&\left(\!\begin{gathered}
			\left(\!\begin{gathered}
			\big(\event'\dataorder\event\big)\;  \vee\;
			\big(\event'\ctrlorder\event\big)\;   \vee\;
			\big(\event'\addressorder\event\big)\\ \vee \\ 
			\Big(\big(\event'\plorder\event\big) \! \wedge\! \big(\newvarof\conf{\event'}\in\set{\newvarof\conf\event,\undetermined}\big) \Big)
			 \end{gathered}\!\right)  \\
			 \implies \\
			\big(\statusof{\event'}=\commit\big) 
		  \end{gathered}\!\right)
		\end{aligned}$
	&  	All events  preceding $\event$
		in $\dataorder$, $\ctrlorder$, $\addressorder$, or
		$\plorder$ (with the same defined variable or undetermined variable)
		 have already been committed. \\ 
\hline
	$\event\in\events:\propsyncndof\conf\event$
 	&  $	\begin{aligned}
	&\forall\event'\in\syncevents: \\		
	&\left(\begin{gathered}	
		\big(\event'\eorder\event\big) \\
		\implies\\
		\big(\forall\proc\in\procset:\event'\in\synpropagatedof\proc\big)
	  \end{gathered}\right)
	 \end{aligned}
		$
	&  	All sync events  preceding $\event$
		in $\eorder$ 
		 have already been propagated to all processes in the system. \\ 
\hline
	$\event\in\events:\comlwsyncndof\conf\event$
 	&  	$
		\forall\event'\in\lwsyncevents: 
		\big(\event'\eorder\event\big) 
		\implies
		\big(\statusof{\event'} = \commit \big) 
		$
	&  	All lwsync events   preceding $\event$
		in $\eorder$ 
		 have already been committed. \\ 		 	 
\hline
	$\event\in\events:\comisyncndof\conf\event$
 	&  
		$
		\forall\event'\in\isyncevents: 
		 \big(\event'\eorder\event\big) 
	          \implies
		  \big(\statusof{\event'} = \commit \big) 
		 $
	&  	All isync events   preceding $\event$
		in $\eorder$ 
		 have already been committed. \\ 
\hline
	$\event\in\events:\allsyncscndof\conf\event$
 	&    $\propsyncndof\conf\event \wedge 
		\comlwsyncndof\conf\event \wedge 
		\comisyncndof\conf\event$
	&  	A conjunction of  $\propsyncndof\conf\event$, 
		$\comlwsyncndof\conf\event$,
		and $\comisyncndof\conf\event$. \\ 
\hline
	$\event\in\syncevents\cup\lwsyncevents:\writegroupcndof\conf\event\pproc$
 	&    $\forall\xvar\in\varset: \astoregroupof\event\xvar \cordereq\propagatedof\pproc\xvar$ 
	&  	For each seen write of $\event$, that write (or some coherence
		successor) has already been propagated to $\proc$.
		\\ 
\hline
	$\event\in\wevents:\syncgroupcndof\conf\event\pproc$
 	&   	$\forall\event'\in\asyngroupof\event:
		\event'\in\synpropagatedof\pproc$	
	&  	All seen synchronizations of $\event$
		have already been propagated to $\proc$. \\ 
\hline
	$\event\in\syncevents\cup\lwsyncevents:\comreadwritecndof\conf\event$
 	&   	$\forall\event'\in\revents\cup\wevents:
		(\event'\eorder\event) 
		\implies
		(\statusof{\event'}=\commit)$	
	&  	All read and write events preceding $\event$ in $\eorder$
		have already been committed. \\ 
\hline
	$\event\in\isyncevents:\defaddresscndof\conf\event$
 	&   	$\begin{aligned}
		\forall\event'&\in\revents\cup\wevents:\\
		&\left(\begin{gathered}
		(\event'\eorder\event) \\
		\implies \\
		(\forall \event''\addressorder\event': \statusof{\event''}=\commit)
		\end{gathered}\right)
		\end{aligned}$
	&  	All events that provide the value
		for address expressions in all read and write events preceding $\event$ in $\eorder$
		have already committed.
		\\  
\hline
 
\end{tabular}

\end{table}

\subsubsection{Committed and Initial Configurations}
We keep the definitions of a {\it committed} configuration and $\confset$ as in Section~\ref{syntax:semantic:section}.
The {\it initial configuration}
$\initconf$ is defined by
$$
\langle\emptyset,\emptyset,\lambda \event.\bot,\lambda \event.\bot,\lambda \event.\bot,\lambda\proc.\lambda\xvar. \initeventof\xvar,\lambda\proc.\emptyset, \lambda\event.\lambda\xvar.\bot,
 \lambda\event.\emptyset, \emptyset\rangle
$$

\subsection{Evaluation Functions}
We   keep the definitions of the functions
$\valof{\conf}{\event,\expr}$ and 
$\valof\conf\event$
as  in
Section~\ref{syntax:semantic:section}.

Let $\event$ be an event  
and $\conf$ be a configuration.
We define
${\sf Var}(\conf,\event)$
to be the variable whose  address is given by $\valof\conf{\event,\addrof{\instrof\event}}$.
Note that 
if $\addrof{\instrof\event}=\myundef$,
then
${\sf Var}(\conf,\event)=\myundef$.
Moreover, if $\addrof{\instrof\event}\neq\myundef$ and $\valof\conf{\event,\addrof{\instrof\event}}=\myundef$,
then ${\sf Var}(\conf,\event)=\undetermined$.
Intuitively, it means that
the event $\event$ is accessing (i.e.\ reading or writing)
to an undetermined variable.

The relations between $\varof{\instrof\event}$ and ${\sf Var}(\conf,\event)$ can be seen 
by considering  different forms of the statement $\stmtof{\instrof{\event}}$ as follows.
\begin{enumerate}
\item
If $\stmtof{\instrof{\event}}$ is of the form
$\reg\assigned\xvar$ or $\xvar\assigned\expr$,
then 
$\varof{\instrof\event} = {\sf Var}(\conf,\event)=\xvar$.
\item
If $\stmtof{\instrof{\event}}$ is of the form
$\reg\assigned[\expr]$ or $[\expr']\assigned\expr$,
then  $\varof{\instrof\event}=\undetermined$ and
${\sf Var}(\conf,\event)\in\varset\cup\set{\undetermined}$.
\item
If  $\stmtof{\instrof{\event}}$ is neither a read nor  write statement,
then $\varof{\instrof\event} = {\sf Var}(\conf,\event)=\myundef$.
\end{enumerate}

\subsection{Transition Relation}
\label{transition:relation:full:section}
The relation $\movesto{}$ taking into account  the address operators and synchronization instructions is defined by the set of inference rules
shown in Figure~\ref{full-rules:fig}.

Analogously to Section~\ref{syntax:semantic:section}, 
we define different transition rules for fetching, initializing, committing, and propagating.
Let $\conf$ be the configuration where we are executing a transition rule.
We keep the rule ${\tt Fetch}$ as in Section~\ref{syntax:semantic:section}.
Below we explain other rules for initializing, committing, and propagating.
Table~\ref{predicate:table:full}
 give all predicates that are extended or introduced.
 
\begin{figure}
\center
\begin{tikzpicture}
\node(dummy){};

\node(n1)[font=\footnotesize] at (dummy) 
{
$\event\not\in\events$,\;\;\;
$\eorder'=\eorder\cup\setcomp{\tuple{\event',\event}}{\event'\in\eventsof\proc}$,\;\;\;
$\instr\in{{\maxinstrof\conf\proc}}$
};
\node(n2)[font=\footnotesize,anchor=north] at (n1.south) 
{
$\conf\movesto\proc\tuple{\events\cup{\event},\eorder',\ilabeling[\event\assigned\instr],\status[\event\assigned\fetch],\rfrom,\propagated,\synpropagated,\astoregroup,\asyngroup,\corder}$
};
\draw (n2.north west) -- (n2.north east);
\node(l)[font=\footnotesize,anchor=west] at  ($(n2.north east)+(5pt,0pt)$)
{
{\tt Fetch}
};
\end{tikzpicture}

\begin{tikzpicture}
\node(dummy){};

\node(n11)[font=\footnotesize] at (dummy) 
{
$\event\in\reventsof\proc$,\;\;\;
$\statusof{\event}=\fetch$,\;\;\;
$\readinitcndof\conf\event$,
};
\node(n12)[font=\footnotesize] at ($(n11.south)+(0pt,-3pt)$) 
{
$\event'=\closestwriteof\conf\event$,\;\;\;
$\statusof{\event'}=\init$,\;\;\;
$\allsyncscndof\conf\event$
};
\node(n2)[font=\footnotesize,anchor=north] at (n12.south) 
{
$\conf\movesto\proc\tuple{\events,\eorder,\ilabeling,\status[\event\assigned\init],\rfrom[\event\assigned\event'],\propagated,\synpropagated,\astoregroup,\asyngroup,\corder}$
};
\draw (n2.north west) -- (n2.north east);
\node(l)[font=\footnotesize,anchor=west] at  ($(n2.north east)+(5pt,0pt)$)
{
{\tt InitReadFromLocal}
};
\end{tikzpicture}

\begin{tikzpicture}
\node(dummy){};

\node(n11)[font=\footnotesize] at (dummy) 
{
$\event\in\reventsof\proc$,\;\;\;
$\statusof{\event}=\fetch$,\;\;\;
$\readinitcndof\conf\event$,\;\;\;
$\allsyncscndof\conf\event$,
};
\node(n12)[font=\footnotesize] at ($(n11.south)+(0pt,-3pt)$) 
{
$(\closestwriteof\conf\event = \myundef) \vee (\event'=\closestwriteof\conf\event \wedge  \statusof{\event'}=\commit)$
};

\node(n2)[font=\footnotesize,anchor=north] at (n12.south) 
{
$
\begin{aligned}
\conf\movesto\proc
   \langle\events,\eorder,\ilabeling,\status[\event\assigned\init],\rfrom[\event\assigned{\propagatedof{\proc}{\newvarof\conf\event}}],\propagated,\synpropagated,\astoregroup,&\\
   \asyngroup,\corder\rangle&
\end{aligned}
$
};
\draw (n2.north west) -- ($(n2.north east)+(0mm,0mm)$);
\node(l)[font=\footnotesize,anchor=west] at  ($(n2.north east)+(3mm,0mm)$)
{
{\tt InitReadFromProp}
};
\end{tikzpicture}

\begin{tikzpicture}
\node(dummy){};

\node(n1)[font=\footnotesize] at (dummy) 
{
$\event\in\reventsof\proc$,\;\;\;
$\statusof{\event}=\init$,\;\;\;
$\commitcndof\conf\event$,\;\;\;
$\readcndof\conf\event$,\;\;\;
$\allsyncscndof\conf\event$
};
\node(n2)[font=\footnotesize,anchor=north] at (n1.south) 
{
$\conf\movesto\proc\tuple{\events,\eorder,\ilabeling,\status[\event\assigned\commit],\rfrom,\propagated,\synpropagated,\astoregroup,\asyngroup,\corder}$
};
\draw (n1.south west) -- (n1.south east);
\node(l)[font=\footnotesize,anchor=west] at  ($(n1.south east)+(5pt,0pt)$)
{
{\tt ComRead}
};
\end{tikzpicture}

\begin{tikzpicture}
\node(dummy){};

\node(n1)[font=\footnotesize] at (dummy) 
{
$\event\in\weventsof\proc$,\;\;\;
$\statusof{\event}=\fetch$,\;\;\;
$\writeinitcndof\conf\event$,\;\;\;
$\allsyncscndof\conf\event$
};
\node(n2)[font=\footnotesize,anchor=north] at (n1.south) 
{
$\conf\movesto\proc\tuple{\events,\eorder,\ilabeling,\status[\event\assigned\init],\rfrom,\propagated,\synpropagated,\astoregroup,\asyngroup,\corder}$
};
\draw (n2.north west) -- (n2.north east);
\node(l)[font=\footnotesize,anchor=west] at  ($(n2.north east)+(5pt,0pt)$)
{
{\tt InitWrite}
};
\end{tikzpicture}

\begin{tikzpicture}
\node(dummy){};

\node(n11)[font=\footnotesize] at (dummy) 
{
$\event\in\weventsof\proc$,\;\;\;
$\xvar=\newvarof\conf\event$,\;\;\;
$\statusof{\event}=\init$,\;\;\;
$\commitcndof\conf\event$,
};
\node(n12)[font=\footnotesize] at ($(n11.south)+(0pt,-3pt)$) 
{
$\allsyncscndof\conf\event$,\;\;\;
$\corder'=\corder\cup
\setcomp{\tuple{\event',\event}}
{\event'\cordereq\propagatedof{\proc}{\xvar}}$
};
\node(n2)[font=\footnotesize,anchor=north] at (n12.south) 
{
$
\begin{aligned}
\conf\movesto\proc
  \langle\events,\eorder,\ilabeling,\status[\event\assigned\commit],\rfrom,\propagated[\tuple{\proc,\xvar}\assigned\event],\synpropagated,\astoregroup,&\\
\asyngroup[\event\assigned\synpropagatedof\proc],
\corder'&\rangle
\end{aligned}
$
};
\draw (n2.north west) -- ($(n2.north east)+(0mm,0mm)$);
\node(l)[font=\footnotesize,anchor=west] at  ($(n2.north east)+(3mm,0mm)$)
{
{\tt ComWrite}
};
\end{tikzpicture}

\begin{tikzpicture}
\node(dummy){};

\node(n1)[font=\footnotesize] at (dummy) 
{
$\qproc\in\procset$,\;\;\;
$\event\in\weventsof\proc$,\;\;\;
$\statusof{\event}=\commit$,\;\;\;
$\propagatedof{\qproc}{\newvarof\conf\event}\corder\event$,
};
\node(n12)[font=\footnotesize] at ($(n11.south)+(0pt,-3pt)$) 
{
$\syncgroupcndof\conf\event\qproc$,\;\;\;
$\corder'=\corder\cup
\setcomp{\tuple{\event',\event}}
{\event'\cordereq\propagatedof{\qproc}{\xvar}}$ 
};

\node(n2)[font=\footnotesize,anchor=north] at (n2.north) 
{
$\conf\movesto\proc\tuple{\events,\eorder,\ilabeling,\status,\rfrom,\propagated[\tuple{\qproc,\newvarof\conf\event}\assigned\event],\synpropagated,\astoregroup,\asyngroup,\corder'}$
};
\draw (n2.north west) -- (n2.north east);
\node(l)[font=\footnotesize,anchor=west] at  ($(n2.north east)+(3pt,0pt)$)
{
{\tt PropWrite}
};
\end{tikzpicture}

\begin{tikzpicture}
\node(dummy){};

\node(n1)[font=\footnotesize] at (dummy) 
{
$\event\in\aeventsof\proc$,\;\;\;
$\statusof{\event}=\fetch$,\;\;\;
$\writeinitcndof\conf\event$,\;\;\;
$\allsyncscndof\conf\event$
};

\node(n2)[font=\footnotesize,anchor=north] at (n1.south) 
{
$\conf\movesto\proc\tuple{\events,\eorder,\ilabeling,\status[\event\assigned\init],\rfrom,\propagated,\corder}$
};
\draw (n1.south west) -- (n1.south east);
\node(l)[font=\footnotesize,anchor=west] at  ($(n1.south east)+(5pt,0pt)$)
{
{\tt InitAssign}
};
\end{tikzpicture}

\begin{tikzpicture}
\node(dummy){};

\node(n1)[font=\footnotesize] at (dummy) 
{
$\event\in\aeventsof\proc$,\;\;\;
$\statusof{\event}=\init$,\;\;\;
$\commitcndof\conf\event$,\;\;\;
$\allsyncscndof\conf\event$
};
\node(n2)[font=\footnotesize,anchor=north] at (n1.south) 
{
$\conf\movesto\proc\tuple{\events,\eorder,\ilabeling,\status[\event\assigned\commit],\rfrom,\propagated,\corder}$
};
\draw (n1.south west) -- (n1.south east);
\node(l)[font=\footnotesize,anchor=west] at  ($(n1.south east)+(5pt,0pt)$)
{
{\tt ComAssign}
};
\end{tikzpicture}

\begin{tikzpicture}
\node(dummy){};

\node(n1)[font=\footnotesize] at (dummy) 
{
$\event\in\acieventsof\proc$,\;\;\;
$\statusof{\event}=\fetch$,\;\;\;
$\commitcndof\conf\event$,\;\;\;
$\validcndof\conf\event$,\;\;\;
$\allsyncscndof\conf\event$
};
\node(n2)[font=\footnotesize,anchor=north] at (n1.south) 
{
$\conf\movesto\proc\tuple{\events,\eorder,\ilabeling,\status[\event\assigned\commit],\rfrom,\propagated,\synpropagated,\astoregroup,\asyngroup,\corder}$
};
\draw (n1.south west) -- (n1.south east);
\node(l)[font=\footnotesize,anchor=west] at  ($(n1.south east)+(5pt,0pt)$)
{
{\tt ComACI}
};
\end{tikzpicture}

\begin{tikzpicture}
\node(dummy){};

\node(n11)[font=\footnotesize] at (dummy) 
{
$\event\in\isynceventsof\proc$,\;\;\;
$\commitcndof\conf\event$,\;\;\;
$\allsyncscndof\conf\event$,\;\;\;
$\defaddresscndof\conf\event$
};
\node(n2)[font=\footnotesize,anchor=north] at (n11.south) 
{
$\conf\movesto\proc\tuple{\events,\eorder,\ilabeling,\status[\event\assigned\commit],\rfrom,\propagated,\synpropagated,\astoregroup,
\asyngroup,
\corder}$
};
\draw (n2.north west) -- (n2.north east);
\node(l)[font=\footnotesize,anchor=west] at  ($(n2.north east)+(5pt,0pt)$)
{
{\tt ComISync}
};
\end{tikzpicture}

\begin{tikzpicture}
\node(dummy){};

\node(n1)[font=\footnotesize] at (dummy) 
{
$\event\in\synceventsof\proc\cup\lwsynceventsof\proc$,\;\;\;
$\commitcndof\conf\event$,\;\;\;
$\allsyncscndof\conf\event$,\;\;\;
$\comreadwritecndof\conf\event$
};
\node(n2)[font=\footnotesize,anchor=north] at (n1.south) 
{
$
\begin{aligned}
\conf\movesto\proc
  \langle\events,\eorder,\ilabeling,\status[\event\assigned\commit],\rfrom,\propagated,\synpropagated[\proc\assigned\synpropagatedof\proc\cup\set{\event}],&\\
  \astoregroup[\tuple{\event,\xvar}\assigned\propagatedof\proc\xvar],
\asyngroup\corder&\rangle
\end{aligned}
$
};
\draw (n2.north west) -- ($(n2.north east)+(0pt,0pt)$);
\node(l)[font=\footnotesize,anchor=west] at  ($(n2.north east)+(5pt,0pt)$)
{
{\tt ComSync}
};
\end{tikzpicture}

\begin{tikzpicture}
\node(dummy){};

\node(n1)[font=\footnotesize] at (dummy) 
{
$\qproc\in\procset$,\;\;\;
$\event\in\synceventsof\proc\cup\lwsynceventsof\proc$,\;\;\;
$\statusof{\event}=\commit$,\;\;\;
$\writegroupcndof\conf\event\qproc$
};
\node(n2)[font=\footnotesize,anchor=north] at (n1.south) 
{
$\conf\movesto\proc\tuple{\events,\eorder,\ilabeling,\status,\rfrom,\propagated,\synpropagated[\qproc\assigned\synpropagatedof\qproc\cup\set{\event}],\astoregroup,\asyngroup,\corder}$
};
\draw (n2.north west) -- (n2.north east);
\node(l)[font=\footnotesize,anchor=west] at  ($(n2.north east)+(5pt,0pt)$)
{
{\tt PropSync}
};
\end{tikzpicture}
\caption{Inference rules with synchronizations and address operators defining the relation $\movesto\proc{}$
where $\proc\in\procset$. We assume that $\conf$ is of the form $\tuple{\events,\eorder,\ilabeling,\status,\rfrom,\propagated,\corder}$.}
\label{full-rules:fig}
\end{figure}

Similar to Section~\ref{syntax:semantic:section}, there are two ways in which read events get their values, namely either
from {\it local} write events  by the rule ${\tt InitReadFromLocal}$
or from write events that are {\it propagated}
 to the process by the rule ${\tt InitReadFromProp}$.
In  the rule ${\tt  InitReadFromLocal}$,
 the process $\proc$ initializes a read
event $\event\in\revents_\proc$ on a  variable $\newvarof\conf\event$
(say $\xvar$), where $\event$ has already been fetched.
We note that
by satisfying predicate 
$\readinitcndof\conf\event$,
if $\varof{\instrof\event}=\undetermined$,
then ${\sf Var}(\conf,\event)\in\varset$,
i.e.\ the variable from which $\event$ is reading has been defined.
Here, the event $\event$ is made to read its value from a
local write event $\event'\in\weventsof\proc$ on $\xvar$ 
such that
\begin{enumerate}
\item
$\event'$
has been 
initialized but not yet committed,
and 
\item 
$\event'$ is the closest write event that precedes $\event$ in $\plorder$ 
(note that we have extended the definition of $\plorder$ to cover the address operators.) 
\end{enumerate}
By condition (2)
$\event'$ is unique if it exists.
To formalize this, we extend the definition of the {\it Closest Write} function
$\closestwriteof\conf\event$ by  taking into account   the address operator.
We define
$\closestwriteof\conf\event:=\event'$ 
where $\event'$ is the unique event such that
\begin{enumerate}
\item
$\event'\in\wevents_\proc$, $\event'\plorder\event$,
\item
$\newvarof\conf{\event'}\in\set{\xvar,\undetermined}$, and
\item
there is no  event $\event''$ such that
$\event''\in\wevents_\proc$, $\event'\plorder\event''\plorder\event$, and $\newvarof\conf{\event''}\in\set{\xvar,\undetermined}$.
\end{enumerate}

If $\closestwriteof\conf\event$ does not exist or it has been committed,
then we use the rule ${\tt InitReadFromProp}$ to let
$\event$ fetch its value from the latest write event on $\xvar$
that has been propagated to $\proc$.
Both rules
${\tt InitReadFromLocal}$ and
${\tt InitReadFromProp}$ 
can only be performed 
for a read event $\event\in\revents_\proc$
if $\event$ satisfies the predicates
$\allsyncscndof\conf\event$\footnote{The semantics of lwsync is formalized as in~\cite{DBLP:conf/pldi/SarkarMOBSMAW12} (page 5): a read event can only be initialized if all
lwsync events preceding it in $\eorder$ have already been committed.} and
$\readinitcndof\conf\event$.

To commit an initialized read event $\event\in\reventsof\proc$, we use the rule 
${\tt ComRead}$.
The rule can be performed if $\event$ satisfies three predicates
in $\conf$:  $\readcndof\conf\event$,
$\commitcndof\conf\event$,
and 
$\allsyncscndof\conf\event$.

A fetched write event  $\event\in\weventsof\proc$ is executed following three rules, namely ${\tt InitWrite}$, ${\tt ComWrite}$, and ${\tt PropWrite}$.
We use the rule 
${\tt InitWrite}$ to initialize the event.
It can be performed if $\event$ satisfies  the predicates
$\writeinitcndof\conf\event$ and
$\allsyncscndof\conf\event$.
The rule ${\tt ComWrite}$ to commit a write event
is similar to the corresponding rule in Section~\ref{syntax:semantic:section},
except that we also keep  information about all
the seen sync and lwsync events of the write event by updating $\asyngroup$.
Write events are propagated to other processes by the ${\tt PropWrite}$ rule.
Taking into account  the synchronization instructions, the rule   requires that
all the seen sync and lwsync events 
of the write event $\event$
have been propagated to process $\qproc$.
This condition is formulated by the predicate
$\syncgroupcndof\conf\event\qproc$.

In a similar way to Section~\ref{transition:relation:section},
a fetched assign event is executed following the rules 
${\tt InitAssign}$ and  ${\tt ComAssign}$.
To reflex the appearance of  synchronizations,
these rules satisfy the predicate
$\allsyncscndof\conf\event$.
Moreover,
an aci event is committed by the rule ${\tt ComACI}$.

Finally, we explain the transition rules for synchronization events.
To commit and propagate a sync or lwsync event,
we use the rules ${\tt ComSync}$ and ${\tt PropSync}$ respectively.
To commit an isync event, we use the rule ${\tt ComISync}$.
These rules
require the five predicates
$\commitcndof\conf\event$,
$\allsyncscndof\conf\event$,
$\defaddresscndof\conf\event$,
$\comreadwritecndof\conf\event$, and
$\writegroupcndof\conf\event\pproc$ to hold.
When a sync or lwsync event in a process $\proc\in\procset$ is committed, it is also 
immediately propagated to $\proc$ itself.
Moreover, we keep  information about all
the seen write events of the  sync or lwsync event by updating $\astoregroup$.

\subsection{Bounded State Reachability}
We keep the definitions of the run $\run$, $\lastof\run$, the complete configuration, the complete run,
$\eprocof\run$, the context, the $\contextnum$-bounded run, 
the state reachability problem,
and the $\contextnum$-bounded state reachability problem
as
in Section~\ref{syntax:semantic:section}.

\subsection{Synchonization Example}
\label{synchonization:example:section}
We give an example of a small concurrent program to illustrate  how sync and lwsync instructions   work under the POWER semantics.

\begin{figure}
 \center
\begin{tikzpicture}
[codeblock/.style={line width=0.5pt, inner xsep=0pt, inner ysep=0pt}]

\node[codeblock,font=\normalsize] (process1) at (current bounding box.north west) 
{
\begingroup
\small{
$
\begin{array}{rcl}
&&\keyword{vars:}\; \xvar, \yvar
\\
&& \keyword{procs:}\; \proc_1, \proc_2 \\
&&\proc_1 \\
&&\hspace{1mm}\; \keyword{regs:}\; \reg_1\\
&&\;\;0:\; \hspace{2mm}  \xvar\assigned1;\\
&&\;\;1:\; \hspace{2mm}  \hwsync;\\
&&\;\;2:\; \hspace{2mm}  \reg_1\!\assigned\!\yvar;\\ 
&&\;\; 3:\; \hspace{2mm} \keyword{assume}\;\; \reg_1=0;\\
&&\;\; 4:\; \hspace{2mm} \text{/* empty line */} \\
&&\;\; 5:\; \hspace{2mm} \keyword\terminated;\\
&&\\
\end{array}
$
}
\endgroup
};

\node[codeblock,font=\normalsize] (process2) at 
($(process1.east)+(28mm,-2mm)$)
{
\begingroup
\small{
$
\begin{array}{rcl}
&& \proc_2\\\
&&\hspace{1mm}\; \keyword{regs:}\; \reg_2\\
&&\;\; 6: \hspace{3mm} \yvar\assigned1;\\
&&\;\;7:\; \hspace{2mm}  \hwsync;\\
&&\;\; 8:\; \hspace{2mm} \reg_2\!\assigned\!\xvar;\\
&&\;\; 9:\; \hspace{2mm} \keyword{assume}\;\; \reg_2=0;\\
&&10:\; \hspace{2mm} \text{/* empty line */} \\
&&11:\; \hspace{2mm} \keyword\terminated;\\
\end{array}
$
}
\endgroup
};
\end{tikzpicture}
\caption{{
A  variant of the SB (Store Buffer)  program~\cite{DBLP:journals/toplas/AlglaveMT14}.
}
}
\label{store-buffer}
\end{figure}

Figure~\ref{store-buffer} illustrates a program that is written following the syntax in Figure~\ref{full_program_syntax}.
The program has two processes $\procset=\set{\proc_1,\proc_2}$ communicating through two variables $\varset=\set{\xvar,\yvar}$. 
Moreover, process $\proc_1$ (resp.\ $\proc_2$)
has a register
$\reg_1$ 
 (resp.\ ${\reg_2}$).
 At the beginning, 
 all the variables and registers are initialized to 0.
Process $\proc_1$ has two  instructions: writing $1$ to $\xvar$ (event $\event_1$) and reading $\yvar$ (event $\event_2$).
Between these two instructions, $\proc_1$ executes a sync instruction (event $\event'$).
Similarly, process $\proc_2$ has two  instructions,
writing $1$ to $\yvar$ (event $\event_3$) and reading $\xvar$  (event $\event_4$), 
and a sync instruction (event $\event''$) between these two  instructions.
In the read operation, process $\proc_3$
loads  the initial value $0$ from $\yvar$ (line 2)  to register
$\reg_1$.
If $\proc_1$ can do that, it reaches  the label of line $4$.
In a similar way to $\proc_1$, process $\proc_2$  loads the initial value $0$ from $\xvar$ to register $\reg_2$.

The state reachability problem under POWER asks whether processes $\proc_1$ and $\proc_2$  can reach  the labels of lines $4$ and $10$ respectively at the same time. 
This state reachability problem has a negative answer according to the POWER semantics~\cite{DBLP:conf/pldi/SarkarSAMW11,DBLP:conf/pldi/SarkarMOBSMAW12}.

\begin{figure}[tb]
 \centering
\begin{minipage}[]{0.45\linewidth}
\centering
\small
 {
\begin{tabular}{| c |  c |}
\hline
 {{\bf Event}} & {\bf Instruction} \\ \hline\hline
 $\event_1$ & $0:\xvar\!\assigned\!1$ \\ 
$\event_2$ & $2:\reg_1\!\assigned\!\yvar$ \\ 
$\event_3$ & $6:\yvar\!\assigned\!1$ \\ 
$\event_4$ & $8:\reg_2\!\assigned\!\xvar$ \\ 
$\event'$ & $1:\hwsync$ \\
$\event''$ & $7:\hwsync$ \\
\hline
\end{tabular}
    } 
\end{minipage}
\caption{
Read, write, and synchronization events in the program in Figure~\ref{store-buffer}.
}
\label{fence_power_run:fig}
\end{figure}

We explain the negative result of the state reachability problem using the transition rules in Figure~\ref{full-rules:fig}.
In order to initialize
the read event $\event_2$,
$\proc_1$ must 
satisfy
the predicate $\allsyncscndof{\conf_2}{\event_2}$
for some $\conf_2\in\confset$
(see the rule
 ${\tt InitReadFromProp}$)
 by 
propagating
its sync event ($\event'$)
to itself and $\proc_2$.
To propagate $\event'$ to $\proc_2$,
all seen write events of $\event'$
must also be propagated to $\proc_2$ (see the rule ${\tt PropSync}$).
The seen write of $\event'$ for $\xvar$
is the write event $\event_1$ 
since $\event_1$ must be committed
and propagated to $\proc_1$ before
$\event'$ can be committed (see the rule ${\tt ComSync}$).
It means that
$\event_2$ can only be initialized after 
the write $\event_1$ has already been propagated to 
$\proc_2$.
 Similarly,
 $\event_4$ can only be initialized after the write event $\event_3$
 has already propagated 
 to $\proc_1$.
 As a consequence, 
 at least one of two processes 
 must see 
 the written value $1$ from the variable
 that it wants to read.
 In other words, it is not possible to allow both processes
 to load the initial values.

If we replace the two sync instructions by two lwsync instructions,
the state reachability problem has a positive answer.
 The reason is that
 in order to intitialize
 $\event_2$,
 $\proc_1$ 
 only need to commit 
 its lwsync event $\event'$  
 without
 propagating it to $\proc_2$ (see the rule ${\tt InitReadFromProp}$).
To commit $\event'$,
$\proc_1$ only need to commit the write event $\event_1$
and can delay the
 propagation of  $\event_1$ to $\proc_2$ (see the rule ${\tt ComSync}$).
It means that $\event_2$ can be initialized before
the write $\event_1$ is propagated to 
$\proc_2$.
 Similarly,
 $\event_4$ can be initialized before
the write $\event_3$ is propagated to 
$\proc_1$.
As a consequence, it is possible for both processes to see the initial value
from variables that they want to read.


\section{Translation with Address Operators and Synchronization Instructions}
\label{full:translation:section}
In this section,
we give the extension of our algorithm
in Section~\ref{translation:section}
that reduces the $\contextnum$-bounded state reachability problem under POWER to the corresponding 
problem under SC
for concurrent programs taking into account of the address operators and synchronization instructions.

Below, we present an extended scheme for the translation, our extended data structures, and 
the translated code for different types of the instructions.

\begin{figure}[tbh!]
\center
\begin{tikzpicture}[codeblock/.style={line width=0.5pt, inner xsep=0pt, inner ysep=0pt}]
\node[codeblock,font=\normalsize] (init) at (current bounding box.north west) {
\small{
$
\begin{array}{rcl}
\translateof{\prog}{\contextnum} &\myeq & \keyword{vars:}\; \hcancel[red]{\xvar^*}\;  \tuple{{\tt addvars}}_\contextnum\\
						&& \keyword{procs:}\; (\translateof{\proc}\contextnum)^*\; \tuple{{\tt initProc}}_\contextnum\; \tuple{{\tt verProc}}_\contextnum \\
\translateof{\proc}\contextnum
						&\myeq& \keyword{regs:}\;\reg^*\\
						&& \keyword{instrs:}\; (\translateof{\instr}\contextnum^\proc)^*\\
\translateof{\instr}\contextnum^\proc 	&\myeq& \lbl : \tuple{{\tt activeCnt}}_\contextnum^\proc\;\; \translateof{\stmt}\contextnum^\proc\;\; \tuple{{\tt closeCnt}}_\contextnum^\proc\\

\translateof{\xvar\assigned\expr}\contextnum^\proc 										&\myeq& \translateof{\xvar\assigned\expr}\contextnum^{\proc, {\tt Write}}\\
\translateof{[\expr']\assigned\expr}\contextnum^\proc 									&\myeq& \translateof{[\expr']\assigned\expr}\contextnum^{\proc, {\tt Write}}\\

\translateof{\reg\assigned[\expr]}\contextnum^\proc 										&\myeq& \translateof{\reg\assigned[\expr]}\contextnum^{\proc, {\tt Read}}\\
\translateof{\reg\assigned\xvar}\contextnum^\proc 										&\myeq& \translateof{\reg\assigned\xvar}\contextnum^{\proc, {\tt Read}}\\

\translateof{\reg\assigned\expr}\contextnum^\proc 										&\myeq& \translateof{\reg\assigned\expr}\contextnum^{\proc, {\tt Assign}}\\

\llbracket\keyword{if} \; \expr\; \keyword{then} \; \instr^*	&\myeq&\keyword{if} \; \expr\; \keyword{then} \; (\translateof{\instr}\contextnum^\proc)^*\\
\; \keyword{else} \; \instr^* \;\;\;\;\;\;\; \rrbracket_\contextnum^\proc && \keyword{else} \; (\translateof{\instr}\contextnum^\proc)^*;\;  \tuple{{\tt control}}_\contextnum^\proc\\
\translateof{\keyword{while} \; \expr\;\keyword{do} \; \instr^*}\contextnum^\proc					&\myeq&\keyword{while} \; \expr\; \keyword{do} \; (\translateof{\instr}\contextnum^\proc)^*;\;  \tuple{{\tt control}}_\contextnum^\proc\\

\translateof{\keyword{assume}~\expr}\contextnum^\proc 									&\myeq&\keyword{assume}~\expr; \; \tuple{{\tt control}}_\contextnum^\proc\\
\translateof{\keyword{assert}~\expr}\contextnum^\proc 									&\myeq&\keyword{assert}~\expr; \; \tuple{{\tt control}}_\contextnum^\proc\\

\translateof{\syncinstr}\contextnum^\proc												&\myeq& \translateof{\syncinstr}\contextnum^{\proc,{\tt Sync}}\\
\translateof{\lwsyncinstr}\contextnum^\proc											&\myeq& \translateof{\lwsyncinstr}\contextnum^{\proc,{\tt Lwsync}}\\
\translateof{\isyncinstr}\contextnum^\proc												&\myeq& \translateof{\isyncinstr}\contextnum^{\proc,{\tt Isync}}	\\

\translateof{\keyword\terminated}\contextnum^\proc										&\myeq&\keyword\terminated\\

\tuple{{\tt addvars}}_\contextnum&\myeq & \cvalof{\sizeof\procset}{\sizeof\varset}\contextnum\; \initcvalof{\sizeof\procset}{\sizeof\varset}\contextnum\\
						&& \utstof{\sizeof\procset}{\sizeof\varset}\contextnum\; \gtstof{\sizeof\procset}{\sizeof\varset}\contextnum\\
						&& \lvalof{\sizeof\procset}{\sizeof\varset}\\
						&& \ireadof{\sizeof\procset}{\sizeof\varset}\; \creadof{\sizeof\procset}{\sizeof\varset}\\
						&& \iwriteof{\sizeof\procset}{\sizeof\varset}\; \cwriteof{\sizeof\procset}{\sizeof\varset}\;\\
						&& \iregof{\sizeof\regset}\; \cregof{\sizeof\regset}\;\\
						&& \ctrlof{\sizeof\procset}\; \activeof\contextnum\; \ccontext\\
						&&{\color{blue}{\usyncststof{\sizeof\procset}\contextnum}\;\;\; \gsyncststof{\sizeof\procset}\contextnum}\\
						&&{\color{blue}{\hwsyncof{\sizeof\procset}\; \lwsyncof{\sizeof\procset}\; \isyncof{\sizeof\procset}}}\\
						&&{\color{blue}{\acksyncof{\sizeof\procset}\; \maxcrcwof{\sizeof{\procset}}}}\\ 
\tuple{{\tt activeCnt}}_\contextnum^\proc 		&\myeq&\assume{\activeof\ccontext=\proc}\\
\tuple{{\tt closeCnt}}_\contextnum^\proc		&\myeq&  \ccontext\assigned\ccontext+\genof{[0..\contextnum-1]};\\
					&&\keyword{assume}(\ccontext \leq \contextnum)\\
\tuple{{\tt control}}_\contextnum^\proc												&\!\myeq\!& \ctrlof\proc\!\assigned\!\ctrlof\proc + \genof{[0..\contextnum\!-\!1]}\!;\\
					&&\keyword{assume}(\ctrlof\proc \leq \contextnum)

\end{array}
$
}
};
\end{tikzpicture}
\caption{{ Translation map $\translateof{.}\contextnum$ with the address operators and synchronization instructions. We omit  the label of an intermediary  instruction when it is irrelevant. The additional variables are written in blue.}
}
\label{full-translation-map}
\end{figure}

\subsection{Scheme}
Figure~\ref{full-translation-map} gives our translation scheme that transforms a program $\prog$ into a program $\sprog$ following the map function $\translateof{.}\contextnum$. Let $\procset$ and $\varset$ be the sets of processes
and (shared) variables in $\prog$.
Similar to Section~\ref{translation:section}, the map  $\translateof{.}\contextnum$ replaces the variables of $\prog$ by $O(\sizeof\procset \cdot \contextnum)$ copies
of the set $\varset$, in addition to
a finite set of  {\it finite-data} structures (explained and formally defined in  Section~\ref{data:structures:section:full}).
The definition of ${\tt initProc}$ and ${\tt verProc}$ will  be given in Section~\ref{initializing:process:section:full} and Section~\ref{verifying:process:section:full} respectively.
The map function  $\translateof{.}\contextnum$  adds for each instruction $\instr$ appearing in $\prog$ the code  ${{\tt activeCnt}}$, the translation for  $\stmtof\instr$, and finally the code  ${{\tt closeCnt}}$.
The translations of write, read, assign, sync, lwsync, isync statements will be described in Section~\ref{write:instructions:full}, Section~\ref{read:instructions:full},
Section~\ref{assign:instructions:full},
Section~\ref{sync:instructions:full}, Section~\ref{lwsync:instructions:full}, and Section~\ref{isync:instructions:full} respectively.

\subsection{Data Structures}
\label{data:structures:section:full}
We keep the data structures
$$
\begin{gathered}
\cvalof{\sizeof\procset}{\sizeof\varset}\contextnum,  \initcvalof{\sizeof\procset}{\sizeof\varset}\contextnum,\utstof{\sizeof\procset}{\sizeof\varset}\contextnum, \gtstof{\sizeof\procset}{\sizeof\varset}\contextnum,\\\lvalof{\sizeof\procset}{\sizeof\varset}, 
\ireadof{\sizeof\procset}{\sizeof\varset}, \creadof{\sizeof\procset}{\sizeof\varset},
\iwriteof{\sizeof\procset}{\sizeof\varset}, \cwriteof{\sizeof\procset}{\sizeof\varset},\\
 \iregof{\sizeof\regset},\cregof{\sizeof\regset},
  \ctrlof{\sizeof\procset},
   \activeof\contextnum,
    \ccontext
\end{gathered}
$$
as in Section~\ref{translation:section}.
The translations of read and write instructions
 taking into account  the address operators
 can be extended from the corresponding translations 
 in Section~\ref{translation:section} by using these data structures.
 Below, we explain our added data structures to handle the synchronization instructions.
The additional data structures are written in blue in Figure~\ref{full-translation-map}.

Similar to the write events, we associate a
{\it timestamp} with each sync or lwsync event.
A synchronization timestamp $\stst$  is a mapping  
$\procset\mapsto[1..\contextnum]$. 
For
a process $\proc\in\procset$, the value of $\ststof\proc$ of
a given sync or lwsync event
represents
the context where the event is propagated to $\proc$.
In contrast to write events,
a sync or lwsync event always be propagated to 
all processes in the system, i.e.\ $1\leq\ststof\proc\leq \contextnum$ for all $\proc\in\procset$.
We use $\tstset$ to denote the set of timestamps  for both write events and synchronization events.
We 
keep the order $\tstorder$ 
and the summary operator $\summary$ on $\tstset$ 
as in Section~\ref{translation:section}.

Our simulation observes the sequence of sync and lwsync events received 
by a process in each context. 
Similar to the write events, the simulation 
will initially {\it guess} and later {\it verify}
the summaries of the timestamps of such a sequence.
This is done using  data structures $\gsyncstst$ and $\usyncstst$.
The mapping
$\gsyncstst:\procset\times[1..\contextnum]\mapsto\mapingsover\procset{[1..\contextnum]}$ stores, 
for process $\proc\in\procset$
and a context $\kk:1\leq\kk\leq\contextnum$, 
an {\it initial guess} $\gsyncststof\proc\kk$ of the summary  
of the timestamps of the sequence of synchronization events
 propagated to $\proc$ up to the 
{\it start} of the context $\kk$.
Starting from a given initial guess for a given context $\kk$, the timestamp is updated successively using the sequence of synchronization events propagated
to $\proc$
in $\kk$. 
The result is stored using the mapping
 $\usyncstst:\procset\times[1..\contextnum]\mapsto\mapingsover\procset{[1..\contextnum]}$.
More precisely, we initially set the value of
$\usyncstst$ to $\gsyncstst$.
Each time a new sync or lwsync event $\event$ 
is created by $\proc$ 
in a context $\kk$, 
we guess the timestamp $\nsyncstst$ of $\event$, and then
update $\usyncststof\proc\kk$ by computing its summary
with $\nsyncstst$.
Thus, given a point in a context $\kk$, $\usyncststof\proc\kk$ contains the summary
of the timestamps of the whole sequence
of synchronization events that
have been propagated to $\proc$ 
up to that point.
At the end of the simulation, we {\it verify}, for each
context $\kk:1\leq\kk<\contextnum$, that 
the value of $\usyncstst$ at the end of  the context $\kk$ is identical to the value
of $\gsyncstst$ for the next context $\kk+1$.

Furthermore, we use four data structures 
to keep track of the contexts
where the synchronization events
are committed and propagated.
The mapping $\hwsync:\procset\mapsto[1..\contextnum]$, 
$\lwsync:\procset\mapsto[1..\contextnum]$, and
$\isync:\procset\mapsto[1..\contextnum]$
give, for a process $\proc\in\procset$, the  committed contexts $\hwsyncof\proc$, $\lwsyncof\proc$, and $\isyncof\proc$
of the latest sync, lwsync, and isync events in $\proc$ respectively.
We  use $\acksync:\procset\mapsto[1..\contextnum]$
to store, for a process $\proc\in\procset$,
the maximal propagating context $\acksyncof\proc$ of all sync events in $\proc$.

We also use $\maxcrcw:\procset\mapsto[1..\contextnum]$
to store, for a process $\proc\in\procset$,
the maximal committing context $\maxcrcwof\proc$ of all read events 
that provide the values 
for some address expressions in some $\eorder$-successor events.
This function will be used to simulate the predicate $\defaddresscnd$
in the rule ${\tt ComISync}$.

\begin{figure}[tb]
\begin{minipage}{0.7\textwidth}
\small{
\begin{algorithm}[H]
\For {\mbox{$\proc\in\procset \wedge \xvar\in\varset$}}{\label{diverseinit}
$\ireadof\proc\xvar\assigned1$;
$\creadof\proc\xvar\assigned1$;
$\iwriteof\proc\xvar\assigned1$;
$\cwriteof\proc\xvar\assigned1$\;
$\lvalof\proc\xvar\assigned\zero$;
$\cvalof\proc\xvar1\assigned\zero$\;
  \For {\mbox{$\qproc\in\procset$}}{
	$\utstof\pproc\xvar1(\qproc)\assigned\tuple{2,1}$;
   }
}

\For {\mbox{$\proc\in\procset$}} {
$\hwsyncof\proc\assigned1$;
$\lwsyncof\proc\assigned1$;
$\isyncof\proc\assigned1$\;
$\ctrlof\proc\assigned1$;
$\acksyncof\proc\assigned1$;
$\maxcrcwof\proc\assigned1$\;
  \For {\mbox{$\qproc\in\procset$}} {
    $\usyncststof\proc1(\qproc)\assigned1$;
  }
}

\For {\mbox{$\reg\in\regset$}}{
$\iregof\reg\assigned1$;~
$\cregof\reg\assigned1$;
}

\For {\mbox{$\proc\in\procset  \wedge \xvar\in\varset \wedge \kk\in[2..\contextnum]$}}{\label{gtstinit}
 \For {\mbox{$\qproc\in\procset$}} {
	$\gtstof\proc\xvar\kk(\qproc)\assigned\genof{\contextnum^{\one\two}}$;
 }
$\utstof\proc\xvar\kk\assigned\gtstof\proc\xvar\kk$\;
$\initcvalof\proc\xvar\kk\assigned\genof\dataset$\;
$\cvalof\proc\xvar\kk\assigned\initcvalof\proc\xvar\kk$\;
}

\For {\mbox{$\proc\in\procset  \wedge \kk\in[2..\contextnum]$}}{\label{gtstinit}
 \For {\mbox{$\qproc\in\procset$}} {
	$\gsyncststof\proc\kk(\qproc)\assigned\genof{[1..\contextnum]}$;
 }
$\usyncststof\proc\kk\assigned\gsyncststof\proc\kk$\;
}

\For{$\kk\in[1..\contextnum]$}{\label{activeproc}
$\activeof\kk\assigned\genof\procset$\;
}

$\ccontext\assigned 1$\;
\caption{$\tuple{{\tt initProc}}_\contextnum$.}
\label{full-init:fig}
\end{algorithm}
}
\end{minipage}
\end{figure}

\subsection{Initializing Process}
\label{initializing:process:section:full}
Algorithm~\ref{full-init:fig} shows the initializating process.
The process initializes all data structures that will be used in the simulation program
$\sprog$ in a similar way to  Section~\ref{translation:section}.

\begin{figure}[tb]
\begin{minipage}{.75\linewidth}
\small{
\begin{algorithm}[H]
 \tcp{Guess}
$\iwriteof\proc\xvar\assigned\genof{[1..\contextnum]}$\;\label{geniw}
$\oldcwrite\assigned\cwriteof\proc\xvar$\;
$\cwriteof\proc\xvar\assigned\genof{[1..\contextnum]}$\;\label{gencw}
\For {$\qproc\in\procset$}{\label{genwtst1}
  $\ntstof\qproc\assigned\genof{\ncontextnum}$\;\label{genwtst2}
}\label{genwtst3}

\tcp{Check}
$\assume{\iwriteof\proc\xvar\geq \ccontext}$\;\label{iwccontext1}
$\assume{\activeof{\iwriteof\proc\xvar}=\proc}$\;\label{iwccontext2}
$\assume{\iwriteof\proc\xvar\geq\iregof\expr}$\;\label{iwdwrinitcnd}
$\assume{\iwriteof\proc\xvar\geq{{\tt max}}\set{\acksyncof\proc,\lwsyncof\proc,\isyncof\proc}}$\;  
$\assume{\cwriteof\proc\xvar\geq\iwriteof\proc\xvar}$\;\label{cwiw1}
${{\tt assume}}(\cwriteof\proc\xvar\geq {{\tt max}}\{\cregof\expr,\ctrlof\proc,\creadof\proc\xvar,\oldcwrite\})$\; \label{cwcmcnd}
\For {$\qproc\in\procset$}{\label{genwcheck}
  \If{$\qproc=\proc$}{\label{ntstcw}
    $\assume{\ntstof\qproc\in\contextnum^{\one} \wedge \projectof{\ntstof\qproc}=\cwriteof\proc\xvar}$\;
    }
  \If{$\qproc\neq\proc$}{\label{ntstcwother}
    ${\tt assume}(\ntstof\qproc\in\contextnum^{\one}\implies \projectof{\ntstof\qproc}\geq \cwriteof\proc\xvar)$\;
   $\assume{\ntstof\qproc\in\contextnum^{\one}\implies \projectof{\ntstof\qproc} \geq \usyncststof\proc{\projectof{\ntstof\proc}}(\qproc)}$\;\label{tstsgcnd}
    }
  \If{$\ntstof\qproc\in\contextnum^{\one}$}{\label{coherencecw}
    $\assume{\utstof\qproc\xvar{\projectof{\ntstof\qproc}}\tstorder\ntst}$\;
     $\assume{\activeof{\projectof{\ntstof\qproc}}=\proc}$\;
    } 
    \lElse {
    $\assume{\exists k: 1 \leq k \leq \contextnum: \ntst \tstorder \utstof\qproc{\xvar}{k}}$
    }
}

\tcp{Update}
\For {$\qproc\in\procset$}{\label{genwupdate}
  \If{$\ntstof\qproc\in\contextnum^{\one}$}{
    $\utstof\qproc\xvar{\projectof{\ntstof\qproc}}\assigned\utstof\qproc\xvar{\projectof{\ntstof\qproc}}\summary\ntst
    $\;
$\cvalof\qproc\xvar{\projectof{\ntstof\qproc}}\assigned\expr$\;\label{valupdate}
 }}
 $\lvalof\proc\xvar\assigned\expr$\;\label{lvalupdate}
 
\caption{{$\translateof{\xvar\assigned\expr}\contextnum^{\proc,{\tt Write}}$.}}
\label{full-write:fig}
\end{algorithm}
}
\end{minipage}
\end{figure}

\begin{figure}[tb]
\begin{minipage}{.9\linewidth}
\small{
\begin{algorithm}[H]
 \tcp{Guess}
$\iwriteof\proc{[\expr']}\assigned\genof{[1..\contextnum]}$\;\label{geniw}
$\oldcwrite\assigned\cwriteof\proc{[\expr']}$\;
$\cwriteof\proc{[\expr']}\assigned\genof{[1..\contextnum]}$\;\label{gencw}
\For {$\qproc\in\procset$}{\label{genwtst1}
  $\ntstof\qproc\assigned\genof{\ncontextnum}$\;\label{genwtst2}
}\label{genwtst3}

\tcp{Check}
$\assume{\iwriteof\proc{[\expr']}\geq \ccontext}$\;\label{iwccontext1}
$\assume{\activeof{\iwriteof\proc{[\expr']}}=\proc}$\;\label{iwccontext2}
$\assume{\iwriteof\proc{[\expr']}\geq\iregof{\expr+\expr'}}$\;\label{iwdwrinitcnd}
$\assume{\iwriteof\proc{[\expr']}\!\geq\!{{\tt max}}\!\set{\acksyncof\proc\!,\!\lwsyncof\proc\!,\!\isyncof\proc}}$\;
$\assume{\cwriteof\proc{[\expr']}\geq\iwriteof\proc{[\expr']}}$\;\label{cwiw1}
${\tt assume}(\cwriteof\proc{[\expr']}\geq\text{\hspace{0em}}{\tt max}\{\cregof{\expr+\expr'},\ctrlof\proc,\creadof\proc{[\expr']},\oldcwrite\})$\;
\For {$\qproc\in\procset$}{\label{genwcheck}
  \If{$\qproc=\proc$}{\label{ntstcw}
    $\assume{\ntstof\qproc\in\contextnum^{\one} \wedge \projectof{\ntstof\qproc}=\cwriteof\proc{[\expr']}}$\;
    }
  \If{$\qproc\neq\proc$}{\label{ntstcwother}
    ${\tt assume}(\ntstof\qproc\in\contextnum^{\one}\implies \projectof{\ntstof\qproc}\geq \cwriteof\proc{[\expr']})$\;
    $\assume{\ntstof\qproc\in\contextnum^{\one}\implies \projectof{\ntstof\qproc} \geq \usyncststof\proc{\projectof{\ntstof\proc}}(\qproc)}$\;\label{tstsgcnd}  	
    } 
  \If{$\ntstof\qproc\in\contextnum^{\one}$}{\label{coherencecw}
    $\assume{\utstof\qproc{[\expr']}{\projectof{\ntstof\qproc}}\tstorder \ntst}$\;
    $\assume{\activeof{\projectof{\ntstof\qproc}}=\proc}$\;
    }
    \lElse {
    $\assume{\exists k: 1 \leq k \leq \contextnum: \ntst \tstorder \utstof\qproc{[\expr']}{k}}$
    }
}

\tcp{Update}
\For {$\qproc\in\procset$}{\label{genwupdate}
  \If{$\ntstof\qproc\in\contextnum^{\one}$}{
    $\utstof\qproc{[\expr']}{\projectof{\ntstof\qproc}}\assigned\utstof\qproc{[\expr']}{\projectof{\ntstof\qproc}}\summary\ntst
    $\;
$\cvalof\qproc{[\expr']}{\projectof{\ntstof\qproc}}\assigned\expr$\;\label{valupdate}
 }}
 $\lvalof\proc{[\expr']}\assigned\expr$\;\label{lvalupdate}
\If{$\maxcrcwof\proc<\cregof{\expr'}$}{
$\maxcrcwof\proc\assigned\cregof{\expr'}$;
}
\caption{{$\translateof{[\expr']\assigned\expr}\contextnum^{\proc,{\tt Write}}$.}}
\label{full-addr-write:fig}
\end{algorithm}
}
\end{minipage}
\end{figure}

\subsection{Write Instructions}
\label{write:instructions:full}

Consider a write instruction $\instr$ of
a process $\proc\in\procset$ whose $\stmtof\instr$ is of the form
$\xvar\assigned\expr$ or $[\expr']\assigned\expr$.
Below we use $\xvar$ to present the variable in the instruction $\instr$ (that can be addressed by the value of $\expr'$).
The translation of this instruction is shown in Algorithm~\ref{full-write:fig} and Algorithm~\ref{full-addr-write:fig}.
Similar to Section~\ref{translation:section}, the code simulates an event $\event$ executing $\instr$, by
preforming of three parts, namely
{\it guessing}, {\it checking}, and {\it update}.

We mention the major changes in the translation of a write instruction in Algorithm~\ref{full-write:fig} and Algorithm~\ref{full-addr-write:fig}. 
In line~\ref{iwdwrinitcnd}, we check whether $\writeinitcnd$
in the rule ${\tt InitWrite}$ holds
by verifying that the  dependencies  $\dataorder$ and    $\addressorder$ is respected.
More precisely, we find, for each register $\reg$ that occurs
in $\regsetof{\instr}$, the initializing context of the latest 
read or assign event loading to $\reg$.
We make sure that the initializing context of $\event$ is later than
the initializing contexts of all these read events.
By definition, the largest of 
all these contexts is 
stored in $\iregof{\expr}$ if $\stmtof\instr$ is $\xvar=\expr$ or $\iregof{\expr+\expr'}$ if $\stmtof\instr$ is $[\expr']=\expr$.
In line~9, we check whether $\allsynscnd$
in the rule ${\tt InitWrite}$ is satisfied.
In 
line 11,
we check that $\commitcnd$ 
in the rule ${\tt ComWrite}$ is satisfied
by verifying that the committing context is larger
than 
\begin{enumerate}
\item
the committing context of all the read and assign events from
which the registers 
in $\regsetof{\instr}$ fetch their values
(to satisfy the  dependencies  $\dataorder$ and    $\addressorder$ in
a similar manner to that described for the initialization rule),
\item
the committing contexts of the latest
read and write events on $\xvar$ in $\proc$, i.e., $\creadof\proc\xvar$ and
$\cwriteof\proc\xvar$
(to satisfy the per-location program order $\plorder$), and
\item
the committing context of the latest
aci event in $\proc$, i.e., $\ctrlof\proc$
(to satisfy the control order $\ctrlorder$).
\end{enumerate}
We note that 
by the checking in lines 9 -- 10,
we guarantee the predicate
$\allsynscnd$
in the rule ${\tt ComWrite}$.
The for-loop of 
line 12
performs three sanity checks on 
$\ntst$ in a similar way  to Section~\ref{translation:section},
except that we add line~17
to guarantee  $\syncgroupcnd$
in the rule ${\tt Prop}$.

If the write instruction contain the address operator,
 we update $\maxcrcwof\proc$ in lines 27 -- 28 in Algorithm~\ref{full-addr-write:fig}
 to keep
  information about the maximal committing context
  of all read events 
 that provide the values for the registers in $\regsetof{\expr'}$.

\begin{figure}[tb]
\begin{minipage}{.8\linewidth}
\small{
\begin{algorithm}[H]
\tcp{Guess}
$\oldiread\assigned\ireadof\proc\xvar$\;\label{storeir}
$\ireadof\proc\xvar\assigned\genof{[1..\contextnum]}$;
$\iregof\reg\assigned\ireadof\proc\xvar$\;\label{genir}
$\oldcread\assigned\creadof\proc\xvar$\;\label{storecr}
$\creadof\proc\xvar\assigned\genof{[1..\contextnum]}$;
$\cregof\reg\assigned\creadof\proc\xvar$\;\label{gencr}

\tcp{Check}
$\assume{\ireadof\proc\xvar\geq\ccontext}$\;\label{irccontext1}
$\assume{\activeof{\ireadof\proc\xvar}=\proc}$\;\label{irccontext2}
$\assume{\ireadof\proc\xvar\geq\iwriteof\proc\xvar}$\;\label{iriwcnd}
\text{{\tt // An intended blank line}}\;
 $\assume{\ireadof\proc\xvar\geq{{\tt max}}\set{\acksyncof\proc,\lwsyncof\proc,\isyncof\proc}}$\;\label{irsyncrinitcnd}
${\tt assume}(\ireadof\proc\xvar\geq\cwriteof\proc\xvar  \implies \utstof\proc\xvar\oldiread\tstorder\utstof\proc\xvar{\ireadof\proc\xvar})$\;\label{rdcwrcnd}
$\assume{\creadof\proc\xvar\geq\ireadof\proc\xvar}$\;\label{crir1}
$\assume{\activeof{\creadof\proc\xvar}=\proc}$\;\label{crir2}
${\tt assume}(\creadof\proc\xvar\geq
{{\tt max}}\set{\ctrlof\proc,\oldcread,\cwriteof\proc\xvar})$\; \label{crcmcnd}
\tcp{Update}
 \lIf{$\ireadof\proc\xvar<\cwriteof\proc\xvar$}{\label{localread}
    $\reg\assigned\lvalof\proc\xvar$
    } \lElse {
    $\reg\assigned\cvalof\proc\xvar{\ireadof\proc\xvar}$
    }

\caption{{$\translateof{\reg\assigned\xvar}\contextnum^{\proc, {\tt Read}}$.}}
\label{full-read:fig}
\end{algorithm}
}
\end{minipage}

\end{figure}

\begin{figure}[tb]
\begin{minipage}{0.95\linewidth}
\small{
\begin{algorithm}[H]
\tcp{Guess}
$\oldiread\assigned\ireadof\proc{[\expr]}$\;\label{storeir}
$\ireadof\proc{[\expr]}\assigned\genof{[1..\contextnum]}$;
$\iregof\reg\assigned\ireadof\proc{[\expr]}$\;\label{genir}
$\oldcread\assigned\creadof\proc{[\expr]}$\;\label{storecr}
$\creadof\proc{[\expr]}\assigned\genof{[1..\contextnum]}$;
$\cregof\reg\assigned\creadof\proc{[\expr]}$\;\label{gencr}

\tcp{Check}
$\assume{\ireadof\proc{[\expr]}\geq\ccontext}$\;\label{irccontext1}
$\assume{\activeof{\ireadof\proc{[\expr]}}=\proc}$\;\label{irccontext2}
$\assume{\ireadof\proc{[\expr]}\geq\iwriteof\proc{[\expr]}}$\;\label{iriwcnd}
$\assume{\ireadof\proc{[\expr]}\geq\iregof\expr}$\;\label{irinitcnd}
$\assume{\ireadof\proc{[\expr]}\!\geq\!{{\tt max}}\!\set{\acksyncof\proc\!,\!\lwsyncof\proc\!,\!\isyncof\proc}}$\;\label{irsyncrinitcnd}
${\tt assume}(\ireadof\proc{[\expr]}\!\geq\!\cwriteof\proc{[\expr]} \!\implies\!  \utstof\proc{[\expr]}\oldiread\!\tstorder\!\utstof\proc{[\expr]}{\ireadof\proc{[\expr]}})$\;\label{rdcwrcnd}
$\assume{\creadof\proc{[\expr]}\geq\ireadof\proc{[\expr]}}$\;\label{crir1}
$\assume{\activeof{\creadof\proc{[\expr]}}=\proc}$\;\label{crir2}
${\tt assume}(\creadof\proc{[\expr]}\geq
{{\tt max}}\set{\cregof\expr, \ctrlof\proc,\oldcread,\cwriteof\proc{[\expr]}})$\; \label{crcmcnd}
\tcp{Update}
 \lIf{$\ireadof\proc{[\expr]}<\cwriteof\proc{[\expr]}$}{\label{localread}
    $\reg\assigned\lvalof\proc{[\expr]}$
    } \lElse {
    $\reg\assigned\cvalof\proc{[\expr]}{\ireadof\proc{[\expr]}}$
    }
\If{$\maxcrcwof\proc<\cregof{\expr}$}{
$\maxcrcwof\proc\assigned\cregof{\expr}$;
}
   
\caption{{$\translateof{\reg\assigned[\expr]}\contextnum^{\proc, {\tt Read}}$.}}
\label{full-addr-read:fig}
\end{algorithm}
}
\end{minipage}

\end{figure}

\subsection{Read Instructions}
\label{read:instructions:full}
Consider a read instruction $\instr$ of
a process $\proc\in\procset$ whose $\stmtof\instr$ is of the form
$\reg\assigned\xvar$ or $\reg\assigned[\expr]$.
Below we use $\xvar$ to present the variable in the instruction $\instr$ 
(that can be addressed by the value of $\expr$).
The translation of this instruction is shown in Algorithm~\ref{full-read:fig} and Algorithm~\ref{full-addr-read:fig}.
In a similar manner to a write instruction,
the code simulates an event $\event$ executing $\instr$
by performing the three parts: 
guessing, checking, and update.

We mention the major changes in the translation of a read instruction in 
Algorithm~\ref{full-read:fig} and Algorithm~\ref{full-addr-read:fig}.
In line  8,
we check whether the predicate
$\readinitcnd$ 
in 
the rules
${\tt InitReadFromLocal}$
and ${\tt InitReadFromProp}$
hold by verifying that the  dependency 
$\addressorder$ is respected
 (this line is empty in Algorithm~\ref{full-read:fig}.)
More precisely, we find, for each register $\reg$ that occurs
in $\regsetof{\instr}$, the initializing context of the latest 
read or assign event loading to $\reg$.
We make sure that the initializing context of $\event$ is later than
the initializing contexts of all these read and assign events.
In line 9,
we check whether 
$\allsynscnd$
in 
the rules
${\tt InitReadFromLocal}$
and ${\tt InitReadFromProp}$
is satisfied.
In 
line 13,
we check that $\commitcnd$ 
in the rule ${\tt ComRead}$ is satisfied
by verifying that the committing context is larger
than 
\begin{enumerate}
\item
the committing context of all the read and assign events from
which the registers 
in $\regsetof{\instr}$ fetch their values
(to satisfy the  dependency  $\addressorder$),
\item
the committing contexts of the latest
read and write events on $\xvar$ in $\proc$, i.e., $\creadof\proc\xvar$ and
$\cwriteof\proc\xvar$
(to satisfy the per-location program order $\plorder$), and
\item
the committing context of the latest
aci event in $\proc$, i.e., $\ctrlof\proc$
(to satisfy the control order $\ctrlorder$).
\end{enumerate}
We note that 
by the checking in lines 9 and 11,
we guarantee the predicate
$\allsynscnd$
in the rule ${\tt ComRead}$.

If the read instruction contain the address operator,
 we update $\maxcrcwof\proc$ in lines 16 -- 17 in Algorithm~\ref{full-addr-read:fig}
 to keep
  information about the maximal committing context
  of all read events 
 that provide the values for the registers in $\regsetof{\expr}$.

\begin{figure}[tb]
\begin{minipage}{0.65\linewidth}
\small{
\begin{algorithm}[H]
\tcp{Guess}
$\iregof\reg\assigned\genof{[1..\contextnum]}$;\label{genireg}\\
$\cregof\reg\assigned\genof{[1..\contextnum]}$\;\label{gencreg}

\tcp{Check}
$\assume{\iregof\reg\geq\ccontext}$\;\label{iregccontext1}
$\assume{\activeof{\iregof\reg}=\proc}$\;\label{iregccontext2}
$\assume{\iregof\reg\geq\iregof\expr}$\;\label{ireginitcnd}
$\assume{\iregof\reg\!\geq\!{{\tt max}}\!\set{\acksyncof\proc\!,\!\lwsyncof\proc\!,\!\isyncof\proc}}$\;\label{iregsyncrinitcnd}
$\assume{\cregof\reg\geq\iregof\reg}$\;\label{crregireg1}
$\assume{\activeof{\cregof\reg}=\proc}$\;\label{cregireg2}
${\tt assume}(\cregof\reg\geq
{{\tt max}}\set{\cregof\expr, \ctrlof\proc})$\; \label{cregcmcnd}
\tcp{Update}
$\reg\assigned\expr$
   
\caption{{$\translateof{\reg\assigned\expr}\contextnum^{\proc, {\tt Assign}}$.}}
\label{full-assign:fig}
\end{algorithm}
}
\end{minipage}

\end{figure}

\subsection{Assign Instructions}
\label{assign:instructions:full}
Consider an assign instruction $\instr$ of
a process $\proc\in\procset$ whose $\stmtof\instr$ is of the form
$\reg\assigned\expr$.
The translation of this instruction is shown in Algorithm~\ref{full-assign:fig}.
The code simulates an event $\event$ executing $\instr$
by performing the three parts: 
guessing, checking, and update.

We mention the major changes in the translation of an assign instruction in 
Algorithm~\ref{full-assign:fig}.
In line 6,
we check whether 
$\allsynscnd$
in 
the rule
${\tt InitAssign}$
is satisfied.
We note that 
by the checking in lines 6 and 7,
we guarantee the predicate
$\allsynscnd$
in the rule ${\tt ComAssign}$.

\begin{figure}
\begin{minipage}{0.65\textwidth}
\small{
\begin{algorithm}[H]
\tcp{Guess}
$\hwsyncof\proc\assigned\genof{[1..\contextnum]}$\;\label{gencsync}
\For {$\qproc\in\procset$}{\label{gensynctst1}
  $\nsyncststof\qproc\assigned\genof{[1..\contextnum]}$\;\label{gensynctst2}
}\label{gensynctst3}

\tcp{Check}
$\assume{\hwsyncof\proc\geq \ccontext}$\;\label{csyncccontext1}
$\assume{\activeof{\hwsyncof\proc}=\proc}$\;
${\tt assume}(\hwsyncof\proc\geq{{\tt max}}\{\ctrlof\proc\})$\; \label{cmcnd}
${\tt assume}(\hwsyncof\proc\geq{{\tt max}}\set{\acksyncof\proc,\lwsyncof\proc,\isyncof\proc})$\; \label{cwsynccnd}
$\assume{\forall \xvar\in\varset: \hwsyncof\proc\geq{{\tt max}}\set{\creadof\proc\xvar,\cwriteof\proc\xvar}}$\;
\For {$\qproc\in\procset$}{\label{genwcheck}
  \lIf{$\qproc=\proc$}{\label{ntstcw}
    $\assume{\nsyncststof\qproc=\hwsyncof\proc}$
    }
  \If{$\qproc\neq\proc$}{\label{ntstcwother}
    ${\tt assume}(\nsyncststof\qproc\geq \hwsyncof\proc)$\;
    \For {$\xvar\in\varset$}{
    	$\assume{ \utstof\proc\xvar{\nsyncststof\proc} \tstorder \utstof\qproc\xvar{\nsyncststof\qproc} }$\;
    }
    $\assume{\activeof{\nsyncststof\qproc}=\proc}$\;
    
    }
}
\tcp{Update}
\For {$\qproc\in\procset$}{\label{genwupdate}
    $\usyncststof\qproc{\nsyncststof\qproc}\assigned\usyncststof\qproc{\nsyncststof\qproc}\summary\nsyncstst$\;
 }
\For {$\qproc\in\procset$}{
 	\lIf{$\acksyncof\proc < \nsyncststof\qproc$}{
 		$\acksyncof\proc \assigned \nsyncststof\qproc$
 	}
 }
   
\caption{{$\translateof{\syncinstr}\contextnum^{\proc, {{\tt Sync}}}$.}}
\label{sync:fig}
\end{algorithm}
}
\end{minipage}

\end{figure}

\subsection{Sync Instructions}
\label{sync:instructions:full}
Consider a sync instruction $\instr$ of
a process $\proc\in\procset$ whose $\stmtof\instr$ is of the form
$\syncinstr$.
The translation of this instruction is shown in Algorithm~\ref{sync:fig}.
The code simulates an event $\event$ running $\instr$
by encoding the two inference
rules  ${\tt ComSync}$ and ${\tt PropSync}$.
In a similar manner to write, read, and assign instructions,
the translation scheme for a sync instruction consists of three parts:
guessing, checking, and   update.

\subsubsection{Guessing}
We guess  the committing  contexts
for the event $\event$, together with its timestamp.
In line~1, we guess the context where the event
$\event$ will be committed.
In the for-loop of line~\ref{genwtst1}, 
we guess a timestamp for $\event$ and store it in $\nsyncstst$.
This means that, for each process
$\qproc\in\procset$, we guess the context where the event $\event$ will
be propagated to $\qproc$ and 
we store this guess in $\nsyncststof\qproc$.

\subsubsection{Checking}
We perform sanity checks on the guessed values
in order to verify that they are consistent with
the POWER semantics.
Lines~4 -- 8 perform the sanity checks
for $\hwsyncof\proc$.
In lines 4 -- 5,  we verify that
the committing context for $\event$
is not smaller than the current context.
This captures the fact that commitment happens after fetching
of $\event$.
It also verifies that commitment happens in
a context where $\proc$ is active.
In line~6, we check whether $\commitcnd$ 
in the rule ${\tt ComSync}$ is satisfied.
To do that, we check that the committing context is larger
than
the committing context of the latest
aci event in $\proc$, i.e., $\ctrlof\proc$
(to satisfy the control dependency order $\ctrlorder$).
Note that  $\dataorder$ and  $\plorder$ (with identical variables)
are not defined for a $\hwsync$ event.
In line 7,  
we check that $\allsynscnd$
in the rule ${\tt ComSync}$ is satisfied.
In line 8,
we check that
$\comreadwritecnd$
in the rule ${\tt ComSync}$ is satisfied.

The for-loop of 
line 9
performs three sanity checks on 
$\ntst$.
In 
line 10, 
we verify that 
$\event$ is propagated to $\proc$ in the same context as
the one where it is committed.
This is consistent with the rule  ${\tt ComSync}$
which requires that when a $\hwsync$ event is committed
then it is immediately propagated to the committing process.
In 
line 11, 
we verify that the context where $\event$ is propagated
to a process $\qproc$ (different from $\proc$) is later than or equal to the one where $\event$ is committed.
This is to be consistent with the fact that a $\hwsync$ event is 
propagated to other processes only after it has been committed.
In 
lines 13 -- 14, 
we check whether $\writegroupcnd$ 
in the rule ${\tt PropSync}$ is satisfied.
Moreover, in line 15, we check that the event is propagated in the contexts where $\proc$ is active.

\subsubsection{Updating}
The for-loop of line 16
uses the timestamp guessed
above for updating the global data structure $\usyncstst$.
More precisely, when the event $\event$ is propagated to a process $\qproc$,
  we add
$\nsyncstst$ to the
summary of the timestamps of the sequence of synchronization events propagated to $\qproc$ 
up to the current point in the context $\nsyncststof\qproc$.
In 
the loop in line 18, we update $\acksyncof\proc$ 
to keep track of the maximal propagating context
of all  sync events of $\proc$.

\begin{figure}[tb]
\begin{minipage}{0.7\textwidth}
\small{
\begin{algorithm}[H]
\tcp{Guess}
$\oldclwsync\assigned\lwsyncof\proc$\;
$\lwsyncof\proc\assigned\genof{[1..\contextnum]}$\;\label{gencsync}
\For {$\qproc\in\procset$}{\label{gensynctst1}
  $\nsyncststof\qproc\assigned\genof{[1..\contextnum]}$\;\label{gensynctst2}
}\label{gensynctst3}

\tcp{Check}
$\assume{\lwsyncof\proc\geq \ccontext}$\;\label{csyncccontext1}
$\assume{\activeof{\lwsyncof\proc}=\proc}$\;
${\tt assume}(\lwsyncof\proc\geq{\tt max}\{\ctrlof\proc\})$\; \label{cmcnd}
${\tt assume}(\lwsyncof\proc\geq{{\tt max}}\set{\acksyncof\proc,\oldclwsync,\isyncof\proc})$\; \label{cwsynccnd}
$\assume{\forall \xvar\in\varset: \lwsyncof\proc\geq{{\tt max}}\set{\creadof\proc\xvar,\cwriteof\proc\xvar}}$\;

\For {$\qproc\in\procset$}{\label{genwcheck}
  \lIf{$\qproc=\proc$}{\label{ntstcw}
    $\assume{\nsyncststof\qproc=\lwsyncof\proc}$
    }
    \If{$\qproc\neq\proc$}{\label{ntstcwother}
    	${\tt assume}(\nsyncststof\qproc\geq \lwsyncof\proc)$\;
    	\For {$\xvar\in\varset$}{
    		$\assume{\utstof\proc\xvar{\nsyncststof\proc} \tstorder \utstof\qproc\xvar{\nsyncststof\qproc}}$\;
    	}
      	$\assume{\activeof{\nsyncststof\qproc}=\proc}$\;     
   }
}

\tcp{Update}
\For {$\qproc\in\procset$}{\label{genwupdate}
    $\usyncststof\qproc{\nsyncststof\qproc}\assigned\usyncststof\qproc{\nsyncststof\qproc}\summary\nsyncstst$\;
}

\caption{{$\translateof{\lwsyncinstr}\contextnum^{\proc,{{\tt Lwsync}}}$.}}
\label{lwsync:fig}
\end{algorithm}
}
\end{minipage}

\end{figure}

\subsection{Lwsync Instructions}
\label{lwsync:instructions:full}
Consider a lwsync instruction $\instr$ of
a process $\proc\in\procset$ whose $\stmtof\instr$ is of the form
$\lwsyncinstr$.
The translation of this instruction is shown in Algorithm~\ref{lwsync:fig}.
The code simulates an event $\event$ executing $\instr$
by encoding the two inference
rules  ${\tt ComSync}$ and ${\tt PropSync}$.
In a similar manner to a sync instruction,
the translation scheme for a lwsync instruction consists of three parts:
guessing, checking, and update.

\subsubsection{Guessing}
We guess  the committing context 
for the event $\event$ together with its timestamp.
In line~2, we guess the context  where the event
$\event$ will be committed
(having stored its old value in the previous line).
In the for-loop of line 3,
we guess a timestamp for $\event$ and store it in $\nsyncstst$.
This means that, for each process
$\qproc\in\procset$, we guess the context where the event $\event$ will
be propagated to $\qproc$ and 
we store this guess in $\nsyncststof\qproc$.

\subsubsection{Checking and Updating}
The checking and update parts 
in the translation for a lwsync instruction
are similar to the corresponding parts in the translation
for a sync instruction, 
except that
we do not need to update
$\acksyncof\proc$ (that is only used  for sync events).

\begin{figure}[tb]
\begin{minipage}{0.7\textwidth}
\small{
\begin{algorithm}[H]
\tcp{Guess}
$\oldcisync\assigned\isyncof\proc$\;
$\isyncof\proc\assigned\genof{[1..\contextnum]}$\;\label{gencsync}

\tcp{Check}
$\assume{\isyncof\proc\geq \ccontext}$\;\label{csyncccontext1}
$\assume{\activeof{\isyncof\proc}=\proc}$\;
${\tt assume}(\isyncof\proc\geq{\tt max}\{\ctrlof\proc\})$\; \label{cmcnd}
${\tt assume}(\isyncof\proc\geq{{\tt max}}\set{\acksyncof\proc,\lwsyncof\proc,\oldcisync})$\; \label{cwsynccnd}
${\tt assume}(\isyncof\proc\geq\maxcrcwof\proc$\;
   
\caption{{$\translateof{\isyncinstr}\contextnum^{\proc,{{\tt Isync}}}$.}}
\label{isync:fig}
\end{algorithm}
}
\end{minipage}

\end{figure}

\subsection{Isync Instructions}
\label{isync:instructions:full}
Consider an isync instruction $\instr$ of
a process $\proc\in\procset$ whose $\stmtof\instr$ is of the form
$\isyncinstr$.
The translation of this instruction is shown in Algorithm~\ref{isync:fig}.
The code simulates an event $\event$ running $\instr$
by encoding the  inference
rule  ${\tt ComISync}$.
In contrast to the transitions for write, read, sync, and lwsync instructions,
the translation scheme for a $\isync$ instruction only consists of two parts:
guessing and checking.

\subsubsection{Guessing}
In line~2, we guess the context where the event
$\event$ will be committed
(having stored its old value in the previous line).

\subsubsection{Checking}
We perform sanity checks on the guessed values
in order to verify that they are consistent with
the POWER semantics.
Lines~3 -- 7 perform the sanity checks
for $\isyncof\proc$.
In lines 3 -- 4,  we verify that
the committing context for $\event$
is not smaller than the current context.
This captures the fact that commitment happens after fetching
of $\event$.
It also verifies that commitment happens in
a context where $\proc$ is active.
In line~5, we check whether $\commitcnd$ 
in the rule ${\tt ComISync}$ is satisfied.
To do that, we check that the committing context is larger
than
the committing context of the latest
aci event in $\proc$, i.e., $\ctrlof\proc$
(to satisfy the control order $\ctrlorder$).
Note that  $\dataorder$ and  $\plorder$ (with identical variables)
are not defined for an $\isync$ event.
In line 6,  
we check that $\allsynscnd$
in the rule ${\tt ComISync}$ is satisfied.
In line 7,
we check that 
$\defaddresscnd$
in the rule ${\tt ComISync}$ is satisfied.

\begin{figure}[tb]
\begin{minipage}{0.55\textwidth}
\small{
\begin{algorithm}[H]
\For {\mbox{$\proc\in\procset \wedge \xvar\in\varset \wedge \kk\in[1..\contextnum-1]$}}{\label{gtstverfiy}
$\assume{\utstof\proc\xvar\kk=\gtstof\proc\xvar{\kk+1}}$\;
$\assume{\cvalof\proc\xvar\kk=\initcvalof\proc\xvar{\kk+1}}$\;
}
\For {\mbox{$\proc\in\procset \wedge \kk\in[1..\contextnum-1]$}}{\label{gtstverfiy}
$\assume{\usyncststof\proc\kk=\gsyncststof\proc{\kk+1}}$\;
}

\lIf {$\lbl$ is reachable} {
 {\it  error}
 }
\caption{{$\tuple{{\tt verProc}}_\contextnum$.}}
\label{full-verify:fig}
\end{algorithm}
}
\end{minipage}
\end{figure}

\subsection{Verifying Process}
\label{verifying:process:section:full}
In Algorithm~\ref{full-verify:fig},
the verifying process makes sure that the updated value $\utst$  of the
timestamp of write events for each pair of process and variable  at the end of a given context is equal
to the guessed value $\gtst$ at the start of the next context.
It also make sure that
the updated value $\usyncstst$ of the
timestamp of synchronization events for each  process at the end of a given context is equal
to the guessed value $\gsyncstst$ at the start of the next context.
Moreover, the verifier process  
 performs the corresponding test for the values
written to the variables (by comparing $\cval$ and $\initcval$).
Finally, it checks whether we reach an error label $\lbl$ (given in the state reachability problem) or not.


\section{Translation Correctness}
From the translation given in Figure~\ref{full-translation-map}
and 
the reasoning followed 
in Section~\ref{translation:section} and Section~\ref{full:translation:section},
we can prove the following theorem.

\begin{thm}
Given an input concurrent program
$\prog$ and a natural number 
$\contextnum$, the code-to-code translation constructs an output concurrent
program $\sprog$ whose size is polynomial
 in $\prog$ and $\contextnum$.
 Moreover,
 for 
a given label 
$\lbl$
and a complete configure
$\conf$, 
there is a complete  $\contextnum$-bounded run
$\run$ of $\prog$ under POWER
such that
$\initconf\movesto\run\conf$ where $\lbl\in\lblof\conf$
if and only if
there is a complete  $\contextnum$-bounded run
$\srun$ of $\sprog$ under SC
such that
$\initconf\movesto\srun\conf$. 
\end{thm}


\section{Experimental Results}
\label{experiment:section}

In order to evaluate the efficiency of our approach,
we have implemented a context-bounded model checker for  concurrent programs running under the POWER semantics, called {\sf Power2SC}. 
We use {\sf CBMC} version 5.1  \cite{DBLP:conf/tacas/ClarkeKL04} as the backend verification tool
because {\sf CBMC} (i) supports our guessing-updating-verifying schema by  a non-deterministic choice of data and (ii) checks state reachability problem of concurrent programs running under the SC semantics.

\subsection{Litmus Tests.}
We have  tested {\sf Power2SC} on small litmus tests.  {\sf Power2SC} manages to successfully run  all  913  litmus tests published in~\cite{DBLP:conf/pldi/SarkarSAMW11}. Furthermore, the output result returned by  {\sf Power2SC}  (with 5 as the maximum number of context switches)  matches the ones returned by the tool \textsf{PPCMEM}~\cite{DBLP:conf/pldi/SarkarSAMW11} in all tests.

\subsection{C/Pthreads Benchmarks.}
 In the following, we present the  evaluation  of {\sf Power2SC} on 24 C/Pthreads
benchmarks 
collected from 
{\sf Goto-instrument}~\cite{DBLP:conf/esop/AlglaveKNT13}, {\sf Nidhugg}~\cite{DBLP:conf/cav/AbdullaAJL16}, {\sf Memorax}~\cite{DBLP:conf/tacas/AbdullaACLR12}, and the SV-COMP17 benchmark suit~\cite{svcomp17}. 
These are widespread medium-sized benchmarks
that are used by many tools for analyzing  concurrent programs running under weak memory models
(e.g.,~\cite{KVY2011,ABP2011,BM2008,AlglaveKT13,Zhang:pldi15,tacas15:tso,fmcad16,DBLP:conf/esop/BouajjaniDM13,DBLP:conf/concur/AbdullaABN16,DBLP:conf/esop/AbdullaAP15,Huang016,DBLP:conf/netys/AbdullaALN15,DBLP:conf/sas/AbdullaACLR12,BAM07}).

We divide our results in  two sets.
The first set concerns unsafe programs  while the second set concerns  safe ones. In both parts, we compare 
results obtained from
{\sf Power2SC}  to the ones obtained from {\sf Goto-instrument}~\cite{DBLP:conf/esop/AlglaveKNT13} and {\sf Nidhugg}~\cite{DBLP:conf/cav/AbdullaAJL16}, which are, to the best of our knowledge,  the only two  tools supporting  C/Pthreads  programs running under the  POWER semantics. We note that CBMC  previously supported POWER~\cite{AlglaveKT13}, but has withdrawn support in later versions. Meanwhile, two recent SMT-based tools \textsf{DARTAGNAN}~\cite{GavrilenkoLFHM19}  and \textsf{PORTHOS}~\cite{LeonFHM17} do not accept C/Pthreads programs as their input files.

All experiments were run on a machine equipped with a 
2.4 Ghz Intel x86-32 Core2 processor and 4 GB RAM.
Furthermore, we set up the time out to 1800 seconds in all experiments.

\begin{table}
\centering
\caption{
Comparing  {\sf Power2SC} to {\sf Goto-instrument}  and  {\sf Nidhugg} on bug detection in unsafe benchmarks.
The LB column indicates whether the tools were instructed to unroll loops up to a certain bound. 
The CB column gives the context bound for {\sf Power2SC}.
The program size is the number of code lines.
A {\it t/o} entry means that the tool failed to complete within 1800 seconds.  
The best running time (in seconds) for each benchmark is given in bold font.
}
\label{table1-fig}
\begin{tabular}{  l |  c |    r    r  r  r  }
\hline
\multirow{2}{*}{{\bf Program/size}} & \multirow{2}{*}{{\bf LB}} &  \multicolumn{1}{c}{ \;\;\;\;{\sf Goto-instrument}}  & \multicolumn{1}{c}{\;{\sf Nidhugg}}  & \multicolumn{2}{c}{\;\;\;{\sf Power2SC}} \\ \cline{3-6} 
                         										&                    	& {\bf time} (s)   & {\bf time} (s)  & \;\;\;\;\;\;{\bf time} (s)  & \;\;{\bf CB}         	\\ \hline\hline
 Bakery/76  \cite{DBLP:conf/tacas/AbdullaACLR12}        		&        8           	&    226         	&   t/o      		&    \bf{1}    		&       3      	\\ 
 Burns/74 \cite{DBLP:conf/tacas/AbdullaACLR12}          		&        8           	&    t/o       	&   t/o            	&    \bf{1}    		&       3     		\\
Dekker/82  \cite{svcomp17}  							&        8           	&    t/o       	&   t/o           	&    \bf{1}     		&       2     		\\ 
 Simple Dekker/69   \cite{DBLP:conf/tacas/AbdullaACLR12}	&        8           	&    12       	&   t/o           	&    {\bf 1}  		&       2     		\\ 
 Dijkstra/82 \cite{DBLP:conf/tacas/AbdullaACLR12}   		&        8           	&    t/o       	&   t/o           	&    {\bf 5}  		&       3       	\\
 Szymanski/83  \cite{svcomp17}						&        8           	&    t/o      		&   t/o      		&    {\bf 1}   		&       4         	\\ 
 Fib\_bench\_0/36 \cite{svcomp17}						&        -           	&    {\bf 2}   	&   1101      	&    4          		&       6        	\\ 
 Lamport/109 \cite{svcomp17} 							&        8           	&    t/o       	&   {\bf 1}    	&    {\bf 1}       		&       3      	\\ 
 Peterson/76 \cite{svcomp17} 							&        8           	&    25      		&   1056   		&    {\bf 1}       		&       3     		\\
 Peterson\_3/96 \cite{DBLP:conf/tacas/AbdullaACLR12}		&        8           	&    t/o       	&   {\bf 1}   	&    3          		&       4        	\\ 
 Pgsql/69	\cite{DBLP:conf/esop/AlglaveKNT13}			&        8           	&    1079   	&   {\bf 1}        	&    {\bf 1}       		&       2           	\\ 
 Pgsql\_bnd/71  \cite{DBLP:conf/cav/AbdullaAJL16} 			&        -           	&    t/o       	&   {\bf 1}    	&    {\bf 1}       		&       2      	\\ 
%
\hline
\end{tabular}
\label{table1-fig}
\end{table}

\subsubsection{Unsafe Benchmarks}
Table~\ref{table1-fig} shows 
the comparision of 
{\sf Power2SC}
to {\sf Goto-instrument}  and  {\sf Nidhugg} on  bug detection in unsafe benchmarks.
We recall that both {\sf Goto-instrument}  and {\sf Power2SC} use {\sf CBMC} as their backend model checker.
Since {\sf CBMC} and {\sf Nidhugg} respectively implement bounded model checking and stateless model checking techniques, they can only work with loop-free programs. Therefore, for any original benchmark containing loops, we have to unroll all the loops  to a certain bound presented by the LB column.
We also note that for the purpose of bug detection, all three tools will be stopped when they hit the first bug.

From Table~\ref{table1-fig}, 
we see that
{\sf Power2SC}  performs  well to detect bugs compared to  {\sf Goto-instrument}  and  {\sf Nidhugg} for most of the unsafe examples. 
The CB column gives the minimum number of context switches for {\sf Power2SC} to detect a bug.
We observe that {\sf Power2SC} manages to find all  the errors using at most 6 contexts
while
{\sf Nidhugg} and {\sf Goto-instrument} time out to return the  errors
for several examples.
This  confirms the observation by Qadeer et al.~\cite{Qadeer08,DBLP:conf/tacas/QadeerR05,MQ07} that few context switches are
normally sufficient to find many bugs in practice.

We believe that the main differences in the performance of {\sf Power2SC}, {\sf Nidhugg}, and {\sf Goto-instrument} in 
Table~\ref{table1-fig} are comes from different strategies used by them to find bugs.
Two tools, {\sf Goto-instrument} and {\sf Nidhugg}, 
consider all possible interleavings between processes\footnote{{\sf Nidhugg}  can reduce the number of explored runs by a dynamic partial order reduction technique.}.
Meanwhile, our tool,  {\sf Power2SC}, limits the number of context switches for each process.
Therefore, 
{\sf Power2SC} can avoid the problem of state explosion in the state reachability problem~\cite{Clarke2012} by  checking  only a subset of the full state space of the input program.

\begin{table}
\centering
\caption{
Comparing {\sf Power2SC} with {\sf Goto-instrument}  and   {\sf Nidhugg} on  safe benchmarks.
The LB and   CB columns have the same meaning as in Table~\ref{table1-fig}.
A {\it t/o} entry means that the tool failed to complete within 1800 seconds.  
The best running time (in seconds) for each benchmark is given in bold font.
}
     \label{table2-fig}
\begin{tabular}{  l |  c |    r    r  r  r  }
\hline
\multirow{2}{*}{{\bf Program/size}} & \multirow{2}{*}{{\bf LB}} &  \multicolumn{1}{c}{ \;\;\;\;{\sf Goto-instrument}}  & \multicolumn{1}{c}{\;{\sf Nidhugg}}  & \multicolumn{2}{c}{\;\;\;{\sf Power2SC}} \\ \cline{3-6} 
                         										&                    	& {\bf time} (s)   & {\bf time} (s)  & \;\;\;\;\;\;{\bf time} (s)  & \;\;{\bf CB}         	\\ \hline\hline
 Bakery/85   \cite{DBLP:conf/tacas/AbdullaACLR12}      	&        8          &    t/o			&  t/o      		&    \bf{20}         	&     3        	\\ 
 Burns/79 \cite{DBLP:conf/tacas/AbdullaACLR12}           	&        8          &    t/o      		&   t/o      		&    {\bf 767}         	&     3       		\\ 
 Dekker/88 \cite{svcomp17}        					&        8          &    t/o       		&   t/o           	&    \bf{1133}         	&     2       		\\
 Simple Dekker/73 \cite{DBLP:conf/tacas/AbdullaACLR12} &        8          &    209       		&   t/o    		&    {\bf 6}          	&     2        	\\
 Dijkstra/88  \cite{DBLP:conf/tacas/AbdullaACLR12}  	&        8          &    t/o       		&   t/o           	&    t/o        		&     3        	\\
Szymanski/93   \cite{svcomp17} 					&        8          &    t/o       		&   t/o           	&     {\bf 36}        	&     4        	\\ 
Fib\_bench\_1/36 \cite{svcomp17} 					&        -           &    9       		&   t/o           	&    \bf{5}        		&     6        	\\ 
 Lamport/119 \cite{svcomp17}						&        8          &    t/o       		&   t/o           	&   t/o        		&     3      		\\ 
Peterson/84 \cite{svcomp17} 						&        8          &    928      		&   t/o           	&     {\bf 4}         	&     3      		\\ 
 Peterson\_3/111 \cite{DBLP:conf/tacas/AbdullaACLR12} 	&        8          &    t/o       		&   t/o           	&     {\bf 50}     	&     4        	\\ 
 Pgsql/73	\cite{DBLP:conf/esop/AlglaveKNT13}		&        8          &    1522       	&   {\bf 2}   	&     18            		&     2         	\\ 
Pgsql\_bnd/75  \cite{DBLP:conf/cav/AbdullaAJL16}  		&        -           &    t/o       		&   t/o           	&     {\bf 4}         	&     2      		\\ 
\hline         
     \end{tabular}

\end{table}

\subsubsection{Safe Benchmarks}
Table~\ref{table2-fig} shows the comparison of 
{\sf Power2SC}
to {\sf Goto-instrument}  and  {\sf Nidhugg} on safe benchmarks.
%
We use the same number of context bounds  for {\sf Power2SC}
as in the case of unsafe examples in Table~\ref{table1-fig}.

We observe that
{\sf Power2SC} manages to run most of the examples (except   {\sf Dijkstra}  and {\sf Lamport}) while {\sf Goto-instrument} and {\sf Nidhugg} time out 
for many examples.
It is important to note that   {\sf Goto-instrument} and {\sf Nidhugg}  do not impose any bound on the number of context switches while {\sf Power2SC}  does.
It means that in the case where   they
return a result for an input program, we can conclude that the program is safe because the whole state space of the program is checked and no error is detected.
For the case of
{\sf Power2SC}, we  only can say that
the input program is safe with respect to a specific context bound.

\subsection{Scaling to Loop Bound and Context Switches}
We perform  more experiments to see how {\sf Power2SC} can be scalable 
with respect to
the numbers of loop bounds and context switches.

\begin{table}
\centering
\caption{
Running  {\sf Power2SC} with different loop bound (LB) on unsafe benchmarks.
We use the same numbers of context bound as in Table~\ref{table1-fig}.
A {\it t/o} entry means that the tool failed to complete within 1800 seconds.  
The best running time (in seconds) for each benchmark is given in bold font.
}
\label{table3-fig}
\begin{tabular}{  l |  c |     r    r  r    }
\hline
{\bf Program/size} & {\bf CB}  & \;\;\;\;\;{\bf LB} = 3 (s)  & \;\;\;\;\;{\bf LB} = 5  (s) & \;\;\;\;\;{\bf LB} = 7 (s)  \\ \hline\hline
 Bakery/76  \cite{DBLP:conf/tacas/AbdullaACLR12}        		&        3           	&    1        	&   1      		&    1   		   	\\ 
 Burns/74 \cite{DBLP:conf/tacas/AbdullaACLR12}          		&        3           	&    1       	&   1           	&    1   		   		\\
Dekker/82  \cite{svcomp17}  							&        2           	&    1        &  1           	&    1     				\\ 
 Simple Dekker/69   \cite{DBLP:conf/tacas/AbdullaACLR12}	&        2           	&    1       	&   1           	&    1  			\\ 
 Dijkstra/82 \cite{DBLP:conf/tacas/AbdullaACLR12}   		&        3           	&    5      	&    5           	&    5 		     	\\
 Szymanski/83  \cite{svcomp17}						&        4           	&    1     	&   1     		&    1		  	\\ 
 Lamport/109 \cite{svcomp17} 							&        3           	&    1     	&    1    		&    1      		   	\\ 
 Peterson/76 \cite{svcomp17} 							&        3           	&    1     	&    1   		&    1       		    		\\
 Peterson\_3/96 \cite{DBLP:conf/tacas/AbdullaACLR12}		&        4           	&    2      	&   2   		&    2          		     	\\ 
 Pgsql/69	\cite{DBLP:conf/esop/AlglaveKNT13}			&        2           	&    1   	&   1        		&    1       		       	\\ 
%
\hline
\end{tabular}
\end{table}

\begin{table}
\centering
\caption{
Running  {\sf Power2SC} with different loop bound (LB) on safe benchmarks.
We use the same numbers of context bound as in Table~\ref{table2-fig}.
A {\it t/o} entry means that the tool failed to complete within 1800 seconds.  
The best running time (in seconds) for each benchmark is given in bold font.
}
\label{table4-fig}
\begin{tabular}{  l |  c |     r    r  r    }
\hline
{\bf Program/size} & {\bf CB}  & \;\;\;\;\;{\bf LB} = 3 (s) & \;\;\;\;\;{\bf LB} = 5 (s) & \;\;\;\;\;{\bf LB} = 7 (s) \\ \hline\hline
 Bakery/76  \cite{DBLP:conf/tacas/AbdullaACLR12}        		&        3           	&    9        	&   24      		&    45   		   	\\ 
 Burns/74 \cite{DBLP:conf/tacas/AbdullaACLR12}          		&        3           	&    8       	&   62           	&    439   		   		\\
Dekker/82  \cite{svcomp17}  							&        2           	&    18       &  138           	&    721     				\\ 
 Simple Dekker/69   \cite{DBLP:conf/tacas/AbdullaACLR12}	&        2           	&    2       	&   5           	&    9  			\\ 
 Dijkstra/82 \cite{DBLP:conf/tacas/AbdullaACLR12}   		&        3           	&    48      	&    t/o           	&    t/o 		     	\\
 Szymanski/83  \cite{svcomp17}						&        4           	&    9     	&   26     		&    57		  	\\ 
 Lamport/109 \cite{svcomp17} 							&        3           	&    277     &    t/o    		&    t/o      		   	\\ 
 Peterson/76 \cite{svcomp17} 							&        3           	&    2     	&    4   		&    6       		    		\\
 Peterson\_3/96 \cite{DBLP:conf/tacas/AbdullaACLR12}		&        4           	&    42      	&   116   		&    237          		     	\\ 
 Pgsql/69	\cite{DBLP:conf/esop/AlglaveKNT13}			&        2           	&    5   	&   14        	&    29       		       	\\ 
%
\hline
\end{tabular}
\end{table}

\subsubsection{Scaling to Loop Bound}
Table~\ref{table3-fig} and Table~\ref{table4-fig}
show the performance of 
{\sf Power2SC} on the same sets of unsafe and safe benchmarks (except \textsf{Fib\_bench\_0} and \textsf{Pgsql\_bnd}  because they does not depend on the loop bound) as in 
Table~\ref{table1-fig} and Table~\ref{table2-fig} 
with different numbers of loop bounds: 3, 5, and 7.

We have different observations for unsafe and safe benchmarks. 
For the unsafe ones in Table~\ref{table3-fig}, 
{\sf Power2SC} behaves quite similar with respect to different loop bounds. 
We conjecture this behaviour  by the characteristic of bugs in these benchmarks: the bugs are swallow in the sense that they can be shown by a small number of context switches and each process does not need to perform many steps to hit the bugs.
In contrast to the case of unsafe benchmarks,
for the case of safe ones in Table~\ref{table4-fig},
we see the relation between loop bounds and the running time of {\sf Power2SC}.
When we increase the number of loop bounds, the size of an input program is bigger, and {\sf Power2SC}
has to spend more time to check all possible behaviours (with respect to specific loop bounds).

\begin{table}
\centering
\caption{
Running  {\sf Power2SC} with different context bound (CB) on unsafe benchmarks.
We use the same numbers of loop bound (LB) as in Table~\ref{table1-fig}.
The  CB + 1 column means that we increase the numbers of context bounds by 1 comparing to the context bounds in Table~\ref{table1-fig}.
A {\it t/o} entry means that the tool failed to complete within 1800 seconds.  
The best running time (in seconds) for each benchmark is given in bold font.
}
\label{table5-fig}
\begin{tabular}{  l |  c |     r    r  r    }
\hline
{\bf Program/size} & {\bf LB}  & \;\;\;\;\;{\bf CB} + 1 (s) & \;\;\;\;\;{\bf CB} + 2 (s) & \;\;\;\;\;{\bf CB} + 3 (s)  \\ \hline\hline
 Bakery/76  \cite{DBLP:conf/tacas/AbdullaACLR12}        		&        8           	&    1        	&   1      		&    1   		   	\\ 
 Burns/74 \cite{DBLP:conf/tacas/AbdullaACLR12}          		&        8           	&    1       	&   1           	&    1   		   		\\
Dekker/82  \cite{svcomp17}  							&        8           	&    1        &  1           	&    1     				\\ 
 Simple Dekker/69   \cite{DBLP:conf/tacas/AbdullaACLR12}	&        8           	&    1       	&   1           	&    1  			\\ 
 Dijkstra/82 \cite{DBLP:conf/tacas/AbdullaACLR12}   		&        8           	&    6      	&    9           	&    9 		     	\\
 Szymanski/83  \cite{svcomp17}						&        8           	&    1     	&   1     		&    1		  	\\ 
Fib\_bench\_1/36 \cite{svcomp17} 						&	  -         	&    4       	&   5           	&    7        		\\ 
 Lamport/109 \cite{svcomp17} 							&        8           	&    2     	&    2    		&    3      		   	\\ 
 Peterson/76 \cite{svcomp17} 							&        8           	&    1     	&    1   		&    1       		    		\\
 Peterson\_3/96 \cite{DBLP:conf/tacas/AbdullaACLR12}		&        8           	&    3      	&   4   		&    6          		     	\\ 
 Pgsql/69	\cite{DBLP:conf/esop/AlglaveKNT13}			&        8           	&    1   	&   1        		&    1       		       	\\ 
 Pgsql\_bnd/75  \cite{DBLP:conf/cav/AbdullaAJL16}  			&        -           	&    1       &   1           	&     1         			\\ 
%
\hline
\end{tabular}
\end{table}

\begin{table}
\centering
\caption{
Running  {\sf Power2SC} with different context bound (CB) on safe benchmarks.
We use the same numbers of loop bound (LB) as in Table~\ref{table2-fig}.
The CB + 1 column means that we increase the numbers of context bounds by 1 comparing to the context bounds in Table~\ref{table2-fig}.
A {\it t/o} entry means that the tool failed to complete within 1800 seconds.  
The best running time (in seconds) for each benchmark is given in bold font.
}
\label{table6-fig}
\begin{tabular}{  l |  c |     r    r  r    }
\hline
{\bf Program/size} & {\bf LB}  & \;\;\;\;\;{\bf CB} + 1 (s) & \;\;\;\;\;{\bf CB} + 2 (s) & \;\;\;\;\;{\bf CB} + 3 (s)  \\ \hline\hline
 Bakery/76  \cite{DBLP:conf/tacas/AbdullaACLR12}        		&        8           	&    83        	&   97      		&    122   		   	\\ 
 Burns/74 \cite{DBLP:conf/tacas/AbdullaACLR12}          		&        8           	&    1193       	&   1237           	&    1288   	  	\\
Dekker/82  \cite{svcomp17}  							&        8           	&    1580        	&  1609           	&    1650     		\\ 
 Simple Dekker/69   \cite{DBLP:conf/tacas/AbdullaACLR12}	&        8           	&    5       		&   5           	&    6  			\\ 
 Dijkstra/82 \cite{DBLP:conf/tacas/AbdullaACLR12}   		&        8           	&    t/o      		&    t/o           	&    t/o		     	\\
 Szymanski/83  \cite{svcomp17}						&        8           	&    90     		&   101     		&    111	  		\\ 
Fib\_bench\_1/36 \cite{svcomp17} 						&	  -         	&    6       		&   6           	&    8        			\\ 
 Lamport/109 \cite{svcomp17} 							&        8           	&    t/o     		&    t/o    		&    t/o   		   	\\ 
 Peterson/76 \cite{svcomp17} 							&        8           	&    7     		&    9   		&    10      		    	\\
 Peterson\_3/96 \cite{DBLP:conf/tacas/AbdullaACLR12}		&        8           	&    421      	&   590   		&    773         		 \\ 
 Pgsql/69	\cite{DBLP:conf/esop/AlglaveKNT13}			&        8           	&    19   		&   21        	&    19       		 \\ 
 Pgsql\_bnd/75  \cite{DBLP:conf/cav/AbdullaAJL16}  			&        -           	&    4       		&  4           	&     5         		\\ 
%
\hline
\end{tabular}
\end{table}

\subsubsection{Scaling to Context Switches}
Table~\ref{table5-fig} and Table~\ref{table6-fig}
show the performance of 
{\sf Power2SC} on the same sets of unsafe and safe benchmarks  as in 
Table~\ref{table1-fig} and Table~\ref{table2-fig} 
with different numbers of context bounds.
We use the notion  {CB} + 1 to indicate that
we 
increase the numbers of context bounds by 1 comparing to the context bounds in
Table~\ref{table1-fig} or Table~\ref{table2-fig}.

There are two main observations. First, we see that while increasing the number of context switches, 
{\sf Power2SC} slows down. This performance decreasing is quite small for unsafe benchmarks, but it becomes clearer for the safe ones. 
Second, we see that
although {\sf Power2SC} slows down when we increase the number of context bounds,
it scales well: {\sf Power2SC} is able to return the results in most of benchmarks.


\section{Conclusions and Future Work}
\label{conclusions:section}
We have presented a method for solving the $\contextnum$-bounded reachability 
problem for concurrent program running under the POWER semantics.
To that end, we have presented a code-to-code scheme that translates the input
program  into an output program whose size is polynomial in the size
of the input program, and that reaches the same set of  process states
when run under the classical SC semantics.
On the theoretical side, this shows the decidability
of the $\contextnum$-bounded state reachability problem under POWER for finite-state programs.
On the practical side, our tool implementation demonstrates that the method
is efficient both in performance and in the ability to detect errors.

We aim at extending our framework to cover other models such as
ARM~\cite{DBLP:conf/popl/FlurGPSSMDS16,DBLP:journals/pacmpl/PulteFDFSS18} and C11~\cite{DBLP:conf/popl/BattyOSSW11}.
Furthermore, our current context-bounded approach for POWER
can  
only work with backend model checking tools that  
supports non-deterministic choice of data.
In order to work with various model checking tools (e.g.\ {\sf SPIN}~\cite{Holzmann:2011:SMC:2029108}),
we might need to consider other under-approximation techniques, and in particular
to consider notions of context and code-to-code translation scheme that are different from the ones we use in this paper.

\section{Acknowledgments}
We thank the LMCS reviewers for helpful feedback.
This work was supported in part by the Swedish Research Council and carried out within the Linnaeus centre of excellence UPMARC, Uppsala Programming for Multicore Architectures Research Center.

\bibliographystyle{plainurl} 
\bibliography{biblio}

\end{document}

%% file: init.tex
\usepackage{amssymb,amsfonts,latexsym,stmaryrd,listings,todonotes,color,pifont,stmaryrd,wrapfig,listings,mathbbol,bbm}
\usepackage[linesnumbered,ruled,vlined]{algorithm2e}

\usepackage{pgf}
\usepackage{tikz}

\usetikzlibrary{automata,positioning}
\usetikzlibrary{trees}
\usetikzlibrary{shapes}
\usetikzlibrary{petri}
\usetikzlibrary{arrows}
\usetikzlibrary{backgrounds}
\usetikzlibrary{calc}
\usetikzlibrary{fit}
\usetikzlibrary{decorations.pathmorphing}
\usetikzlibrary{decorations.text}
\usetikzlibrary{shapes.callouts}

\usepackage{hyperref}

\usepackage[noend]{algpseudocode}

\usepackage{multirow}
\usepackage{soul}
\usepackage{subfig}
\usepackage[makeroom]{cancel}

\usepackage{marvosym}
\usepackage{xcolor,cancel}
\usepackage{ifpdf}
\usepackage{float}
\usepackage{array}


%% file: comm.tex
\newcommand\set[1]{\left\{#1\right\}}
\newcommand\setcomp[2]{\left\{#1\ | \ #2\right\}}
\newcommand\tuple[1]{\left\langle{#1}\right\rangle}
\newcommand\gen{{\tt gen}}
\newcommand\genof[1]{\gen\left(#1\right)}
\newcommand\mapingsover[2]{\left[{#1}\rightarrow{#2}\right]}
\newcommand\fun{f}

\newcommand\myundef\bot
\newcommand\aset{A}
\newcommand\bset{B}
\newcommand\gfun{g}
\newcommand\aelem{a}
\newcommand\belem{b}
\newcommand\ii{i}
\newcommand\jj{j}
\newcommand\kk{k}
\newcommand\nn{n}
\newcommand\xx{x}
\newcommand\wordsover[1]{{#1}^*}
\newcommand\word{w}
\newcommand\app\cdot
\newcommand\sizeof[1]{|{#1}|}
\newcommand\conf{{\mathbb c}}
\newcommand\initconf{{\mathbb c}_{\it init}}
\newcommand\confset{{\mathbb C}}

\newcommand\eventset{{\mathcal E}}
\newcommand\events{{\mathbbm E}}
\newcommand\event{{\mathbbm e}}
\newcommand\revents{\events^{{\tt R}}}
\newcommand\reventsof[1]{\events^{{\tt R}}_{#1}}
\newcommand\aevents{\events^{{\tt A}}}
\newcommand\aeventsof[1]{\events^{{\tt A}}_{#1}}
\newcommand\wevents{\events^{{\tt W}}}
\newcommand\weventsof[1]{\events^{{\tt W}}_{#1}}

\newcommand\initeventof[1]{\event^{{\tt init}}_{#1}}
\newcommand\initeventset{\eventset^{{\tt init}}}
\newcommand\acievents{\events^{{\tt ACI}}}
\newcommand\acieventsof[1]{\events^{{\tt ACI}}_{#1}}
\newcommand\eventsof[1]{\events_{#1}}
\newcommand\eorder{\prec}
\newcommand\plorder{\prec_{{\tt poloc}}}
\newcommand\dataorder{\prec_{{\tt data}}}
\newcommand\ctrlorder{\prec_{{\tt ctrl}}}

\newcommand\closestwrite{{\tt CW}}
\newcommand\closestwriteof[2]{\closestwrite\left(#1,#2\right)}
\newcommand\lblset\Lambda

\newcommand\labeling{{\tt lbl}}
\newcommand\labelingof[1]{\labeling\left(#1\right)}
\newcommand\lbl{l}
\newcommand\lblof[1]{{\tt lbl}\left(#1\right)}
\newcommand\termlblof[1]{\lbl^{\keyword\terminated}_{#1}}
\newcommand\instrset{\mathfrak{I}}
\newcommand\instrsetof[1]{\instrset_{#1}}
\newcommand\prog{{\it Prog}}
\newcommand\sprog{\prog^\bullet}
\newcommand\procset{{\mathcal P}}
\newcommand\procof[1]{{\tt proc}\left(#1\right)}
\newcommand\proc{{p}}

\newcommand\sproc{p^\bullet}
\newcommand\pproc\proc
\newcommand\qproc{{q}}
\newcommand\rfrom{{\tt rf}}
\newcommand\rfromof[1]{\rfrom\left(#1\right)}
\newcommand\status{{\tt status}}
\newcommand\statusof[1]{\status\left({#1}\right)}
\newcommand\fetch{{\tt fetch}}
\newcommand\init{{\tt init}}
\newcommand\commit{{\tt com}}
\newcommand\propagated{{\tt Prop}}
\newcommand\propagatedof[2]{\propagated\left(#1,#2\right)}
\newcommand\allregs{{\mathcal R}}

\newcommand\regset{{\mathcal R}}
\newcommand\regsetof[1]{\regset\left(#1\right)}
\newcommand\reg{\$r}
\newcommand\varset{{\mathcal X}}
\newcommand\xvar{x}
\newcommand\yvar{y}
\newcommand\dataset{{\mathcal D}}

\newcommand\zero{0}
\newcommand\expr{{\it exp}}
\newcommand\keyword[1]{{\tt #1}}
\newcommand\varof[1]{{\tt var}\left(#1\right)}
\newcommand\exprof[1]{{\tt exp}\left(#1\right)}
\newcommand\addrof[1]{{\tt addr}\left(#1\right)}
\newcommand\terminated{term}
\newcommand\corder{\prec_{\tt co}}
\newcommand\cordereq{\preceq_{\tt co}}
\newcommand\addtocorder\oplus
\newcommand\nextof[1]{{\tt next}\left(#1\right)}
\newcommand\tnextof[1]{{\tt Tnext}\left(#1\right)}
\newcommand\fnextof[1]{{\tt Fnext}\left(#1\right)}
\newcommand\assigned\leftarrow
\newcommand{\movesto}[1]{\xrightarrow{#1}{}}

\newcommand\maxinstrof[2]{{\tt MaxI}\left(#1,#2\right)}
\newcommand\instr{\mathfrak{i}}

\newcommand\initinstrof[1]{\instr^{{\it init}}_{#1}}
\newcommand\stmt{\mathfrak{s}}
\newcommand\ilabeling{{\tt ins}}
\newcommand\instrof[1]{\ilabeling\left(#1\right)}
\newcommand\stmtof[1]{{\tt stmt}\left(#1\right)}
\newcommand\commitcnd{{\tt ComCnd}}
\newcommand\commitcndof[2]{\commitcnd\left(#1,#2\right)}
\newcommand\validcnd{{\tt ValidCnd}}
\newcommand\validcndof[2]{\validcnd\left(#1,#2\right)}
\newcommand\readcnd{{\tt RdCnd}}
\newcommand\readcndof[2]{\readcnd\left(#1,#2\right)}
\newcommand\writeinitcnd{{\tt InitCnd}}
\newcommand\writeinitcndof[2]{\writeinitcnd\left(#1,#2\right)}
\newcommand\assume[1]{{\tt assume}\left(#1\right)}

\newcommand\run\pi
\newcommand\srun{\run^\bullet}
\newcommand\procseq\sigma
\newcommand\eprocof[1]{{#1}\!\uparrow}
\newcommand\lastof[1]{{\tt last}\left(#1\right)}

\newcommand\contextnum{{\mathbbm K}}
\newcommand\ncontextnum{{\mathbbm \contextnum^{\one\two}}}
\newcommand\stst{\tau_{{\tt sync}}}
\newcommand\ststof[1]{\stst\left(#1\right)}
\newcommand\tst\tau
\newcommand\tstof[1]{\tst\left(#1\right)}
\newcommand\tstset{{\mathbb T}}
\newcommand\tstorder\sqsubseteq
\newcommand\ststorder\sqsubset
\newcommand\never\undefined
\newcommand\tstlub\sqcup
\newcommand\tstseq\sigma
\newcommand\gtst{\alpha^{\it init}}
\newcommand\gtstof[3]{\gtst\left(#1,#2,#3\right)}
\newcommand\utst{\alpha}
\newcommand\utstof[3]{\utst\left(#1,#2,#3\right)}
\newcommand\summary\oplus
\newcommand\ntst{\beta}
\newcommand\ntstof[1]{\ntst\left(#1\right)}
\newcommand\cval{\mu}
\newcommand\cvalof[3]{\cval\left(#1,#2,#3\right)}
\newcommand\initcval{\mu^{\it init}}
\newcommand\initcvalof[3]{\initcval\left(#1,#2,#3\right)}
\newcommand\lval{\nu}
\newcommand\lvalof[2]{\lval\left(#1,#2\right)}
\newcommand\myactive{{\tt active}}
\newcommand\activeof[1]{\myactive\left(#1\right)}
\newcommand\iwrite{{\tt iW}}
\newcommand\iwriteof[2]{\iwrite\left(#1,#2\right)}
\newcommand\cwrite{{\tt cW}}
\newcommand\cwriteof[2]{\cwrite\left(#1,#2\right)}
\newcommand\oldiread{{\tt old\text{-}iR}}

\newcommand\iread{{\tt iR}}
\newcommand\ireadof[2]{\iread\left(#1,#2\right)}
\newcommand\cread{{\tt cR}}
\newcommand\oldcread{{\tt old\text{-}cR}}

\newcommand\creadof[2]{\cread\left(#1,#2\right)}
\newcommand\ireg{{\tt iReg}}
\newcommand\iregof[1]{\ireg\left(#1\right)}
\newcommand\creg{{\tt cReg}}
\newcommand\cregof[1]{\creg\left(#1\right)}
\newcommand\ctrl{{\tt ctrl}}
\newcommand\ctrlof[1]{\ctrl\left(#1\right)}

\newcommand\oldcwrite{{\tt old\text{-}cW}}

\newcommand\ccontext{{\it cnt}}
\newcommand\true{{\it true}}
\newcommand\false{{\it false}}
\newcommand\val{{\tt Val}}
\newcommand\valof[2]{\val\left(#1,#2\right)}

\makeatletter
\newcommand{\removelatexerror}{\let\@latex@error\@gobble}
\makeatother

\newcommand{\translateof}[2]{\llbracket{#1}\rrbracket_{#2}}
\newcommand\myeq{\mathrel{\overset{\makebox[0pt]{\mbox{\normalfont\tiny\sffamily def}}}{=}}}

\newcommand{\one}{\mbox{\ding{192}}}
\newcommand{\two}{\mbox{\ding{193}}}

\tikzset{background rectangle/.style={fill=gray!10,rounded corners}}

\newcommand\hcancel[2][black]{\setbox0=\hbox{$#2$}%
\rlap{\raisebox{.20\ht0}{\textcolor{#1}{\rule{\wd0}{1pt}}}}#2}

\newcommand\addressorder{\prec_{{\tt addr}}}
\newcommand\synpropagated{{\tt SyncProp}}
\newcommand\synpropagatedof[1]{\synpropagated\left(#1\right)}

\newcommand\asyngroup{{\tt SeenSyncs}}
\newcommand\asyngroupof[1]{\asyngroup\left(#1\right)}

\newcommand\astoregroup{{\tt SeenWr}}
\newcommand\astoregroupof[2]{\astoregroup\left(#1,#2\right)}

\newcommand\syncevents{\events^{{\tt SS}}}
\newcommand\synceventsof[1]{\syncevents_{#1}}
\newcommand\lwsyncevents{\events^{{\tt LS}}}
\newcommand\lwsynceventsof[1]{\lwsyncevents_{#1}}
\newcommand\isyncevents{\events^{{\tt IS}}}
\newcommand\isynceventsof[1]{\isyncevents_{#1}}

\newcommand\syncinstr{{\tt sync}}
\newcommand\lwsyncinstr{{\tt lwsync}}
\newcommand\isyncinstr{{\tt isync}}

\newcommand\allsynscnd{{\tt AllSyncCnd}}
\newcommand\allsyncscndof[2]{\allsynscnd\left(#1,#2\right)}

\newcommand\propsyncnd{{\tt PropSyncs}}
\newcommand\propsyncndof[2]{\propsyncnd\left(#1,#2\right)}
\newcommand\comlwsyncnd{{\tt ComLwsyncs}}
\newcommand\comlwsyncndof[2]{\comlwsyncnd\left(#1,#2\right)}
\newcommand\comisyncnd{{\tt ComIsyncs}}
\newcommand\comisyncndof[2]{\comisyncnd\left(#1,#2\right)}

\newcommand\comreadwritecnd{{\tt ComRdWrCnd}}
\newcommand\comreadwritecndof[2]{\comreadwritecnd\left(#1,#2\right)}

\newcommand\defaddresscnd{{\tt AddrRdWrCnd}}
\newcommand\defaddresscndof[2]{\defaddresscnd\left(#1,#2\right)}

\newcommand\syncgroupcnd{{\tt SeenSyncCnd}}
\newcommand\syncgroupcndof[3]{\syncgroupcnd\left(#1,#2,#3\right)}

\newcommand\writegroupcnd{{\tt SeenWrCnd}}
\newcommand\writegroupcndof[3]{\writegroupcnd\left(#1,#2,#3\right)}

\newcommand\newvarof[2]{{\tt Var}\left(#1,#2\right)}

\newcommand\undefined{\bot}
\newcommand\undetermined{\top}

\newcommand\readinitcnd{{\tt RdInitCnd}}
\newcommand\readinitcndof[2]{\readinitcnd\left(#1,#2\right)}

\newcommand\gsyncstst{\gamma^{{\it init}}}
\newcommand\gsyncststof[2]{\gsyncstst\left(#1,#2\right)}
\newcommand\usyncstst{\gamma}
\newcommand\usyncststof[2]{\usyncstst\left(#1,#2\right)}

\newcommand\nsyncstst{\delta}
\newcommand\nsyncststof[1]{\nsyncstst\left(#1\right)}
\newcommand\isync{{\tt isync}}
\newcommand\isyncof[1]{\isync\left(#1\right)}
\newcommand\hwsync{{\tt sync}}
\newcommand\hwsyncof[1]{\hwsync\left(#1\right)}
\newcommand\lwsync{{\tt lwsync}}
\newcommand\lwsyncof[1]{{\tt lsync}\left(#1\right)}

\newcommand\maxcrcw{{\tt maxAddrCR}}
\newcommand\maxcrcwof[1]{\maxcrcw\left(#1\right)}

\newcommand\acksync{{\tt ack}}
\newcommand\acksyncof[1]{\acksync\left(#1\right)}

\newcommand\oldclwsync{{\tt old\text{-}lsync}}
\newcommand\oldcisync{{\tt old\text{-}isync}}

\newcolumntype{L}[1]{>{\raggedright\let\newline\\\arraybackslash\hspace{0pt}}m{#1}}
\newcolumntype{C}[1]{>{\centering\let\newline\\\arraybackslash\hspace{0pt}}m{#1}}
\newcolumntype{R}[1]{>{\raggedleft\let\newline\\\arraybackslash\hspace{0pt}}m{#1}}

\newcommand{\projectof}[1]{{#1\!\downarrow}}

%% file: main.bbl
\begin{thebibliography}{10}

\bibitem{gotoinstrument}
{Goto-Instrument}.
\newblock https://www.cs.ox.ac.uk/people/vincent.nimal/instrument/manual.shtml,
  2013.

\bibitem{svcomp17}
{SV-COM17 benchmark suit}.
\newblock https://sv-comp.sosy-lab.org/2017/benchmarks.php, 2017.

\bibitem{tacas15:tso}
Parosh~Aziz Abdulla, Stavros Aronis, Mohamed~Faouzi Atig, Bengt Jonsson, Carl
  Leonardsson, and Konstantinos~F. Sagonas.
\newblock Stateless model checking for {TSO} and {PSO}.
\newblock In {\em TACAS}, volume 9035 of {\em LNCS}, pages 353--367. Springer,
  2015.

\bibitem{DBLP:conf/pldi/AbdullaAAK19}
Parosh~Aziz Abdulla, Jatin Arora, Mohamed~Faouzi Atig, and Shankara~Narayanan
  Krishna.
\newblock Verification of programs under the release-acquire semantics.
\newblock In {\em {PLDI} 2019}, pages 1117--1132, 2019.

\bibitem{DBLP:conf/concur/AbdullaABN16}
Parosh~Aziz Abdulla, Mohamed~Faouzi Atig, Ahmed Bouajjani, and Tuan~Phong Ngo.
\newblock The benefits of duality in verifying concurrent programs under {TSO}.
\newblock In {\em {CONCUR}}, volume~59 of {\em LIPIcs}, pages 5:1--5:15.
  Schloss Dagstuhl - Leibniz-Zentrum fuer Informatik, 2016.

\bibitem{abdullaABN17}
Parosh~Aziz Abdulla, Mohamed~Faouzi Atig, Ahmed Bouajjani, and Tuan~Phong Ngo.
\newblock Context-bounded analysis for {POWER}.
\newblock In {\em {TACAS} 2017}, pages 56--74, 2017.

\bibitem{lmcs:4228}
Parosh~Aziz Abdulla, Mohamed~Faouzi Atig, Ahmed Bouajjani, and Tuan~Phong Ngo.
\newblock {A Load-Buffer Semantics for Total Store Ordering}.
\newblock {\em {LMCS} 2018}, {Volume 14, Issue 1}, January 2018.

\bibitem{DBLP:conf/sas/AbdullaACLR12}
Parosh~Aziz Abdulla, Mohamed~Faouzi Atig, Yu{-}Fang Chen, Carl Leonardsson, and
  Ahmed Rezine.
\newblock Automatic fence insertion in integer programs via predicate
  abstraction.
\newblock In {\em {SAS} 2012}, pages 164--180, 2012.

\bibitem{DBLP:conf/tacas/AbdullaACLR12}
Parosh~Aziz Abdulla, Mohamed~Faouzi Atig, Yu{-}Fang Chen, Carl Leonardsson, and
  Ahmed Rezine.
\newblock Counter-example guided fence insertion under {TSO}.
\newblock In {\em {TACAS} 2012}, volume 7214 of {\em LNCS}, pages 204--219.
  Springer, 2012.

\bibitem{DBLP:conf/cav/AbdullaAJL16}
Parosh~Aziz Abdulla, Mohamed~Faouzi Atig, Bengt Jonsson, and Carl Leonardsson.
\newblock Stateless model checking for {POWER}.
\newblock In {\em {CAV}}, volume 9780 of {\em LNCS}, pages 134--156. Springer,
  2016.

\bibitem{AbdullaAJN18}
Parosh~Aziz Abdulla, Mohamed~Faouzi Atig, Bengt Jonsson, and Tuan~Phong Ngo.
\newblock Optimal stateless model checking under the release-acquire semantics.
\newblock {\em {PACMPL}}, 2({OOPSLA}):135:1--135:29, 2018.

\bibitem{DBLP:journals/pacmpl/AbdullaAJN18}
Parosh~Aziz Abdulla, Mohamed~Faouzi Atig, Bengt Jonsson, and Tuan~Phong Ngo.
\newblock Optimal stateless model checking under the release-acquire semantics.
\newblock {\em {PACMPL}}, 2({OOPSLA}):135:1--135:29, 2018.

\bibitem{DBLP:conf/netys/AbdullaALN15}
Parosh~Aziz Abdulla, Mohamed~Faouzi Atig, Magnus L{\aa}ng, and Tuan~Phong Ngo.
\newblock Precise and sound automatic fence insertion procedure under {PSO}.
\newblock In {\em {NETYS} 2015}, pages 32--47, 2015.

\bibitem{DBLP:conf/esop/AbdullaAP15}
Parosh~Aziz Abdulla, Mohamed~Faouzi Atig, and Ngo~Tuan Phong.
\newblock The best of both worlds: Trading efficiency and optimality in fence
  insertion for {TSO}.
\newblock In {\em {ESOP}}, volume 9032 of {\em LNCS}, pages 308--332. Springer,
  2015.

\bibitem{DBLP:conf/esop/AlglaveKNT13}
J.~Alglave, D.~Kroening, V.~Nimal, and M.~Tautschnig.
\newblock Software verification for weak memory via program transformation.
\newblock In {\em ESOP}, volume 7792 of {\em LNCS}, pages 512--532. Springer,
  2013.

\bibitem{AlglaveKT13}
J.~Alglave, D.~Kroening, and M.~Tautschnig.
\newblock Partial orders for efficient bounded model checking of concurrent
  software.
\newblock In {\em CAV}, volume 8044 of {\em LNCS}, pages 141--157, 2013.

\bibitem{DBLP:journals/toplas/AlglaveMT14}
Jade Alglave, Luc Maranget, and Michael Tautschnig.
\newblock Herding cats: Modelling, simulation, testing, and data mining for
  weak memory.
\newblock {\em {ACM} TOPLAS}, 36(2):7:1--7:74, 2014.

\bibitem{ABP2011}
M.~F. Atig, A.~Bouajjani, and G.~Parlato.
\newblock Getting rid of store-buffers in {TSO} analysis.
\newblock In {\em CAV}, volume 6806 of {\em LNCS}, pages 99--115. Springer,
  2011.

\bibitem{DBLP:conf/popl/BattyOSSW11}
Mark Batty, Scott Owens, Susmit Sarkar, Peter Sewell, and Tjark Weber.
\newblock Mathematizing {C++} concurrency.
\newblock In {\em Proceedings of the 38th {ACM} {SIGPLAN-SIGACT} Symposium on
  Principles of Programming Languages, {POPL} 2011, Austin, TX, USA, January
  26-28, 2011}, pages 55--66, 2011.

\bibitem{DBLP:conf/esop/BouajjaniDM13}
Ahmed Bouajjani, Egor Derevenetc, and Roland Meyer.
\newblock Checking and enforcing robustness against {TSO}.
\newblock In {\em ESOP}, volume 7792 of {\em LNCS}, pages 533--553. Springer,
  2013.

\bibitem{BAM07}
S.~Burckhardt, R.~Alur, and M.~M.~K. Martin.
\newblock {CheckFence}: checking consistency of concurrent data types on
  relaxed memory models.
\newblock In {\em PLDI}, pages 12--21. ACM, 2007.

\bibitem{BM2008}
Sebastian Burckhardt and Madanlal Musuvathi.
\newblock Effective program verification for relaxed memory models.
\newblock In {\em CAV}, volume 5123 of {\em LNCS}, pages 107--120. Springer,
  2008.

\bibitem{BSS2011}
Jacob Burnim, Koushik Sen, and Christos Stergiou.
\newblock Testing concurrent programs on relaxed memory models.
\newblock In {\em ISSTA}, pages 122--132. ACM, 2011.

\bibitem{Clarke2012}
Edmund~M. Clarke, William Klieber, Milo{\v{s}} Nov{\'a}{\v{c}}ek, and Paolo
  Zuliani.
\newblock Model checking and the state explosion problem.
\newblock In Bertrand Meyer and Martin Nordio, editors, {\em {LASER}}, pages
  1--30, 2012.

\bibitem{DBLP:conf/tacas/ClarkeKL04}
Edmund~M. Clarke, Daniel Kroening, and Flavio Lerda.
\newblock A tool for checking {ANSI-C} programs.
\newblock In {\em {TACAS}}, volume 2988 of {\em LNCS}, pages 168--176.
  Springer, 2004.

\bibitem{DBLP:conf/sas/DanMVY13}
A.~Marian Dan, Y.~Meshman, M.~T. Vechev, and E.~Yahav.
\newblock Predicate abstraction for relaxed memory models.
\newblock In {\em SAS}, volume 7935 of {\em LNCS}, pages 84--104. Springer,
  2013.

\bibitem{Dan201762}
Andrei Dan, Yuri Meshman, Martin Vechev, and Eran Yahav.
\newblock Effective abstractions for verification under relaxed memory models.
\newblock {\em CLSS}, 47, Part 1:62--76, 2017.

\bibitem{DBLP:conf/oopsla/DemskyL15}
Brian Demsky and Patrick Lam.
\newblock Satcheck: Sat-directed stateless model checking for {SC} and {TSO}.
\newblock In {\em {OOPSLA} 2015}, pages 20--36. {ACM}, 2015.

\bibitem{DM14}
Egor Derevenetc and Roland Meyer.
\newblock Robustness against {Power} is {PSpace}-complete.
\newblock In {\em ICALP (2)}, volume 8573 of {\em LNCS}, pages 158--170.
  Springer, 2014.

\bibitem{DBLP:conf/popl/FlurGPSSMDS16}
Shaked Flur, Kathryn~E. Gray, Christopher Pulte, Susmit Sarkar, Ali Sezgin, Luc
  Maranget, Will Deacon, and Peter Sewell.
\newblock Modelling the armv8 architecture, operationally: concurrency and
  {ISA}.
\newblock In {\em {POPL} 2016}, pages 608--621, 2016.

\bibitem{GavrilenkoLFHM19}
Natalia Gavrilenko, Hern{'{a}}n {Ponce de Le{'{o}}n}, Florian Furbach, Keijo
  Heljanko, and Roland Meyer.
\newblock {BMC} for weak memory models: Relation analysis for compact {SMT}
  encodings.
\newblock In {\em {CAV} 2019}, pages 355--365, 2019.

\bibitem{Holzmann:2011:SMC:2029108}
Gerard Holzmann.
\newblock {\em The SPIN Model Checker: Primer and Reference Manual}.
\newblock Addison-Wesley Professional, 1st edition, 2011.

\bibitem{Huang016}
Shiyou Huang and Jeff Huang.
\newblock Maximal causality reduction for {TSO} and {PSO}.
\newblock In {\em {OOPSLA} 2016}, pages 447--461, 2016.

\bibitem{InversoT0TP14}
Omar Inverso, Ermenegildo Tomasco, Bernd Fischer, Salvatore {La Torre}, and
  Gennaro Parlato.
\newblock Bounded model checking of multi-threaded {C} programs via lazy
  sequentialization.
\newblock In {\em {CAV} 2014}, pages 585--602, 2014.

\bibitem{Kokologiannakis18}
Michalis Kokologiannakis, Ori Lahav, Konstantinos Sagonas, and Viktor
  Vafeiadis.
\newblock Effective stateless model checking for {C/C++} concurrency.
\newblock {\em {PACMPL} 2018}, 2({POPL}):17:1--17:32, 2018.

\bibitem{DBLP:conf/pldi/Kokologiannakis19}
Michalis Kokologiannakis, Azalea Raad, and Viktor Vafeiadis.
\newblock Model checking for weakly consistent libraries.
\newblock In {\em {PLDI} 2019}, pages 96--110, 2019.

\bibitem{KVY2010}
Michael Kuperstein, Martin~T. Vechev, and Eran Yahav.
\newblock Automatic inference of memory fences.
\newblock In {\em FMCAD}, pages 111--119. IEEE, 2010.

\bibitem{KVY2011}
Michael Kuperstein, Martin~T. Vechev, and Eran Yahav.
\newblock Partial-coherence abstractions for relaxed memory models.
\newblock In {\em PLDI}, pages 187--198. ACM, 2011.

\bibitem{DBLP:conf/cav/TorreMP09}
Salvatore {La Torre}, P.~Madhusudan, and Gennaro Parlato.
\newblock Reducing context-bounded concurrent reachability to sequential
  reachability.
\newblock In {\em {CAV}}, volume 5643 of {\em LNCS}, pages 477--492. Springer,
  2009.

\bibitem{DBLP:conf/pldi/LahavM19}
Ori Lahav and Roy Margalit.
\newblock Robustness against release/acquire semantics.
\newblock In {\em {PLDI} 2019}, pages 126--141, 2019.

\bibitem{DBLP:conf/fm/LahavV16}
Ori Lahav and Viktor Vafeiadis.
\newblock Explaining relaxed memory models with program transformations.
\newblock In {\em {FM} 2016}, pages 479--495, 2016.

\bibitem{DBLP:journals/fmsd/LalR09}
Akash Lal and Thomas~W. Reps.
\newblock Reducing concurrent analysis under a context bound to sequential
  analysis.
\newblock {\em FMSD}, 35(1):73--97, 2009.

\bibitem{lamport-79}
L.~Lamport.
\newblock How to make a multiprocessor computer that correctly executes
  multiprocess programs.
\newblock {\em IEEE Trans. Comp.}, C-28(9), 1979.

\bibitem{DBLP:conf/pldi/LiuNPVY12}
Feng Liu, Nayden Nedev, Nedyalko Prisadnikov, Martin~T. Vechev, and Eran Yahav.
\newblock Dynamic synthesis for relaxed memory models.
\newblock In {\em {PLDI} 2012}, pages 429--440. {ACM}, 2012.

\bibitem{DBLP:conf/cav/Mador-HaimMSMAOAMSW12}
Sela Mador{-}Haim, Luc Maranget, Susmit Sarkar, Kayvan Memarian, Jade Alglave,
  Scott Owens, Rajeev Alur, Milo M.~K. Martin, Peter Sewell, and Derek
  Williams.
\newblock An axiomatic memory model for {POWER} multiprocessors.
\newblock In {\em {CAV}}, volume 7358, pages 495--512. Springer, 2012.

\bibitem{MQ07}
Madanlal Musuvathi and Shaz Qadeer.
\newblock Iterative context bounding for systematic testing of multithreaded
  programs.
\newblock In {\em PLDI}, pages 446--455. ACM, 2007.

\bibitem{Nguyen0TP16}
Truc~L. Nguyen, Bernd Fischer, Salvatore {La Torre}, and Gennaro Parlato.
\newblock Lazy sequentialization for the safety verification of unbounded
  concurrent programs.
\newblock In {\em {ATVA} 2016}, pages 174--191, 2016.

\bibitem{NorrisD16}
Brian Norris and Brian Demsky.
\newblock A practical approach for model checking {C/C++11} code.
\newblock {\em {TOPLAS} 2016}, 38(3):10:1--10:51, 2016.

\bibitem{DBLP:journals/toplas/NorrisD16}
Brian Norris and Brian Demsky.
\newblock A practical approach for model checking {C/C++11} code.
\newblock {\em {TOPLAS}}, 38(3):10:1--10:51, 2016.

\bibitem{DBLP:conf/ppopp/OuD17}
Peizhao Ou and Brian Demsky.
\newblock Checking concurrent data structures under the {C/C++11} memory model.
\newblock In {\em {PPOPP} 2017}, pages 45--59, 2017.

\bibitem{DBLP:conf/tphol/OwensSS09}
Scott Owens, Susmit Sarkar, and Peter Sewell.
\newblock A better x86 memory model: x86-tso.
\newblock In {\em TPHOLs}, volume 5674 of {\em LNCS}, pages 391--407. Springer,
  2009.

\bibitem{LeonFHM17}
Hern{'{a}}n {Ponce de Le{'{o}}n}, Florian Furbach, Keijo Heljanko, and Roland
  Meyer.
\newblock Portability analysis for weak memory models. {PORTHOS:} one tool for
  all models.
\newblock In {\em {SAS} 2017}, pages 299--320, 2017.

\bibitem{LeonFHM18}
Hern{'{a}}n {Ponce de Le{'{o}}n}, Florian Furbach, Keijo Heljanko, and Roland
  Meyer.
\newblock {BMC} with memory models as modules.
\newblock In {\em {FMCAD} 2018}, pages 1--9, 2018.

\bibitem{DBLP:journals/pacmpl/PulteFDFSS18}
Christopher Pulte, Shaked Flur, Will Deacon, Jon French, Susmit Sarkar, and
  Peter Sewell.
\newblock Simplifying {ARM} concurrency: multicopy-atomic axiomatic and
  operational models for armv8.
\newblock {\em {PACMPL}}, 2({POPL}):19:1--19:29, 2018.

\bibitem{Qadeer08}
S.~Qadeer.
\newblock The case for context-bounded verification of concurrent programs.
\newblock In {\em SPIN}, volume 5156 of {\em Lecture Notes in Computer
  Science}. Springer, 2008.

\bibitem{DBLP:conf/tacas/QadeerR05}
Shaz Qadeer and Jakob Rehof.
\newblock Context-bounded model checking of concurrent software.
\newblock In {\em {TACAS}}, volume 3440 of {\em LNCS}, pages 93--107. Springer,
  2005.

\bibitem{DBLP:journals/pacmpl/RaadDRLV19}
Azalea Raad, Marko Doko, Lovro Rozic, Ori Lahav, and Viktor Vafeiadis.
\newblock On library correctness under weak memory consistency: specifying and
  verifying concurrent libraries under declarative consistency models.
\newblock {\em {PACMPL}}, 3({POPL}):68:1--68:31, 2019.

\bibitem{DBLP:conf/pldi/SarkarMOBSMAW12}
Susmit Sarkar, Kayvan Memarian, Scott Owens, Mark Batty, Peter Sewell, Luc
  Maranget, Jade Alglave, and Derek Williams.
\newblock Synchronising {C/C++} and {POWER}.
\newblock In {\em {PLDI} 2012}, pages 311--322, 2012.

\bibitem{DBLP:conf/pldi/SarkarSAMW11}
Susmit Sarkar, Peter Sewell, Jade Alglave, Luc Maranget, and Derek Williams.
\newblock Understanding {POWER} multiprocessors.
\newblock In {\em {PLDI}}, pages 175--186. {ACM}, 2011.

\bibitem{SSONM2010}
P.~Sewell, S.~Sarkar, S.~Owens, F.~Z. Nardelli, and M.~O. Myreen.
\newblock x86-tso: A rigorous and usable programmer's model for x86
  multiprocessors.
\newblock {\em CACM}, 53, 2010.

\bibitem{TomascoI0TP15a}
Ermenegildo Tomasco, Omar Inverso, Bernd Fischer, Salvatore {La Torre}, and
  Gennaro Parlato.
\newblock Verifying concurrent programs by memory unwinding.
\newblock In {\em {TACAS} 2015}, pages 551--565, 2015.

\bibitem{eps402285}
Ermenegildo Tomasco, Truc~Nguyen Lam, Bernd Fischer, Salvatore~La Torre, and
  Gennaro Parlato.
\newblock Embedding weak memory models within eager sequentialization.
\newblock October 2016.

\bibitem{fmcad16}
Ermenegildo Tomasco, Truc~Nguyen Lam, Omar Inverso, Bernd Fischer, Salvatore~La
  Torre, and Gennaro Parlato.
\newblock Lazy sequentialization for tso and pso via shared memory
  abstractions.
\newblock In {\em {FMCAD}16}, pages 193--200, 2016.

\bibitem{TomascoN0TP17}
Ermenegildo Tomasco, Truc~Lam Nguyen, Bernd Fischer, Salvatore {La Torre}, and
  Gennaro Parlato.
\newblock Using shared memory abstractions to design eager sequentializations
  for weak memory models.
\newblock In {\em {SEFM} 2017}, pages 185--202, 2017.

\bibitem{DBLP:conf/ictac/TravkinW16}
Oleg Travkin and Heike Wehrheim.
\newblock Verification of concurrent programs on weak memory models.
\newblock In {\em {ICTAC} 2016}, pages 3--24, 2016.

\bibitem{yang-gopalakrishnan-PDPS04}
Y.~Yang, G.~Gopalakrishnan, G.~Lindstrom, and K.~Slind.
\newblock Nemos: A framework for axiomatic and executable specifications of
  memory consistency models.
\newblock In {\em IPDPS}. IEEE, 2004.

\bibitem{Zhang:pldi15}
N.~Zhang, M.~Kusano, and C.~Wang.
\newblock Dynamic partial order reduction for relaxed memory models.
\newblock In {\em PLDI}, pages 250--259. ACM, 2015.

\end{thebibliography}
